\documentclass[11pt]{article}
\pdfoutput=1
\usepackage{dcolumn}% Align table columns on decimal point
\usepackage{bm}% bold math
\usepackage{graphicx}% Include figure files
\usepackage{amssymb,amsmath}
\usepackage{multirow}
\usepackage{cite,color,url}
\usepackage[colorlinks=true
,urlcolor=blue
,anchorcolor=blue
,citecolor=blue
,filecolor=blue
,linkcolor=blue
,menucolor=blue
,linktocpage=true
,pdfproducer=medialab
,pdfa=true
]{hyperref}

\usepackage{slashed}
\usepackage{epsfig,psfrag,rotating,soul}
\usepackage{rotfloat}
\usepackage[font={small}]{caption}
\usepackage{colortbl}
\evensidemargin \oddsidemargin
\allowdisplaybreaks

\let\OLDthebibliography\thebibliography
\renewcommand\thebibliography[1]{
  \OLDthebibliography{#1}
  \setlength{\parskip}{0pt}
  \setlength{\itemsep}{0pt plus 0.3ex}
}
\def\Etmiss{E_T\hspace{-16pt}\not{}\hspace{12pt}}
\def\Etmiss{E_T^{\rm miss}}

\def\stau{\widetilde{\tau}}
\def\avstau{\overline{m}_{\stau_{12}}}

\definecolor{mygray}{gray}{0.85} 
\definecolor{myblue}{cmyk}{0.65, 0.37, 0.0, 0.19}

\begin{document}
\thispagestyle{empty}

\def\thefootnote{\fnsymbol{footnote}}

\begin{flushright}
IFT-UAM/CSIC-21-4
\end{flushright}

\vspace*{1cm}

\begin{center}

\begin{Large}
\textbf{\textsc{Discovery and exclusion prospects for staus \\[.25em] produced by heavy Higgs bosons decays at the LHC}}
\end{Large}

\vspace{1cm}

{\sc
Ernesto~Arganda$^{1, 2}$%
\footnote{{\tt \href{mailto:ernesto.arganda@csic.es}{ernesto.arganda@csic.es}}}%
, Victor~Mart\'{	\i}n-Lozano$^{3}$%
\footnote{{\tt \href{mailto:victor.lozano@desy.de}{victor.lozano@desy.de}}}, Anibal D.~Medina$^{2}$%
\footnote{{\tt \href{mailto:anibal.medina@fisica.unlp.edu.ar}{anibal.medina@fisica.unlp.edu.ar}}}%
and Nicolas~I.~Mileo$^{2}$%
\footnote{{\tt \href{mailto:mileo@fisica.unlp.edu.ar}{mileo@fisica.unlp.edu.ar}}}%

}

\vspace*{.7cm}

{\sl
$^1$Instituto de F\'{\i}sica Te\'orica UAM/CSIC, \\
C/ Nicol\'as Cabrera 13-15, Campus de Cantoblanco, 28049, Madrid, Spain

\vspace*{0.1cm}

$^2$IFLP, CONICET - Dpto. de F\'{\i}sica, Universidad Nacional de La Plata, \\ 
C.C. 67, 1900 La Plata, Argentina

\vspace*{0.1cm}

$^3$DESY,  Notkestra{\ss}e  85,  22607  Hamburg,  Germany
}

\end{center}

\vspace{0.1cm}

\begin{abstract}
\noindent
In a previous work we developed a search strategy for staus produced by the decay of the heavy CP-even Higgs boson $H$ within the context of the large $\tan\beta$ regime of the minimal supersymmetric standard model (MSSM) in an scenario of large stau mixing. Here we study the performance of such search strategy by confronting it with the complementary mixing pattern in which decays of both the CP-even and CP-odd heavy Higgs bosons contribute to the production of $\widetilde{\tau}_1\widetilde{\tau}_2^{*} \;+\; c.c$ pairs. Again, we focus on final states with two opposite-sign tau leptons and large missing transverse energy. We find that our proposed search strategy, although optimized for the large stau mixing scenario, is still quite sensitive to the complementary mixing pattern. For instance, with a total integrated luminosity of only 100 fb$^{-1}$ we are able to exclude heavy Higgs masses above 850 GeV for average stau masses higher than 290 GeV. We also extend the results reported in the preceding work for the large mixing scenario by including now the exclusion limits at 100 fb$^{-1}$ and the prospects both for exclusion and discovery in a potential high luminosity phase of the LHC (1000 fb$^{-1}$). Finally, we discuss the possibility to distinguish the two mixing scenarios when they share the same relevant mass spectrum and both reach the discovery level with our search strategy.

\end{abstract}

\def\thefootnote{\arabic{footnote}}
\setcounter{page}{0}
\setcounter{footnote}{0}

\newpage

\section{Introduction}
\label{intro}

Among the theories that extend  the standard model (SM) of particle physics, supersymmetry (SUSY) remains as one of the most interesting and promising candidates (for reviews, see, e.g.,~\cite{Haber:1993wf,Martin:1997ns}). From a phenomenological point of view, its minimal version with $R$-parity conservation~\cite{Farrar:1978xj}, the minimal supersymmetric standard model (MSSM)~\cite{Nilles:1983ge,Haber:1984rc,Gunion:1984yn}, has as its main virtues  a solution to the gauge hierarchy problem, a potential unification of SM gauge couplings at high energies and a viable dark-matter candidate, the lightest supersymmetric particle (LSP)~\cite{Goldberg:1983nd,Ellis:1983ew}. The MSSM predicts the existence of superpartners (sparticles) for each SM particle: squarks/sleptons, gauginos and higgsinos are the companions of quarks/leptons and gauge and Higgs bosons, respectively. The MSSM Higgs sector contains two scalar doublets that under the assumption of a CP-conserving potential leads to a physical spectrum that includes three neutral Higgs bosons (a light scalar $h$, a heavy scalar $H$, and a heavy pseudoscalar $A$) and a pair of charged Higgs bosons ($H^\pm$), of which the 125-GeV SM-like Higgs boson~\cite{Aad:2012tfa,Chatrchyan:2012ufa} can be easily accommodated as the lightest  CP-even Higgs boson $h$ (see for instance~\cite{Bahl:2018zmf}). Together with the phenomenological signals of these additional Higgs bosons, the existence of sparticles produces a rich phenomenology with characteristic collider signals that are being searched for at the Large Hadron Collider (LHC). 

A particular interesting example are the supersymmetric scalar partners of the tau leptons, the staus,  which are being intensely searched for at the LHC by the ATLAS and CMS collaborations~\cite{Aad:2014yka,Aad:2015eda,ATLAS:2016fhh,CMS:2017wox,CMS:2017rio,CMS:2019eln,Sirunyan:2019mlu,Aad:2019byo}, since in many scenarios where SM gauged  mediators are responsible for the transmission  of SUSY breaking from a hidden sector to the visible sector, staus could be among the lightest sparticles. In a previous work~\cite{Arganda:2018hdn}, we have demonstrated that stau-pair production at the LHC that are originated from the decay of a heavy scalar Higgs boson, and  where the staus subsequently mainly decay into a tau lepton and the LSP neutralino, can be very promising in the large-tan$\beta$ regime within MSSM scenarios with large stau mixing~\cite{Medina:2017bke}.  Indeed, we found that in these regions of the MSSM parameter space, resonant stau-pair production cross sections are 1-2 orders of magnitude larger than the usually considered electroweak (EW) production mechanism. The search strategy we developed was dedicated to scenarios with a stau mixing pattern for which the only relevant Higgs decay channel into staus was $H \to$ $\widetilde{\tau}_1^*\widetilde{\tau}_1$, being $\widetilde{\tau}_1$ the lightest stau. This class of stau mixing pattern occurs when the stau mixing angle is large and the diagonal entries of the stau mass matrix are of similar value. By means of a set of basic cuts, we obtained signal-to-background significances at the discovery level for a LHC center-of-mass energy of 14 TeV and a total integrated luminosity of 100 fb$^{-1}$. For this new work we would like to extend our previous analysis and  show that our search strategy also works very well for scenarios with a complementary stau mixing pattern in which the stau mixing angle is small but the non-diagonal entries of the stau mass matrix are large, mainly due to a  sufficiently large stau trilinear coupling. In these latter scenarios, contrary to the ones analyzed in our previous work, not only the CP-even Higgs contributes to the stau pair production but there is also the CP-odd Higgs contribution which in this case  is non-vanishing, potentially increasing the phenomenological signals. Furthermore, mixed combinations of heaviest and lightest staus are preferable produced  via the heavy Higgs boson decays and thus we have the decay patterns, $H/A \to \widetilde{\tau}_1^*\widetilde{\tau}_2, \widetilde{\tau}_2^*\widetilde{\tau}_1$, where $\widetilde{\tau}_2$ is the heaviest stau, as the main source of staus. Since in this mixing pattern $m_{\widetilde{\tau}_2}$ can be much larger than $m_{\widetilde{\tau}_1}$, we expect that the the collider phenomenology of the stau decays products to be quite different from the patterns analyzed by us before and can potentially help in distinguishing both mixing patterns from each other. From a more general approach, this search strategy could be applied to any process at the LHC with an identical topology, that is, the resonant production of a pair of charged scalars which decay into a tau lepton and an invisible particle, consisting of final states with a $\tau$-lepton pair plus a large amount of missing transverse energy ($E_T^\text{miss}$).

%In other words, we will demonstrate that our search strategy is really efficient for the discovery or exclusion of heavy (scalar or pseudoscalar) Higgs bosons decaying into a pair of staus. 

The paper is organized as follows: Section~\ref{staumssm} is devoted to review the theoretical features of the large stau mixing MSSM scenarios we work with, paying special attention to the main characteristics of the $H/A$ decays into a stau pair. The collider analysis we develop along this work is presented in Section~\ref{sec:collider}, together with the description of our search strategy for stau pairs, originated from heavy Higgs boson resonances and decaying into the lightest neutralino and a tau lepton. A compendium of the obtained results is presented in Section~\ref{sec:results}, for both classes of stau scenarios and with a final study of the potential discrimination between them, leaving Section~\ref{sec:conclusions} for a discussion of our main conclusions.

\section{$H/A$ decays into stau leptons within the MSSM}
\label{staumssm}

As mentioned in the introduction, we work in the context of the MSSM in the large $\tan\beta \gg 1$ limit in which the couplings of the heavy Higgs bosons, $H_k=H,A$ to the down-type sfermions  of mixed chiralities are given by, 
\begin{eqnarray}
g_{H\widetilde{d}_L\widetilde{d}_R}= -\frac{1}{2} m_d [-\mu + A_d\tan\beta], \label{eq:ghdd} \\
g_{A\widetilde{d}_L\widetilde{d}_R}= -\frac{1}{2} m_d [\mu + A_d\tan\beta],\label{eq:gadd}
\end{eqnarray}
where $m_d$ is the mass of the down-type fermion, $\mu$ is the Higgsino mass and $A_d$ is the trilinear coupling given in the soft SUSY breaking Lagrangian. The couplings involving the same chiral states are proportional to SM fermion and gauge boson masses, do not involve soft SUSY breaking parameters and therefore cannot be enhanced~\cite{Djouadi:1996pj}. In contrast, the couplings involving different chiral states depend not only on the SM fermion mass but also on SUSY parameters, namely, the trilinear soft breaking parameter $A_{d}$ and the $\mu$ parameter. In Ref.~\cite{Medina:2017bke} it was shown that in a regime where $\tan\beta \gg 1$ and for large values of $A_d$, the couplings of Eqs.~\eqref{eq:ghdd} and \eqref{eq:gadd} can in fact be enhanced in the case of staus in particular. This is translated into larger branching fractions to staus, ${\rm BR}(H_k\to \sum_{i,j=1,2} \widetilde{\tau}_i^*\widetilde{\tau}_j)$, implying consequentially that the branching ratio to taus, ${\rm BR}(H_k\to \tau^+\tau^-)$, decreases. It is this  what allows for the resurrection of certain regions of the MSSM that seem at first sight to be excluded from di-tau searches at the LHC.  In this regime we can find two scenarios with sizable branching fraction into staus: in one of them $\widetilde{\tau}_1^*\widetilde{\tau}_1$ is the dominant decay mode (Scenario I), while in the other the $\widetilde{\tau}_1^*\widetilde{\tau}_2$ + $c.c.$ mode gives the dominant contribution (Scenario II). Let us take a closer look at these scenarios in terms of the stau mass matrix and the corresponding mixing patterns.

The stau mass matrix is defined as,
\begin{eqnarray}
&\mathcal{M}^2_{\stau}=\left(
\begin{array}{cc}
m_{\stau_{11}} & m_{\stau_{12}} \\
m^{*}_{\stau_{12}} & m_{\stau_{22}}
\end{array}
\right)=&\notag\\
&\left(
\begin{array}{cc}
m_{\tilde L_3}^2 + m_\tau^2 + (-1/2+1/3\sin^2\theta_w)m^2_Z\cos 2\beta & m_\tau(A_\tau -\mu\tan\beta) \\
(m_\tau(A_\tau -\mu\tan\beta))^* & m_{\tilde E_3}^2 + m_\tau^2 + 1/3\sin^2\theta_wm^2_Z\cos 2\beta
\end{array}
\right),&
\end{eqnarray}
where the trilinear coupling, $A_\tau$, comes from the soft Lagrangian term $\mathcal{L}_{\rm soft}\supset y_\tau A_\tau \tilde{\bar{e}}_3 \tilde{L}_3 H_d + c.c.$
The diagonalization of the mass matrix leads to the following mass eigenvalues and eigenstates: 
\begin{eqnarray}
m_{\stau_{1},\stau_{2}}^2= \frac{1}{2}(m_{\stau_{11}}+m_{\stau_{22}}\pm \Delta_{\stau}), \\
\stau_{1}=\stau_L\cos\theta_{\stau} + \stau_R\sin\theta_{\stau}, \\
\stau_{2}=-\stau_L\sin\theta_{\stau} + \stau_R\cos\theta_{\stau}.
\end{eqnarray}
where $\Delta_{\stau}\equiv \sqrt{(m_{\stau_{11}}-m_{\stau_{22}})+4m_{\stau_{12}}}$, assuming that $m^{*}_{\stau_{12}}=m_{\stau_{12}}$, and the mixing angle can be written as
\begin{eqnarray}
\tan 2\theta_{\stau}=\frac{2m_{\stau_{12}}}{m_{\stau_{11}}-m_{\stau_{22}}}.
\label{eq:mixingangle}
\end{eqnarray}

Now that we have defined the stau sector, let us study the decay of a heavy Higgs boson, $H_k=H, A$, into staus. The decay width is given by
\begin{eqnarray}
\Gamma(H_k\to \widetilde{\tau}_i^*\widetilde{\tau}_j)=\frac{G_F}{2\sqrt{2}M_{H_k}}\lambda^{1/2}_{\stau_i\stau_jH_k}g^2_{H_k\stau_i\stau_j}\quad (i,j=1,2),
\end{eqnarray}
where 
\begin{eqnarray}
\lambda_{ijk}=\left(1-\frac{m^2_i}{m_k^2}-\frac{m^2_j}{m_k^2}\right)^2-4\frac{m_i^2 m_j^2}{m^4_k},
\end{eqnarray}
is the kinematic factor in a two-body decay, and $g_{H_k\stau_i\stau_j}$ is the coupling of the Higgs $H_k$ to the staus $\stau_i$ and $\stau_j$. It is important to note that this coupling is a combination of the chiral couplings given in Eqs.~(\ref{eq:ghdd}) and~(\ref{eq:gadd}), and then can  be written as
\begin{eqnarray}
g_{H_k\stau_i\stau_j}=\sum_{\alpha,\beta=L,R} T_{ij\alpha\beta}g_{H_k\stau_\alpha\stau_\beta},
\end{eqnarray}
where the $T_{ij\alpha\beta}$ are the elements of a $4\times 4$ matrix that relates the mass eigenstates with the interaction states.

Scenario I is achieved, as commented above, when the mixing angle is maximal, $\theta_{\stau}\sim \pi/4$. In this case, according to Eq.~\eqref{eq:mixingangle}, it is not only important that $m_{\stau_{12}}$ is large but it is also required that $m_{\stau_{11}}\sim m_{\stau_{22}}$. In such case, there is a cancellation of the contributions that mix chiralities in the coupling $g_{H \widetilde{\tau}_1 \widetilde{\tau}_2}$ involving different staus, leaving them only proportional to the chiral diagonal couplings that as we mentioned before cannot be enhanced. The mass diagonal couplings $g_{H \widetilde{\tau}_1 \widetilde{\tau}_1}$ and $g_{H \widetilde{\tau}_2 \widetilde{\tau}_2}$  on the other hand do not present this cancellation and depend on the parameters that mixes chiralities, which can be increased by our choice of parameters. Therefore, the Higgs couplings to $\widetilde{\tau}_1^*\widetilde{\tau}_2$ and $\widetilde{\tau}_2^*\widetilde{\tau}_1$ suffer a cancellation, while the couplings to $\widetilde{\tau}_1^*\widetilde{\tau}_1$ and $\widetilde{\tau}_2^*\widetilde{\tau}_2$ are maximal. Since in this situation the decays into pairs of heavier staus $\widetilde{\tau}_2$ are usually not kinematically available, the decay of $H$ is dominated by the decays into $\widetilde{\tau}_1^*\widetilde{\tau}_1$. Furthermore, as a consequence of the fact that the mixed-mass couplings are suppressed, the couplings involving the CP-odd Higgs $A$ to staus are also suppressed, and the stau production via Higgs boson decays is dominated by the heaviest CP-even Higgs $H$.

The situation that characterizes Scenario II arises when the mixing angle is small but $m^2_{\stau_{12}}$ is large due to the $A_\tau$ term~\cite{Djouadi:1996pj}. In this case the mixed chiral couplings are maximized, such that the left-right part of the coupling of $H$ to $\widetilde{\tau}_1^*\widetilde{\tau}_2$ and the right-left part of the coupling of $H$ to $\widetilde{\tau}_2^*\widetilde{\tau}_1$ are maximal. This latter pattern of decay also shows up for the supersymmetric decays of the CP-odd Higgs $A$ to staus due to CP conservation.

\section{Collider analysis}
\label{sec:collider}

In Ref.~\cite{Arganda:2018hdn} we developed a search strategy that proved to be very efficient as a discovery tool within the context of what we call here Scenario I. In this section we will apply the same analysis which was optimized for Scenario I to both scenarios in order to test its discovery potential not only for Scenario I but also for Scenario II. In the two scenarios the staus are resonantly produced through a heavy Higgs boson: $H$, in the case of Scenario I, and $H/A$ in the case of Scenario II. As it was mentioned in Ref.~\cite{Arganda:2018hdn}, the cross section of the resonant production is significantly larger than that corresponding to the  EW pair production in the mass range $m_{H/A}\in [800-1200]$ GeV with values of $\tan \beta \in [25 - 50]$. This fact relies on the relatively low masses for the scalar and pseudoscalar resonances, the larger production via bottom fusion for the heavy Higgs bosons in the large $\tan\beta$ regions compared to the EW-size production via SM electroweak gauge bosons, and the large values of $A_{\tau}$ that enhance the decays of the heavy Higgs bosons, allowing non-negligible values of the branching ratios to staus, BR$(H\to \stau_{1}^*\stau_1)\sim 0.1-0.2$ in the case of Scenario I, and BR$(H/A\to \stau_{i}^*\stau_{j})\sim 0.1-0.4$ ($i,j=1,2$ and $i\neq j$) for Scenario II.

Our analysis focuses on the process $b\bar{b}\to H/A \to \stau_{i}^*\stau_{j}\to \tau^+\widetilde{\chi}_1^0\tau^-\widetilde{\chi}_1^0$, with taus decaying hadronically for both scenarios. Regardless on which stau mass state is produced, the final state involves two opposite-sign tau leptons and a large amount of missing energy, that comes from the two LSP neutralinos, $\widetilde{\chi}_1^0$. In order to test the two stau mixing scenarios, we have taken the benchmarks points used in Ref.~\cite{Arganda:2018hdn} for Scenario I and new ones produced in such a way that fulfill the requirements of Scenario II, but were included already in~ Ref.~\cite{Medina:2017bke}. A caveat is in order given that the parameter space considered for our scenarios I and II may be already excluded by the most recent LHC searches for Higgs bosons decaying into two tau leptons~\cite{Sirunyan:2018zut,Aad:2020zxo}, even when considering the additional decays into staus~\cite{Medina:2017bke}. However, as mentioned in the conclusion of Ref.~\cite{Medina:2017bke}, points of parameter space that have just been excluded by the recent di-tau searches can once again be resurrected considering the supersymmetric decays of the heavy Higgs bosons into staus in a completely analogous manner. We will therefore continue our analysis with these older points given that our conclusions regarding the strength of the search strategies proposed and the possible phenomenological signals will remain the same as for newly equivalent resurrected parameter points that are allowed to evade the latest ditau searches.
       
All the points were computed using {\ttfamily SPheno 3.3.8}~\cite{Porod:2003um, Porod:2011nf}, from which we obtain all the spectra and phenomenological properties, like the branching ratios. To test the different points, we produce Monte Carlo events for the $b$-quark fusion process that dominates the heavy Higgs production in the large $\tan\beta$ limit,  $b\bar{b}\to H/A \to \stau_{i}^*\stau_{j}\to \tau^+\widetilde{\chi}_1^0\tau^-\widetilde{\chi}_1^0$, at a center-of-mass energy of $\sqrt{s}=14$ TeV using the tool {\tt MadGraph\_aMC@NLO 2.6}~\cite{Alwall:2014hca}. In order to compute the signal cross section we make use of the tool {\tt SusHi}~\cite{Harlander:2012pb, Harlander:2016hcx} that gives the results for Higgs boson production cross sections at NNLO for the different production modes. The obtained values confirm that the dominant production mode is $b$-quark annihilation, with a cross section at least two orders of magnitude larger than the gluon fusion mode. 

\renewcommand{\arraystretch}{1.3}
\begin{table}[ht]
	\begin{center}
		\begin{tabular}{ccc}
			\hline
			\textbf{Parameter}	&  \textbf{Scenario I} & \textbf{Scenario II} \\ \hline
			$m_A$ [GeV]	&  947.5 & 1148.9  \\
			$\tan\beta$	&  33.8  & 45.33  \\
			$M_1$ [GeV]	&  100 & 100  \\
			$M_2,\,M_3$ [GeV]	    &  2200 & 2200 \\
			$\mu$ [GeV]	    &  -327.2 & -273.13 \\
			$A_{\tau}$ [GeV]	    &  -859.4 & 1125  \\
			$m_{\tilde{L}_{3}}$ [GeV]	    &  412.9 & 591.9  \\
			$m_{\tilde{E}_{3}}$ [GeV]	    &  393.8 & 363.1 \\
			$m_H$ [GeV]	    &  947.6 & 1149 \\  
			$m_{\stau_{1}}$ [GeV]	    &  367.5 & 350.7 \\
			$m_{\stau_{2}}$ [GeV]	    &  408.4 &  583.9 \\
			$m_{\widetilde{\chi}_1^0}$ [GeV]	    &  99 & 98.2  \\ \hline
		\end{tabular}
	\end{center}
	\caption{Benchmark points of Scenario I and Scenario II that are used to develop the collider analysis.}
	\label{tab:benchmark}
\end{table}

In Table~\ref{tab:benchmark} we show two particular benchmark points that are representative of each scenario and that we use to prove our search strategy. In Scenario I we use a point with a heavy Higgs mass $m_H=947.6$ GeV, $\tan \beta = 33.8$, and a lightest stau mass $m_{\stau_{1}}=367.5$ GeV. With these values the production cross section for the heavy Higgs boson $H$ is $\sigma_{bbH}=194.2$ fb for $b$-quark annihilation and $\sigma_{ggH}=3.2$ fb for gluon fusion. As we remarked before, the large $\tan\beta$ value makes the $b$-quark annihilation cross section larger than gluon fusion. The values of the branching fraction of the heavy Higgs boson $H$ decaying into staus and tau leptons are 0.17 and 0.09, respectively. As we can see here, the enhancement of the decay into staus leads to a decrease in the branching ratio to tau leptons. The branching fraction of the lightest stau into a tau lepton and the lightest neutralino is ${\rm BR}(\stau_{1}\to \tau \widetilde{\chi}_1^0)=0.98$. Thus the total cross section for the process $pp\to H \to \stau_{1}^*\stau_{1}\to \tau\widetilde{\chi}_1^0\tau\widetilde{\chi}_1^0$ at $\sqrt{s}=14$ TeV is $\sigma^{\rm total}_{\rm S-I}=31.7$ fb. In the case of Scenario II, we have $m_{H}=1149$ GeV, $m_A=1148.9$ GeV, and $\tan\beta=45.33$. In this benchmark point the production cross section is roughly $\sigma_{bbH}\sim\sigma_{bbA}\sim 120$ fb. The relatively large production cross section despite the value of the masses is due to the large value of $\tan\beta$ that enhances the $b$-quark annihilation production for both CP-even and CP-odd Higgs bosons. For this scenario the stau masses are $m_{\stau_{1}}=350.7$ GeV and $m_{\stau_{2}}=583.9$ GeV, and the branching fraction of the heavy neutral Higgs bosons are ${\rm BR}(A\to \sum_{i,j=1,2}^{i\neq j} \widetilde{\tau}_i^*\widetilde{\tau}_j)=0.25$ and ${\rm BR}(H\to \sum_{i,j=1,2}^{i\neq j} \widetilde{\tau}_i^*\widetilde{\tau}_j)=0.22$. We can notice that for this benchmark point the decay of the heavy Higgs bosons to a lightest stau pair is zero in the case of the CP-odd Higgs boson and ${\rm BR}(H\to\stau_{1}^*\stau_{1})=0.01$ for the CP-even one. This is just a realization of the properties of this scenario where there is an enhancement of the chiral couplings and the non-mixed states of the staus. The branching ratios for both stau states are ${\rm BR}(\stau_{2}\to \tau \widetilde{\chi}_1^0)=0.28$ and ${\rm BR}(\stau_{1}\to \tau \widetilde{\chi}_1^0)=0.82$ respectively. Therefore, the total cross section of the process $pp\to H/A \to \sum_{i,j=1,2}\stau_{i}^*\stau_{j}\to \tau\widetilde{\chi}_1^0\tau\widetilde{\chi}_1^0$ for this Scenario-II benchmark point at $\sqrt{s}=14$ TeV is $\sigma^{\rm total}_{\rm S-II}=13.2$ fb. This cross section is smaller than the one obtained in Scenario I, even when in Scenario II there are two resonant states contributing to the production of the staus. However, this was expected since in Scenario I $m_{H/A}$ is lighter than it is in Scenario II and, on top of that, the branching ratio of staus is almost saturated by the $\tau\widetilde\chi^0_1$ channel, which is not the case in Scenario II.

\renewcommand{\arraystretch}{1.4}
\begin{table}[ht]
	\begin{center}
	\begin{tabular}{cc}
	\hline
	\textbf{Background}	&  \textbf{Cross section [fb]} \\ \hline
	$t\bar{t}$	&  10125    \\
	$W$+jets	&  6.257$\times 10^{6}$    \\
	$Z$+jets	&  4.254$\times 10^{6}$    \\
	$WW$	    &  1188.6   \\
	$ZZ$	    &  183.3  \\ \hline
	\end{tabular}
	\end{center}
    \caption{List of backgrounds and their cross sections to the process $b\bar{b}\to H/A \to \stau_{i}^*\stau_{j}\to \tau^+\widetilde{\chi}_1^0\tau^-\widetilde{\chi}_1^0$ at the LHC at a center-of-mass energy of $\sqrt{s}=14$ TeV.}
    \label{tab:backgrounds}
\end{table}

The main backgrounds of the considered process are $t\bar{t}$, $W$+jets, $Z$+jets, $WW$ and $ZZ$, and they are listed in Table~\ref{tab:backgrounds} with their cross sections at $\sqrt{s}=14$ TeV. In principle, one has to include the QCD multijet background as well, however this is highly suppressed once the cuts involving large amounts of $\Etmiss$ are applied, as shown in Ref.~\cite{Arganda:2018hdn}. Although all the events corresponding to the background processes have been generated at leading order, the cross sections for $t\bar{t}$, $WW$ and $ZZ$ have been rescaled with K-factors of 1.5, 1.4, and 1.3, respectively, extracted from Ref.~\cite{Alwall:2014hca}.  In  addition,  the  cross  sections  for  the $W$+jets  and $Z$+jets backgrounds have been estimated by considering up to two light jets. It is important to note that for the $t\bar{t}$, $W$+jets, and $WW$ backgrounds we have included only the decay of the $W$ boson into $\tau\nu_{\tau}$, while in the case of the $ZZ$ and $Z$+jets backgrounds, we have considered the decays $ZZ\to \tau^+\tau^-\nu \bar{\nu}$ and $Z\to \tau^+\tau^-$, respectively. Both the signal and the different backgrounds have been generated with {\tt MadGraph\_aMC@NLO 2.6}~\cite{Alwall:2014hca} and showered with {\tt PYTHIA 8}~\cite{Sjostrand:2014zea},  while the detector response has been simulated with {\tt Delphes 3}~\cite{deFavereau:2013fsa}. The implementation of the different cuts of the search strategy that we present below have been carried out with {\tt MadAnalysis 5}~\cite{Conte:2014zja} in the expert mode.  

\subsection{Search strategy}
\label{sec:searchstrategy}

We will describe here the search strategy that we follow which was first proposed in Ref.~\cite{Arganda:2018hdn}. First of all, we apply some basic selection cuts that define the final state that we are searching for. We require that both signal and background events exhibit two opposite-sign tau leptons and we also demand that they have the following properties:
\begin{eqnarray}
p_T^{\tau_1}>50\, {\rm GeV}\, ,\, p_T^{\tau_2}>40\, {\rm GeV}\, ,\, |\eta^{\tau}|< 2.47\, .
\end{eqnarray}
Here we define $\tau_1$ and $\tau_2$ as the leading and subleading tau leptons, respectively, $p_T$ is the transverse momentum of the corresponding tau lepton and $\eta^{\tau}$ is its pseudorapidity. Given the topology of the signal process, one must rely on the large amount of transverse missing energy, $\Etmiss$, coming from the two LSP neutralinos escaping the detector in order to discriminate it from the background. For such reason, in this analysis we take into account kinematic variables that depend directly on $\Etmiss$:
\begin{enumerate}
\item The transverse mass $m_T$, defined as 
\begin{eqnarray}
m_{T}(\vec{p}_T^{\,\,i},\, \vec{p}_T^{\,\,\rm inv})=\sqrt{m^2_i+2\left(\sqrt{m^2_i + |\vec{p}_T^{\,\,i}|^2}\Etmiss - \vec{p}_T^{\,\,i}\cdot\vec{p}_T^{\,\,\rm inv} \right)},
\end{eqnarray}
where $i$ denotes the detected particle with transverse momentum $\vec{p}^{\,\, i}_T$ and mass $m_i$, and $\vec{p}_T^{\,\,\rm inv}$ is the total missing transverse momentum.
\item The stransverse mass $m_{T2}$, which is designed to target events with two sources of missing transverse momentum:
\begin{eqnarray}
m_{T2}= \min\limits_{\slashed{\vec{p}}_1 + \slashed{\vec{p}}_2=\vec{p}_T^{\,\,\rm inv}}\left\{\max\left[m_T(\vec{p}^{i}_T,\slashed{\vec{p}}_1),m_T(\vec{p}^{j}_T,\slashed{\vec{p}}_2)\right]\right\},
\end{eqnarray}
where $i$ and $j$ are the two visible states from the parent decays, and $\slashed{\vec{p}}_1$ and $\slashed{\vec{p}}_2$ are the corresponding missing transverse momenta. The power of the $m_{T2}$ variable comes from the fact that its distribution presents an endpoint around the mass of the parent decaying particle. This feature makes this variable quite efficient to discriminate between the signal and the $t\bar{t}$ and $WW$ backgrounds.
\end{enumerate}

\begin{figure}[ht!]
	\begin{center}
		\begin{tabular}{cc}
			\centering
			\hspace*{-3mm}
			\includegraphics[scale=0.40]{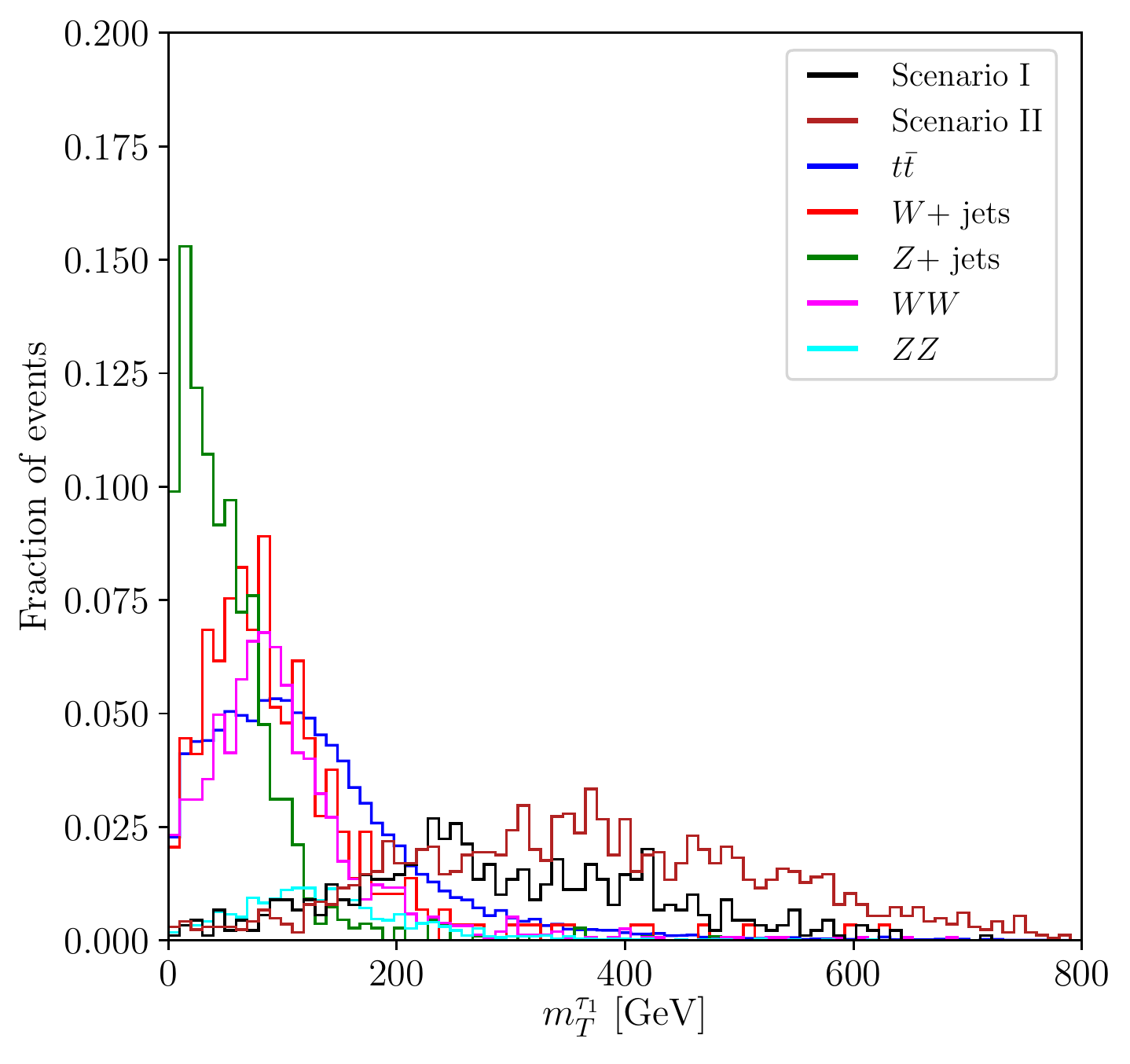} &
			\includegraphics[scale=0.40]{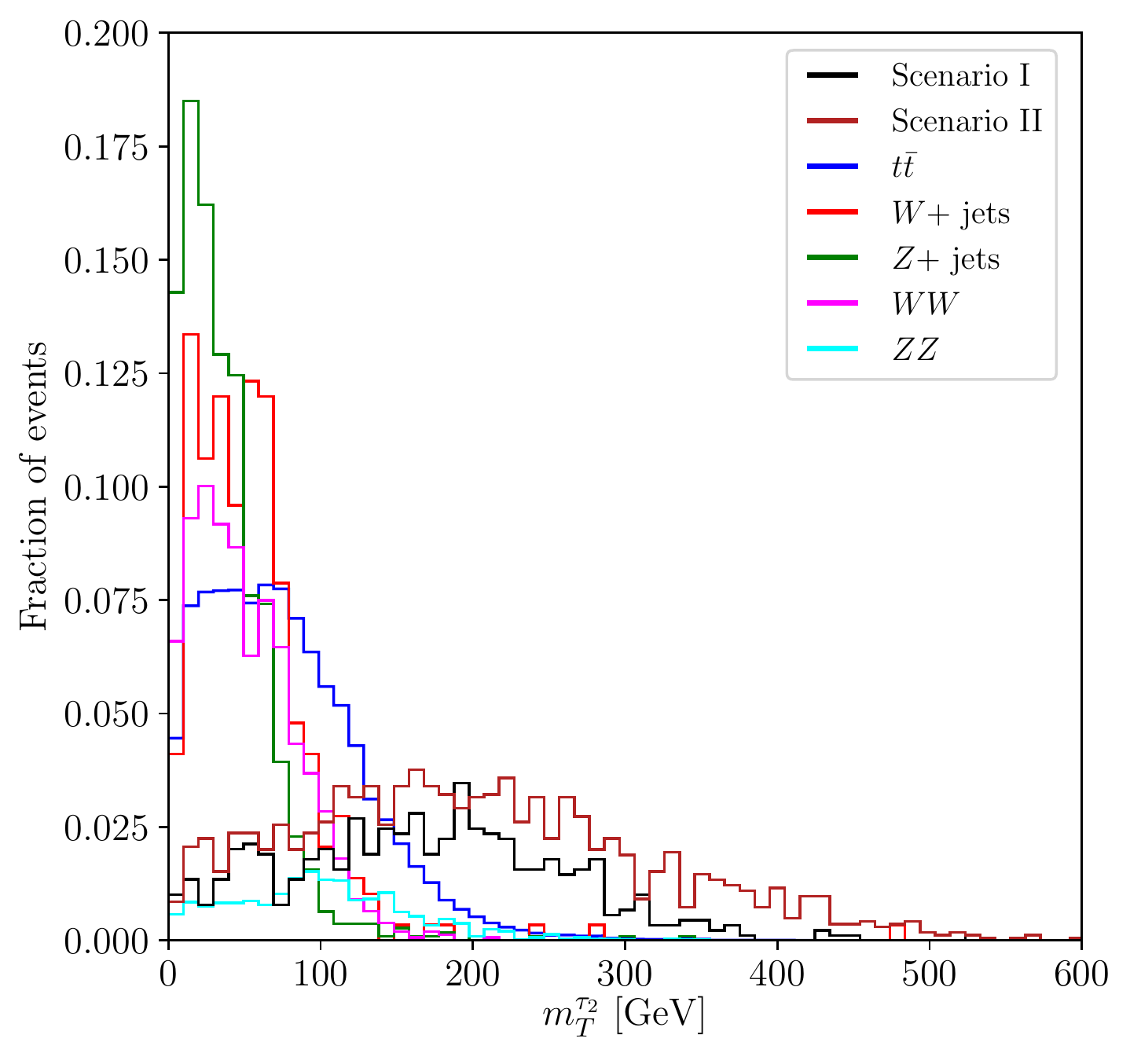}\\
			\includegraphics[scale=0.40]{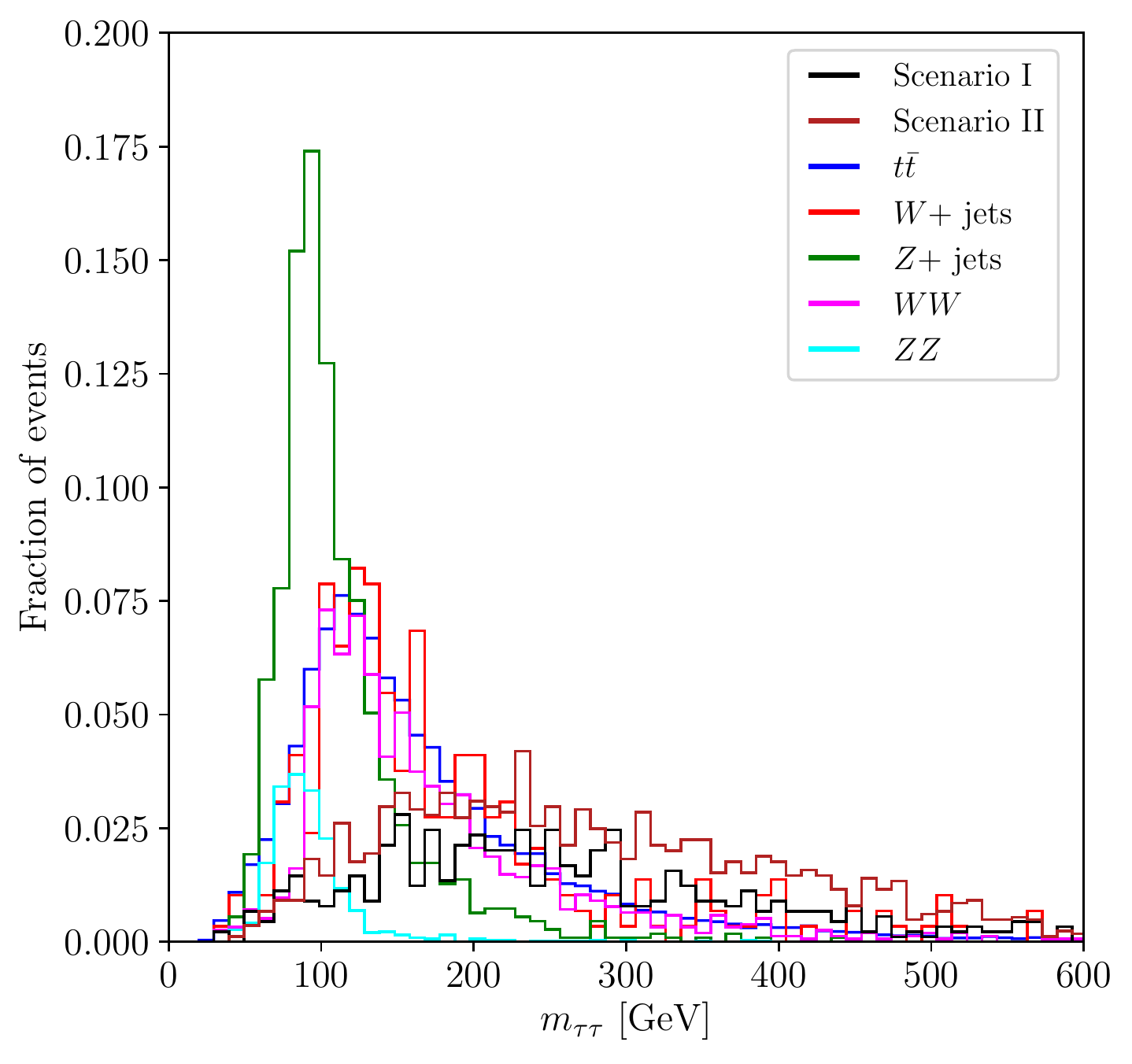} &
			\includegraphics[scale=0.40]{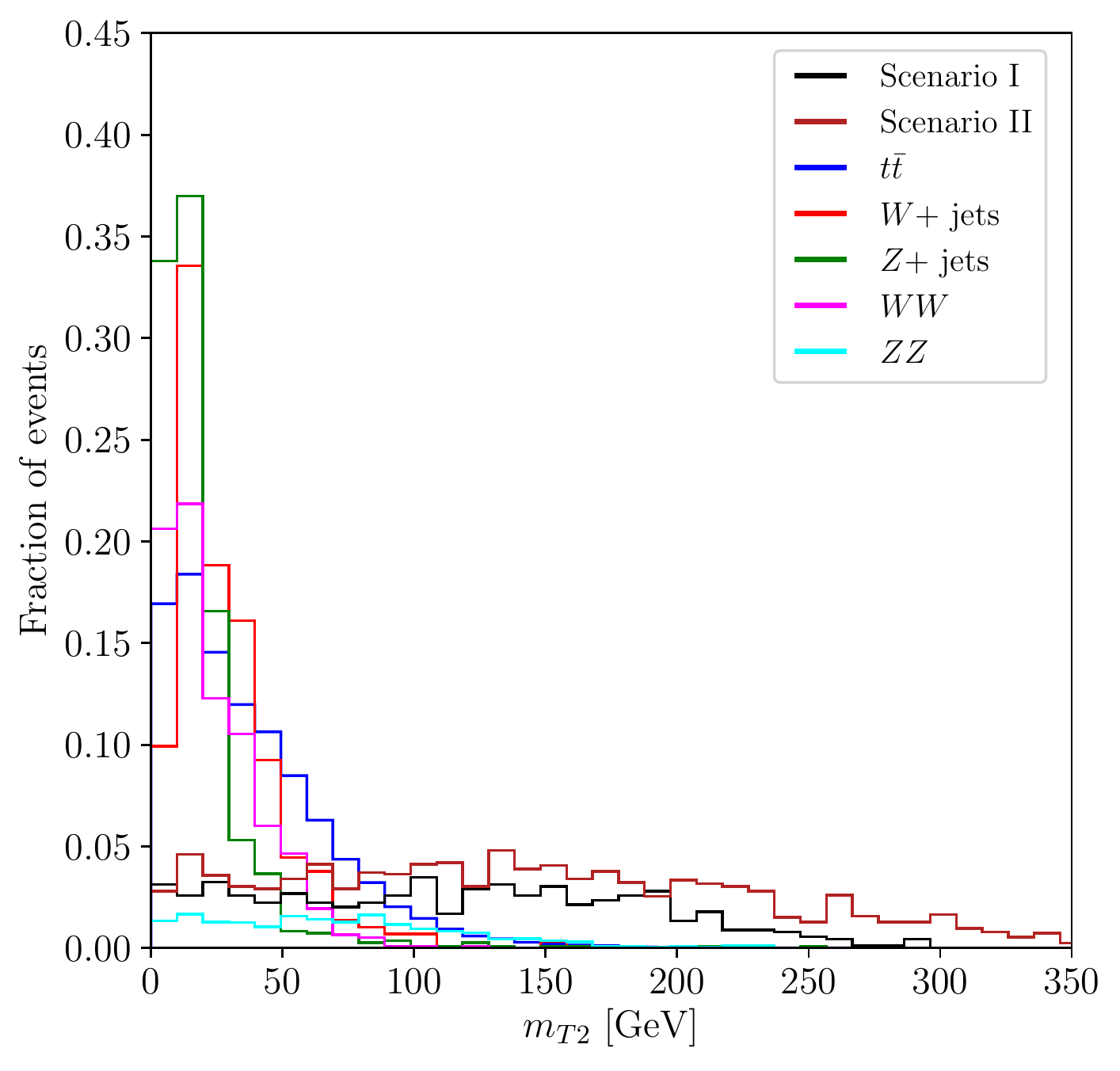}	
		\end{tabular}
		\caption{Four different distributions of kinematic variables after the selection cuts are applied to the signal and background events. On each case we show the distribution of the signal for the two scenarios along with those corresponding to the backgrounds listed in Table~\ref{tab:backgrounds}. Top left panel: Transverse mass of the leading tau lepton, $m_{T}^{\tau_{1}}$. Top right panel: Transverse mass of the subleading tau lepton, $m_{T}^{\tau_{2}}$. Bottom left panel: Invariant mass of the two tau leptons, $m_{\tau\tau}$. Bottom right panel: Stransverse mass $m_{T2}$.}
		\label{fig:cuts}
	\end{center}
\end{figure}

\newpage

In Fig.~\ref{fig:cuts} we depict the distributions of several variables after applying the selection cuts defined above to the signal and background events. The two top panels represent the transverse mass of the leading and the subleading tau leptons. From these two distributions we can see that the background events concentrate in the low transverse mass region, $m_{T}^{\tau}\lesssim 200$ GeV for the leading tau lepton and $m_{T}^{\tau}\lesssim 120$ GeV in the case of the subleading tau lepton. On the other hand, the distribution of the signal events reaches heavier transverse masses due to the fact that there is more missing transverse energy coming from the neutralinos. For that reason we require the transverse mass of the two tau leptons to be greater than 120 GeV. The bottom left panel shows the invariant mass of the tau lepton pair. We see that it is easy to discriminate the events coming from the $ZZ$ and $Z+$jets backgrounds  since they peak at $m_{\tau\tau}\sim m_{Z}$. Therefore, we set a cut on the invariant mass of the two tau leptons of $m_{\tau\tau}>100$ GeV. The bottom right panel depicts the $m_{T2}$ variable. From the shape of the distribution, it is clear that this variable is crucial to discriminate between the signal and the background events. The aim of the $m_{T2}$ variable is to select the processes in which there is a large amount of missing transverse energy, $\Etmiss$, coming from at least two sources. Moreover, recall that this variable exhibit an endpoint around the mass of the parent decaying particle. These features explain the quick decrease for the background distributions, which are mostly concentrated at $m_{T2}<150$ GeV. On the other hand, the signal distribution extends towards higher values of $m_{T2}$. Based on this we impose the cut $m_{T2}>180$ GeV that has a significant impact on the background events. In addition to the cuts already explained, we have also included a cut in the angular separation of the two tau leptons, $\Delta R(\tau_1,\,\tau_2)$, and imposed a $b$-jet veto\footnote{Given that in our scenarios the main production mechanism of the resonances is the $b$-quark annihilation, it could be interesting to study the signal process without imposing the $b$-jet veto. In principle, one could define different signal regions according to the number of $b$-tagged jets,  allowing for a better discovery rate and efficiency. However, the proper identification of $b$-jets arising from the initial state is a complex task and would require a detailed study on its own, which is out of the scope of this paper.} along with the requirement that the number of light jets is smaller than 2.  All the cuts are summarized in Table~\ref{tab:cuts}, where the left column contains the selection cuts and the right column includes the cuts that define our signal region. It is important to emphasize here that the fact that our search strategy for stau pairs is based on their production through heavy resonances ($H$ and $A$ Higgs bosons) allows us to impose a much more restrictive cut on the $m_{T2}$ variable than in strategies based on the usual electroweak stau production (see for example Table 1 of~\cite{Aad:2019byo}). In the latter case, the background is much less suppressed and unfortunately mimics the signal better.

\renewcommand{\arraystretch}{1.4}
\begin{table}[ht]
	\begin{center}
		\begin{tabular}{c||c}
			\hline
			\textbf{Selection Cuts}	&  \textbf{Signal Region Cuts}  \\ \hline
			2 OS taus &  $N_b=0$ \& $N_j<2$   \\
			$p_T^{\tau_1}>50\, {\rm GeV}$&  $\Delta R (\tau_{1}, \tau_{2})<3.5$   \\
			$p_T^{\tau_2}>40\, {\rm GeV}$&  $m_{T}^{\tau_1},m_{T}^{\tau_2}>120$ GeV   \\
		    $|\eta^{\tau}|< 2.47  $&  $m_{\tau\tau}>100$ GeV  \\
				    &  $m_{T2}>180$ GeV   \\ \hline
		\end{tabular}
	\end{center}
	\caption{List of cuts performed in the collider analysis. The first column shows the selection cuts that define the process. In the second column we depict the selected cuts in order to discriminate the signal from the background.}
	\label{tab:cuts}
\end{table}

We can now test the efficiency of the search strategy by applying it to the benchmark points of Scenarios I and II given in Table~\ref{tab:benchmark}. In order to simulate the background, we have followed the same procedure as in Ref.~\cite{Arganda:2018hdn}, generating the same number of events as it is indicated in Table~\ref{tab:significance} for every background source. Following similar searches~\cite{Aad:2014yka, ATLAS:2016fhh} we assume a systematic uncertainty of 30\% on the estimated sum of all backgrounds. In order to compute the significance of the signal events, $S$, with respect to the background events, $B$, including the potential systematic uncertainties we use~\cite{Cowan:2012}
\begin{eqnarray}
\mathcal{S}_{\rm dis}=\sqrt{2\left((B+S)\log\left(\frac{(S+B)(B+\sigma_{B}^2)}{B^2+(S+B)\sigma_{B}^2}\right)-\frac{B^2}{\sigma_{B}^2}\log\left(1+\frac{\sigma_{B}^2S}{B(B+\sigma_{B}^2)}\right)\right)},
\label{eq:discovery}
\end{eqnarray}
where $\sigma_{B}=(\Delta B)B$, with $\Delta B$ being  the  relative  systematic  uncertainty, in our case $\Delta B=$ 30\%. In Table~\ref{tab:significance} we show the number of events for every source of background and the signal events of both scenarios at a LHC center-of-mass energy of $\sqrt{s}=14$ TeV and for a total integrated luminosity of $\mathcal{L}=100$ fb$^{-1}$. For each scenario we show a column with the number of events if no cut is applied and a second column with the number of events after applying the cuts. We see that with a 30\% of systematic uncertainties, a signal significance of 6.62$\sigma$ for Scenario I and 5.24$\sigma$ for Scenario II are obtained for a luminosity of 100 fb$^{-1}$. In spite of the differences between the two scenarios in terms of the nature of the stau states, the decaying resonance and the Higgs-stau coupling, the analysis appears to be efficient in both cases.

\renewcommand{\arraystretch}{1.5}
\begin{table}[ht]
	\begin{center}
		\begin{tabular}{c|c|c|c|c}

				&  \multicolumn{2}{ >{\columncolor{myblue}} c}{\textbf{Scenario I}} &\multicolumn{2}{ | >{\columncolor{myblue}} c|}{\textbf{Scenario II}} \\ \hline
				& \multicolumn{1}{| >{\columncolor{mygray}} c|}{\textbf{No Cuts}}&     \multicolumn{1}{ >{\columncolor{mygray}} c|}{\textbf{SR}}& \multicolumn{1}{ >{\columncolor{mygray}} c|}{\textbf{No Cuts}}&\multicolumn{1}{ >{\columncolor{mygray}} c|}{\textbf{SR}} \\ \hline
			\multicolumn{1}{ >{\columncolor{mygray}} c|}{Signal}	&  3171 &28.78&1317& 21.16  \\ \hline
			\multicolumn{1}{ >{\columncolor{mygray}} c|}{$t\bar{t}$}	&  1012500 &2.03&1012500&2.03   \\ \hline
			\multicolumn{1}{ >{\columncolor{mygray}} c|}{$W$+jets}	&  $6.257\times 10^8$ &0.65&$6.257\times 10^8$&0.65   \\ \hline
			\multicolumn{1}{ >{\columncolor{mygray}} c|}{$Z$+jets}	&  $4.254\times 10^8$ &1.01&$4.254\times 10^8$&1.01   \\ \hline
			\multicolumn{1}{ >{\columncolor{mygray}} c|}{$WW$}	&  118860 &0&118860&0   \\ \hline
			\multicolumn{1}{ >{\columncolor{mygray}} c|}{$ZZ$}	&  18330 &0.37&18330&0.37   \\ \hline \hline
			\multicolumn{1}{ >{\columncolor{mygray}} c|}{$\mathcal{S}_{\rm dis}$}&$\mathcal{O}(10^{-5})$&6.62& $\mathcal{O}(10^{-5})$&5.24
		\end{tabular}
	\end{center}
	\caption{Number of signal and background events at $\sqrt{s}=14$ TeV for an integrated luminosity of $\mathcal{L}=100$ fb$^{-1}$ before and after applying cuts. The last line represents the signal significance obtained using Eq.~\eqref{eq:discovery}.}
	\label{tab:significance}
\end{table}

One can also think the other way around and imagine that no significant signal events are found for a given luminosity. In that situation, one can set 95\% C.L. exclusion limits by using the exclusion significance as follows~\cite{Cowan:2012}:
\begin{eqnarray}
\mathcal{S}_{\rm exc}=\sqrt{2\left(B\log\left(\frac{B}{S+B}\right)+S\right)} \leq 1.64 ,
\label{eq:exclusion}
\end{eqnarray}
where $B$ is the total number of background events and $S$ is the number of signal events at a given luminosity $\mathcal{L}$. In the next section we will analyze a set of points for each scenario in terms of exclusion and discovery significances using this search strategy.

\section{Results}

\label{sec:results}

In this section we use the search strategy described above and test it against several benchmark points from both Scenario I and Scenario II.  For each benchmark point we study the efficiency of the analysis in terms of potential exclusion at 95\% C.L. and discovery signal significance by considering two values of integrated luminosity at $\sqrt{s}=14$ TeV, $\mathcal{L}=100$ fb$^{-1}$, to be easily reached at the next run of the LHC, and $\mathcal{L}=1000$ fb$^{-1}$, corresponding to the high-luminosity LHC (HL-LHC). With these two values of the luminosity we can explore the future prospects and the reach of this search strategy in terms of physical parameters such as the mass of a new heavy (pseudo) scalar.

\subsection{Scenario I: $H\to \stau_1^{} \stau_1^*\to \tau^+\widetilde{\chi}_1^0\tau^-\widetilde{\chi}_1^0$}
\label{subsec:scenarioi}

\begin{figure}[ht]
	\begin{center}
		\begin{tabular}{cc}
			\centering
			\hspace*{-3mm}
			\includegraphics[scale=0.40]{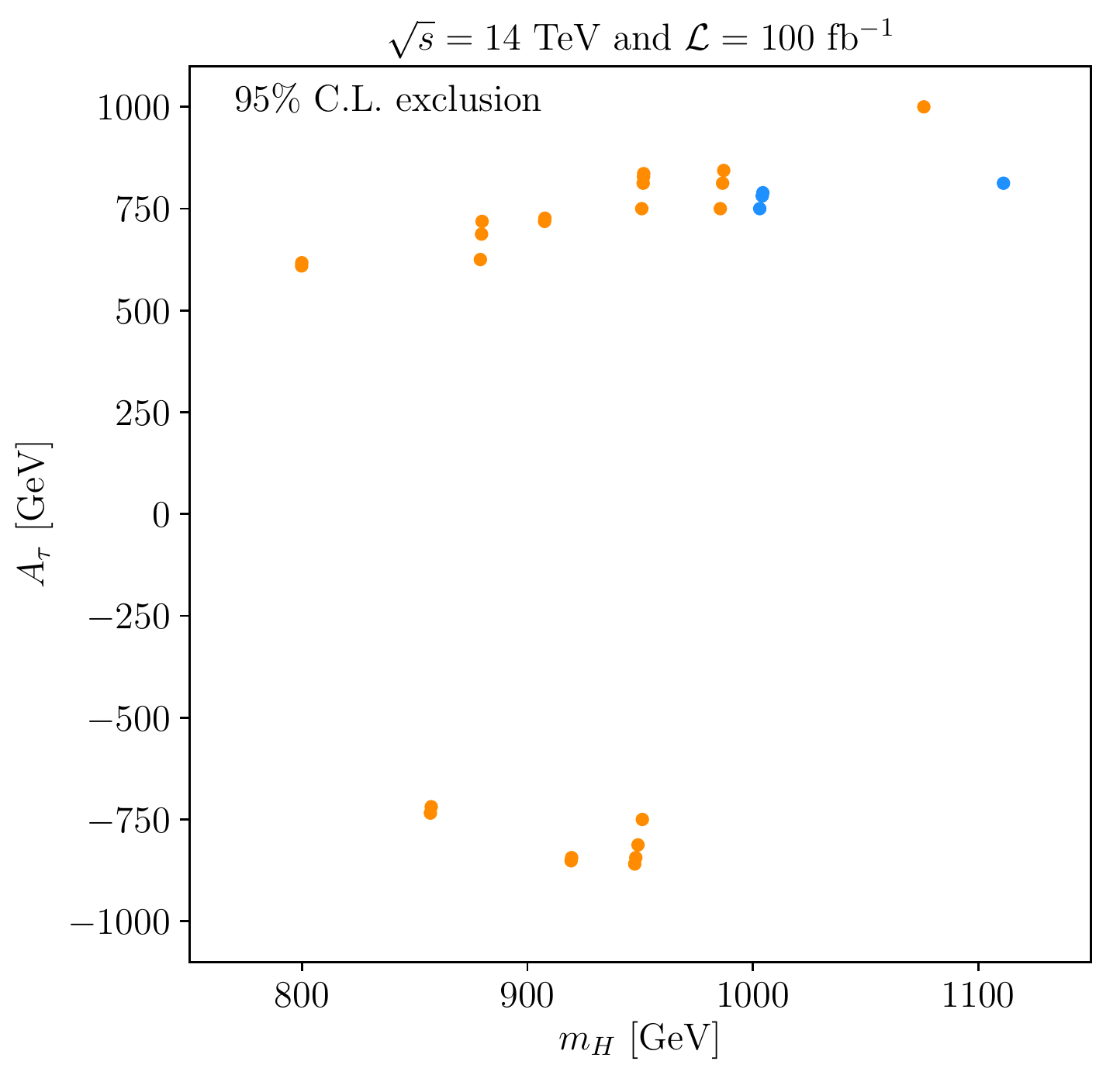} &
			\includegraphics[scale=0.40]{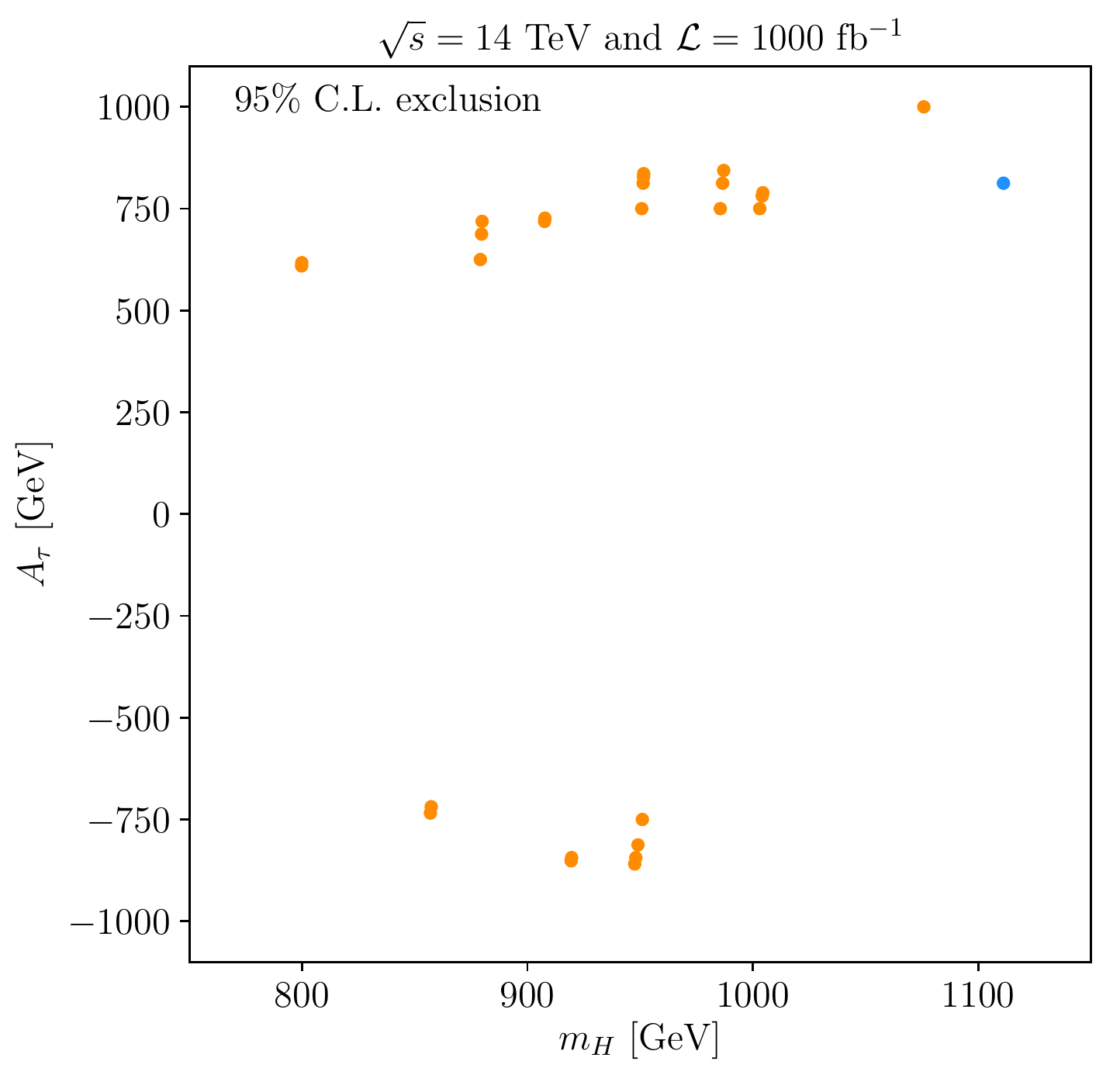}\\
			\includegraphics[scale=0.40]{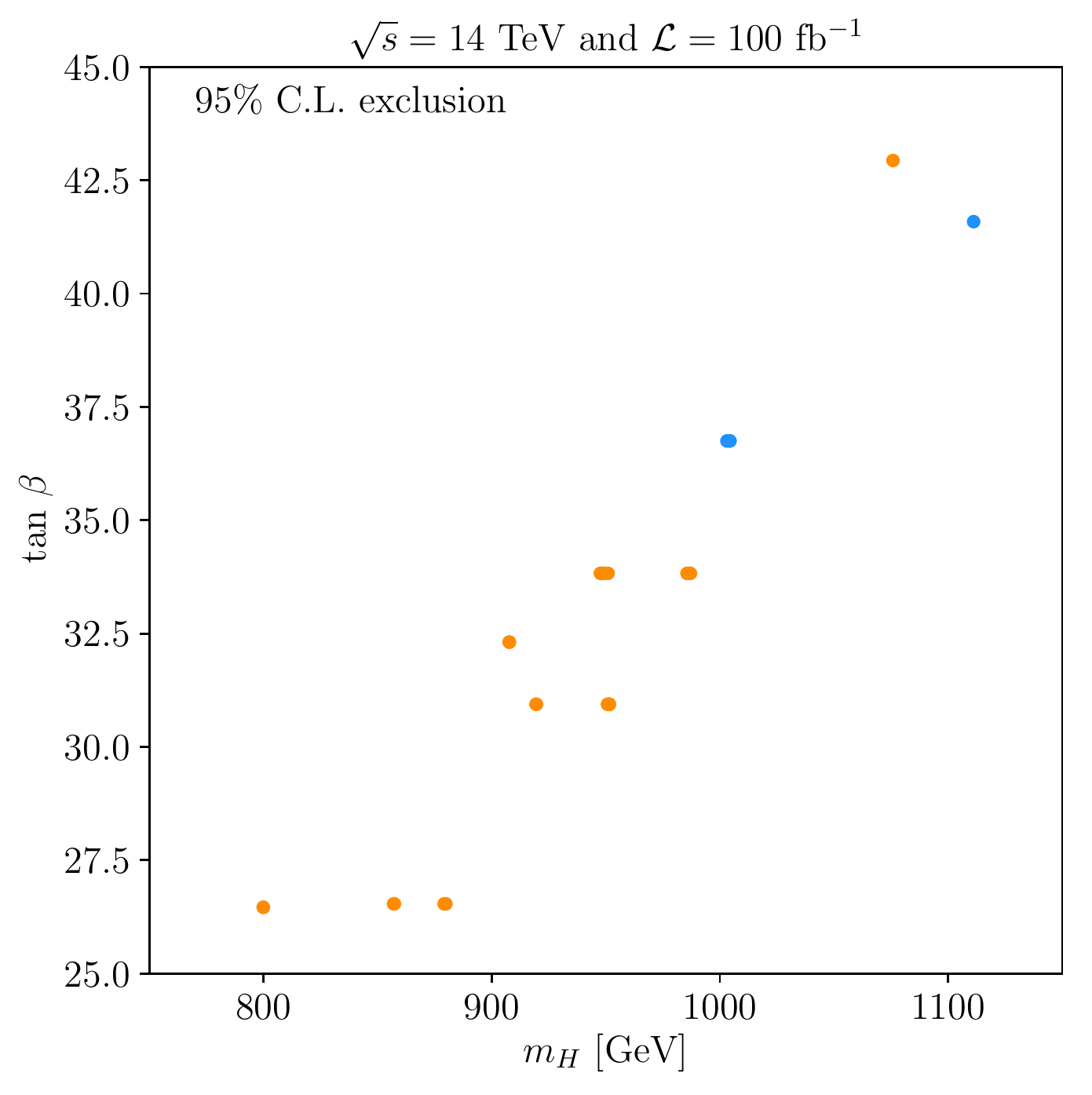} &
			\includegraphics[scale=0.40]{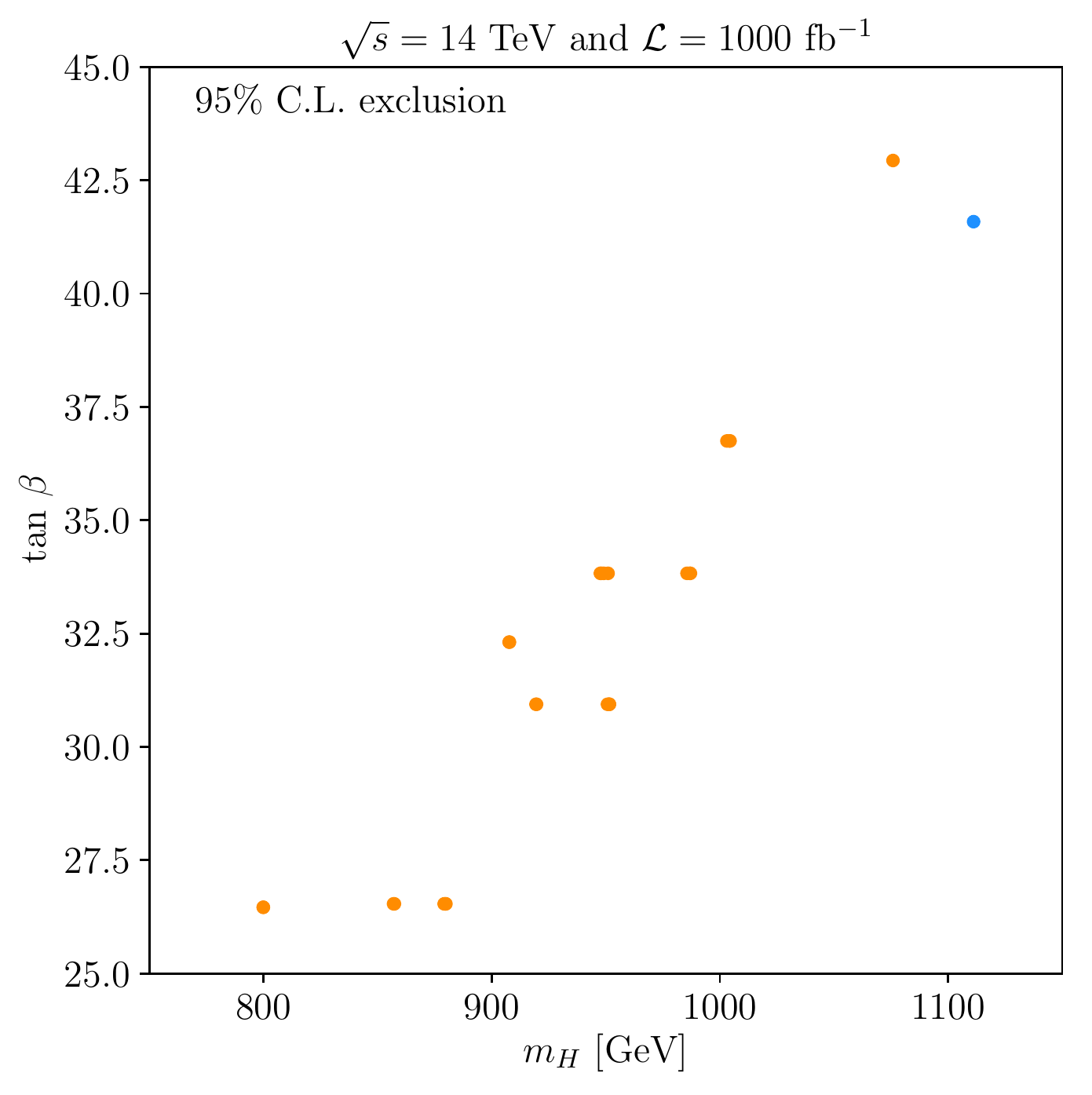}
		\end{tabular}
		\caption{Potential exclusion at 95$\%$ C.L. obtained from Eq.~\eqref{eq:exclusion} in the [$m_H$, $A_{\tau}$] plane (top) and [$\tan\beta$, $A_{\tau}$] plane (bottom), within Scenario I, for a center-of-mass energy of $\sqrt{s} =$ 14 TeV and total integrated luminosities of 100 fb$^{-1}$ (left panel) and 1000 fb$^{-1}$ (right panel). We display in orange the benchmark points that are excluded at 95\% C.L. and in blue those that are allowed.}
		\label{fig:ataumH-exc}
	\end{center}
\end{figure}

From Scenario I we take 27 benchmark points that were described in Ref.~\cite{Arganda:2018hdn}. These points are characterized by different values of $m_H$, $\tan\beta$, $A_\tau$, and $m_{\stau_{1}}$. We have applied the search strategy described in Section~\ref{sec:searchstrategy} and we have studied the exclusion power of the analysis as well as the signal significance of discovery. The results are shown in Fig.~\ref{fig:ataumH-exc}, where the orange points correspond to excluded benchmarks and the blue ones to the benchmarks that cannot be ruled out at 95\% C.L. by our analysis.

For $\mathcal{L}=100$ fb$^{-1}$ 23 of the 27 benchmarks points are excluded. All the points with heavy Higgs boson mass smaller than 1000 GeV are ruled out. Above this value, $m_H >$ 1000 GeV, there is still one benchmark with $m_H$= 1075 GeV that can be excluded, while the remaining 4 benchmarks in this mass region are allowed.  This is due to the fact that the benchmark point with $m_ H =$ 1075 GeV has also a value of $A_\tau$ large enough to enhance the coupling $g_{Hdd}$ and then the branching ratio into staus, which is $\sim$16\%. Moreover, this point has a negative value of $\mu$ and then its contribution to the coupling in Eq.~\eqref{eq:ghdd} adds to that corresponding to the $A_\tau$. Finally, as can be seen in the plane [$\tan\beta$, $A_\tau$], this benchmark point includes a large value of $\tan\beta$ which increases the production cross section compensating the suppression due to the large value of $m_H$. Furthermore, a large value of $\tan\beta$ also enhances $g_{Hdd}$.

Within the high-luminosity phase of the LHC, it could be possible to exclude benchmark points with heavy Higgs boson masses above 1 TeV. However, as can be seen from the right panel of Fig.~\ref{fig:ataumH-exc}, it seems that for trilinear couplings smaller than 1 TeV, our analysis cannot probe heavy Higgs boson masses above 1.1 TeV, even with values of $\tan\beta$ as large as 42.

\begin{figure}[ht]
	\begin{center}
		\begin{tabular}{cc}
			\centering
			\hspace*{-3mm}
			\includegraphics[scale=0.40]{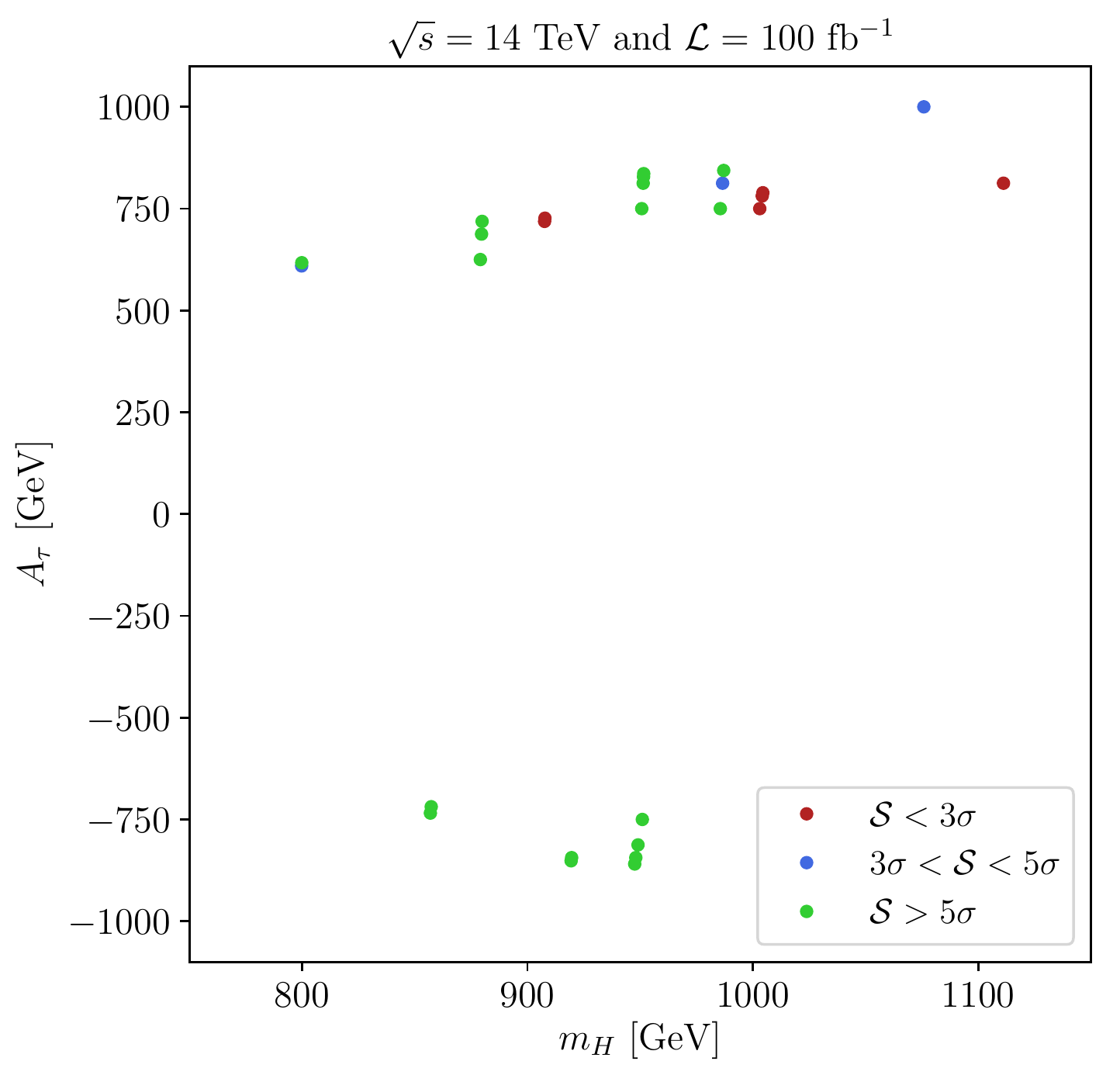} &
			\includegraphics[scale=0.40]{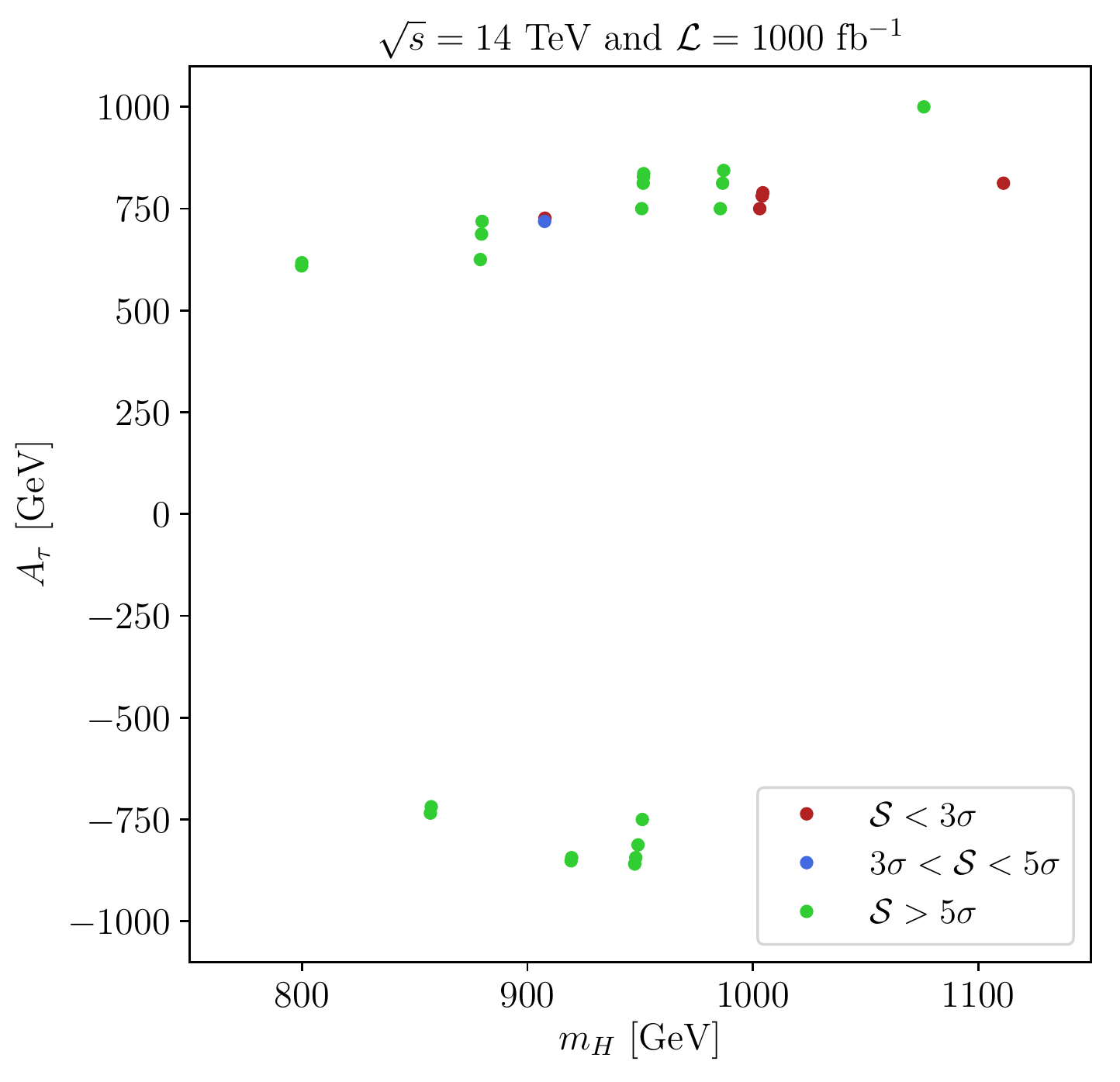}\\
			\includegraphics[scale=0.40]{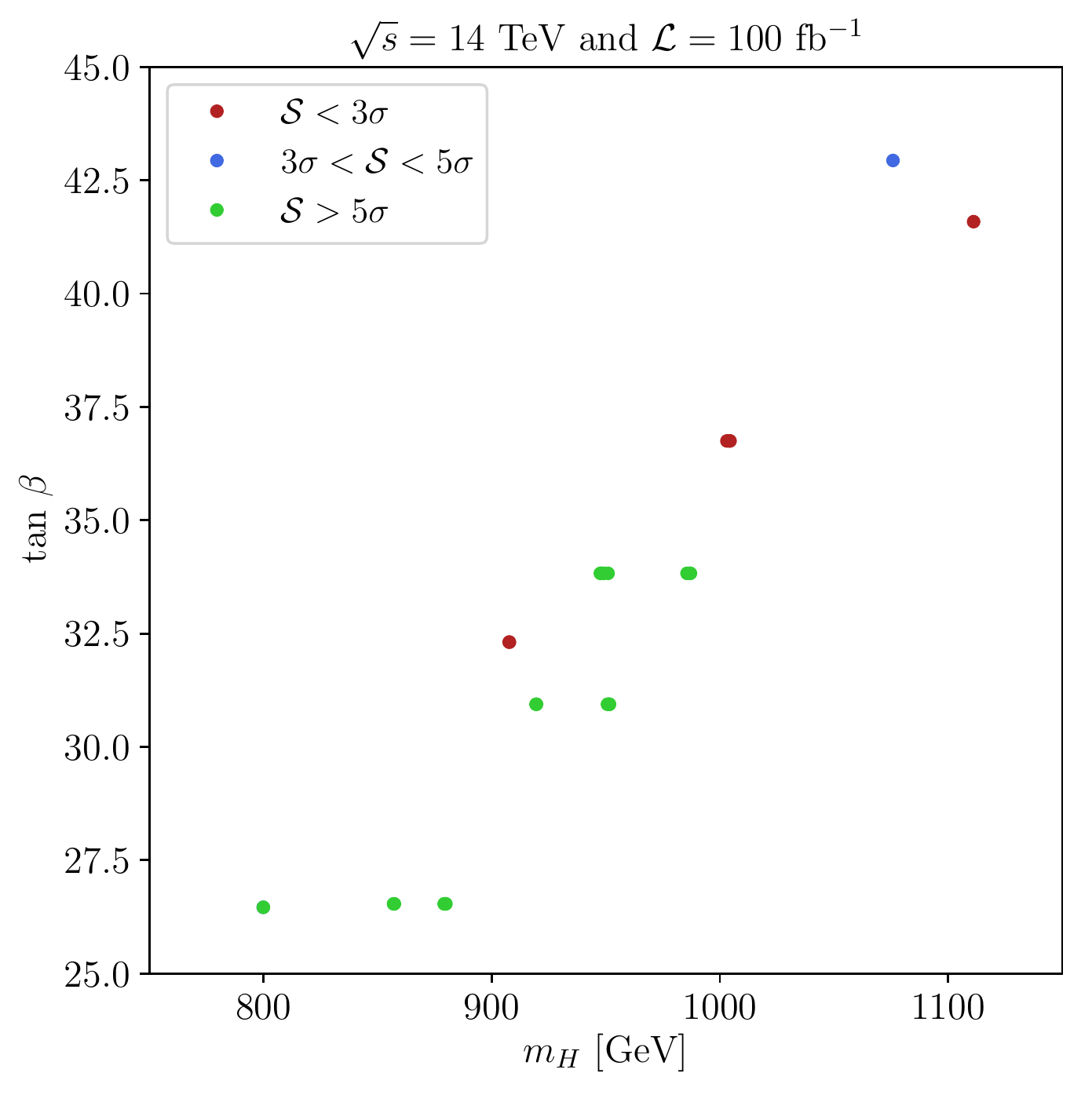} &
			\includegraphics[scale=0.40]{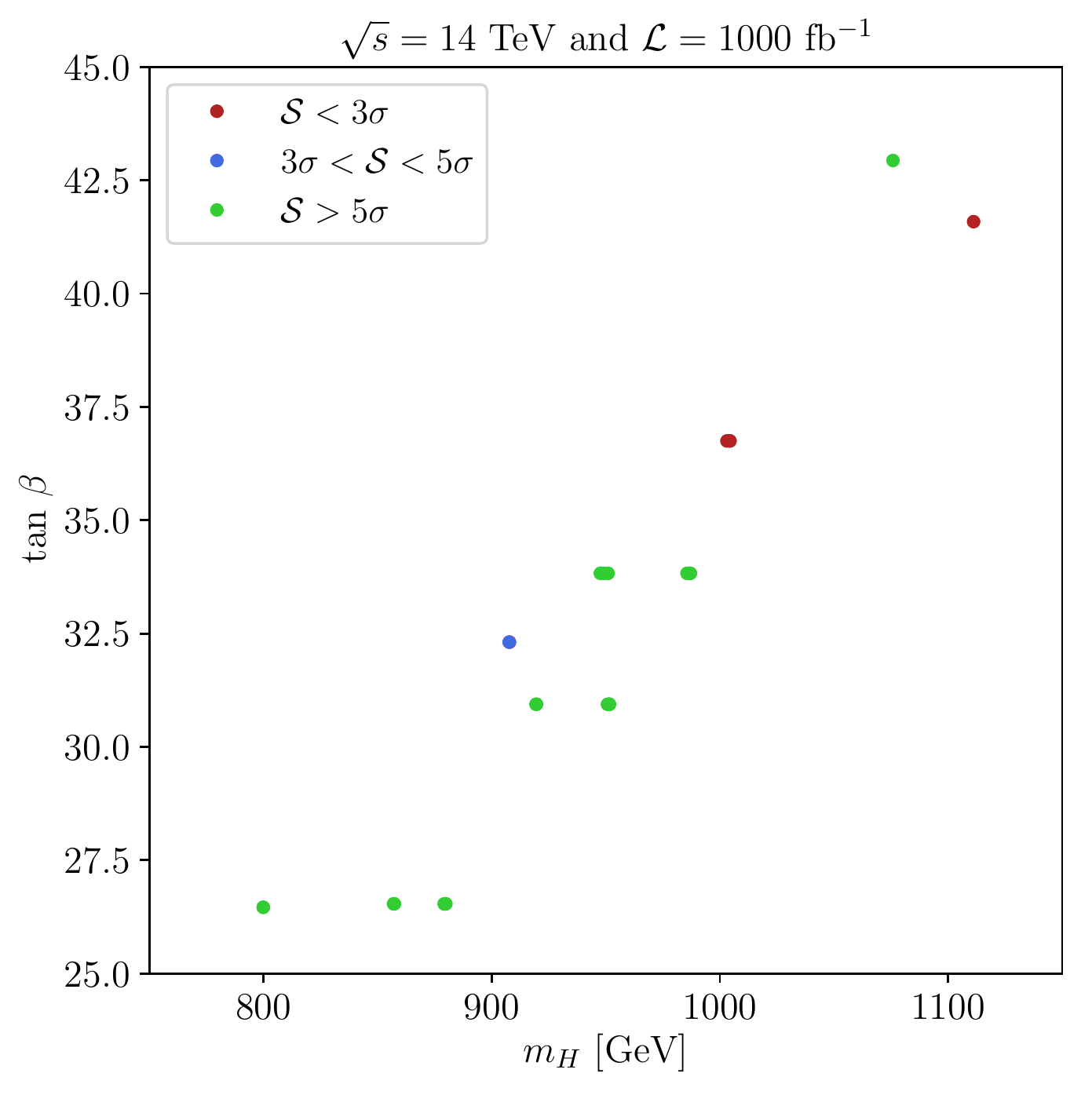}
		\end{tabular}
		\caption{Signal significance in the [$m_H$, $A_{\tau}$] plane (top) and [$\tan\beta$, $A_{\tau}$] plane (bottom), within Scenario I for a center-of-mass energy of $\sqrt{s} =$ 14 TeV and total integrated luminosities of 100 fb$^{-1}$ (left panel) and 1000 fb$^{-1}$ (right panel). Red circles correspond to significances below the evidence level ($\mathcal{S}<3\sigma$), blue circles to significances between the evidence level and the discovery one ($3\sigma <\mathcal{S}<5\sigma$), and green circles to significances larger than the discovery level ($\mathcal{S}>5\sigma$).}
		\label{fig:ataumH-dis}
	\end{center}
\end{figure}

In Fig.~\ref{fig:ataumH-dis} we show the discovery prospects for each of the 27 benchmarks in the [$m_H$, $A_\tau$] and [$m_H$, $\tan\beta$] planes. Signal significances in the ranges $\mathcal{S}<3\sigma$, $3\sigma <\mathcal{S}<5\sigma$ (evidence level) and $\mathcal{S}>5\sigma$ (discovery level) are displayed in red, blue and green, respectively. We see that most of the benchmarks with $m_H$ below 1 TeV lie in the evidence or the discovery level. Only those around $m_H =$ 910 GeV with a trilinear coupling $A_\tau =$ 720 GeV are below the evidence level due to the small branching ratio of $H$ into staus, which is almost 10\%. Among these benchmark points, solely one could be tested by increasing the luminosity to 1000 fb$^{-1}$. The benchmarks in the mass region $m_H >$ 1 TeV are difficult to probe even at $\mathcal{L}=1000$ fb$^{-1}$. Again the exception is the point with $m_H =$ 1075 GeV, due to the combination of a large trilinear coupling ($A_\tau =$ 1 TeV) that leads to a branching ratio of 16\%, and a value of $\tan\beta =$ 43 that is large enough to increase the production cross section despite the large heavy Higgs boson mass.  From the low panels of Fig.~\ref{fig:ataumH-dis}, we see that there is a region with $\tan\beta$ $\in$ (37-41) and $m_H \geq$ 1 TeV in which the search strategy is not efficient. This is due to the fact that the production cross section decreases as $m_H$ grows, and the benchmarks in the mass region above 1 TeV correspond to $\tan\beta$ values that are not large enough to compensate this trough their impact on the production cross section and the decay rate. It seems that values of $\tan\beta$ above 41 are required in order to test the region $m_H >$ 1 TeV with our search strategy.

\begin{figure}[ht]
	\begin{center}
			\includegraphics[scale=0.5]{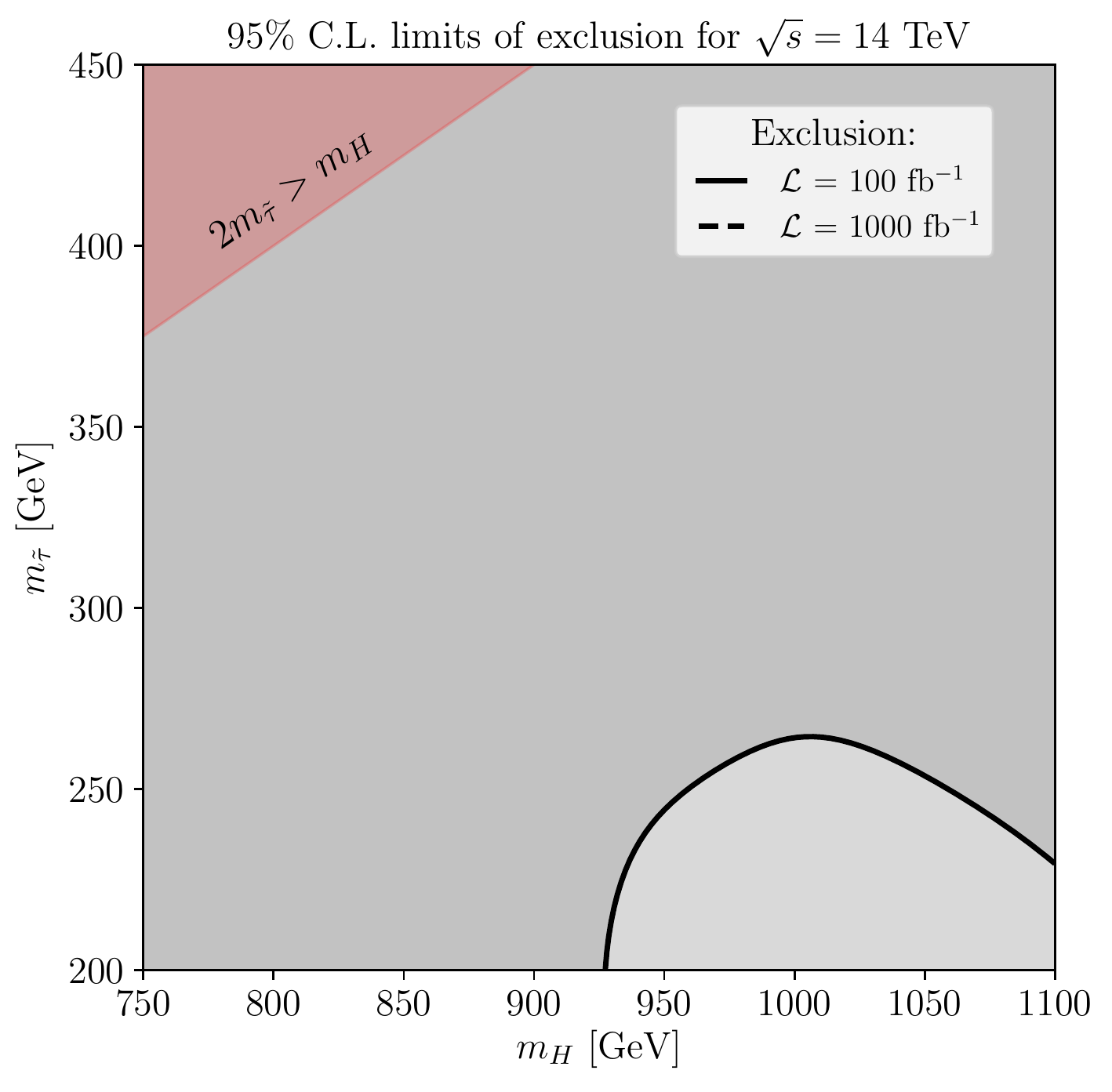} 
		\caption{Potential areas of exclusion at 95$\%$ C.L. within Scenario I for a center-of-mass energy of $\sqrt{s} =$ 14 TeV. The area above the full line, here in dark gray, represents the one that could be possibly excluded by this analysis at 100 fb$^{-1}$ of integrated luminosity if no evidence of signal is found. The area below it, here in light gray, that is defined by the dashed line shows the potentially excluded range at 1000 fb$^{-1}$ of integrated luminosity. However, the dashed line here is not visible since it goes below masses of the lightest stau of $m_{\stau_1}<200$ GeV. The shaded red area is forbidden because the decay mode $H \to \tilde\tau_1 \tilde\tau_1^\ast$ is kinematically closed.}
		\label{fig:mstaumH-exc}
	\end{center}
\end{figure}

By performing an interpolation based on the 27 benchmark points studied above, we can now interpret the obtained results in the [$m_H$, $m_{\stau_{1}}$] plane. This is shown in Fig.~\ref{fig:mstaumH-exc}, where we display the 95\% C.L. exclusion limits for $\mathcal{L}=100$ fb$^{-1}$ (dark gray) and $\mathcal{L}=1000$ fb$^{-1}$ (light gray). The red area on the left top corner is kinematically forbidden. We see that the search strategy is able to probe most of the [$m_H$, $m_{\stau_{1}}$] plane with a luminosity of $\mathcal{L}=100$ fb$^{-1}$. On the other hand, a higher luminosity is required to gain sensitivity in the region at $m_H >$ 930 GeV and $m_{\stau_{1}} <$ 260 GeV because the large values of $m_H$ reduces the production cross section and also the small values of $m_{\stau_{1}}$ lead to a substantial decrease in the amount of $\Etmiss$, which makes the $m_{T2}$ cut less powerful. The last explains the fact that for a given heavy Higgs boson mass above 930 GeV we can move from the allowed to the excluded region by increasing the stau mass. With $\mathcal{L}=1000$ fb$^{-1}$, the search strategy becomes sensitive to the whole area comprised by heavy Higgs boson masses between 750 GeV and 1100 GeV and stau masses between 200 GeV and 450 GeV.

\begin{figure}[ht]
	\begin{center}
		\begin{tabular}{cc}
			\centering
			\hspace*{-3mm}
			\includegraphics[scale=0.5]{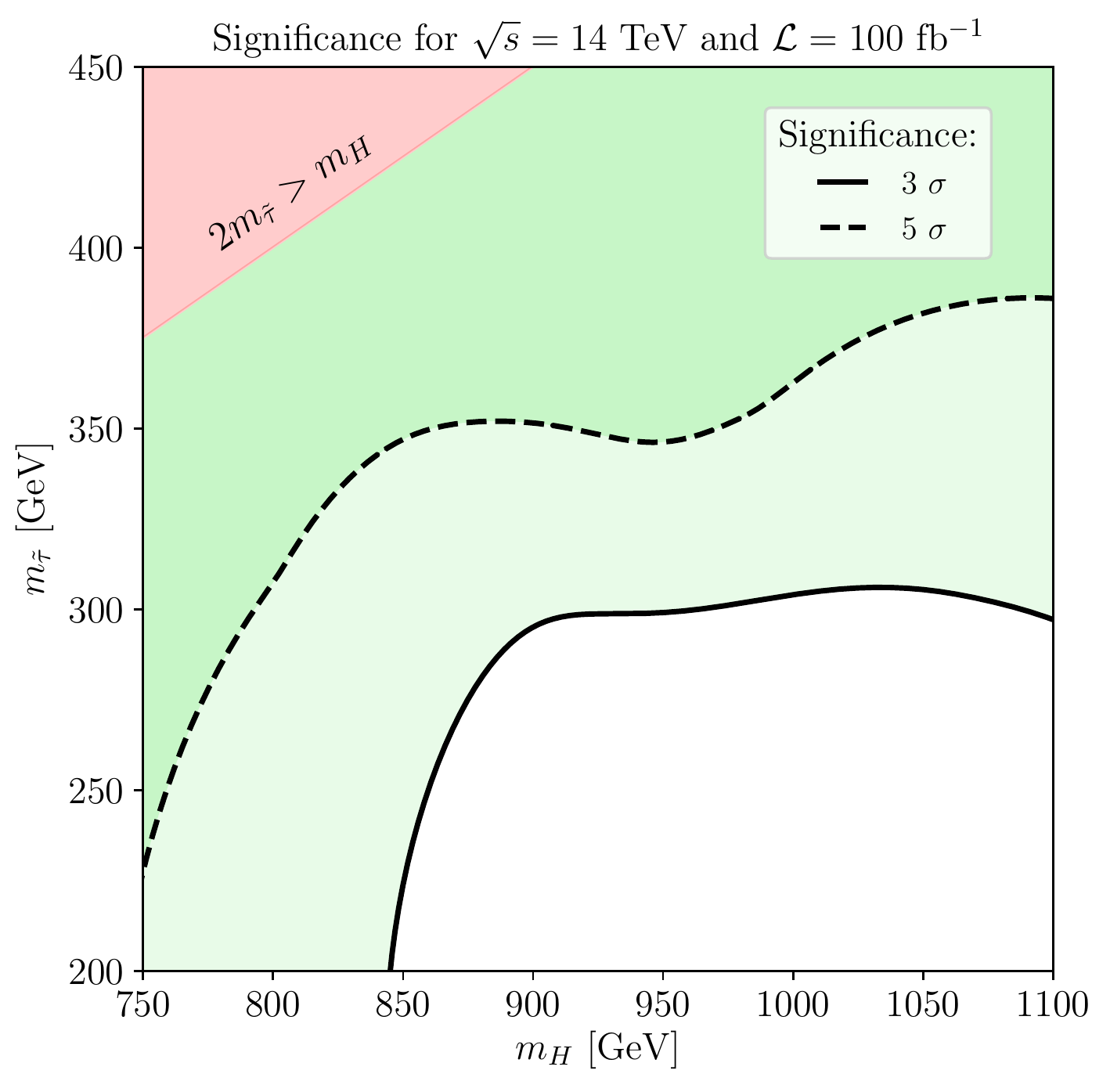} &
			\includegraphics[scale=0.5]{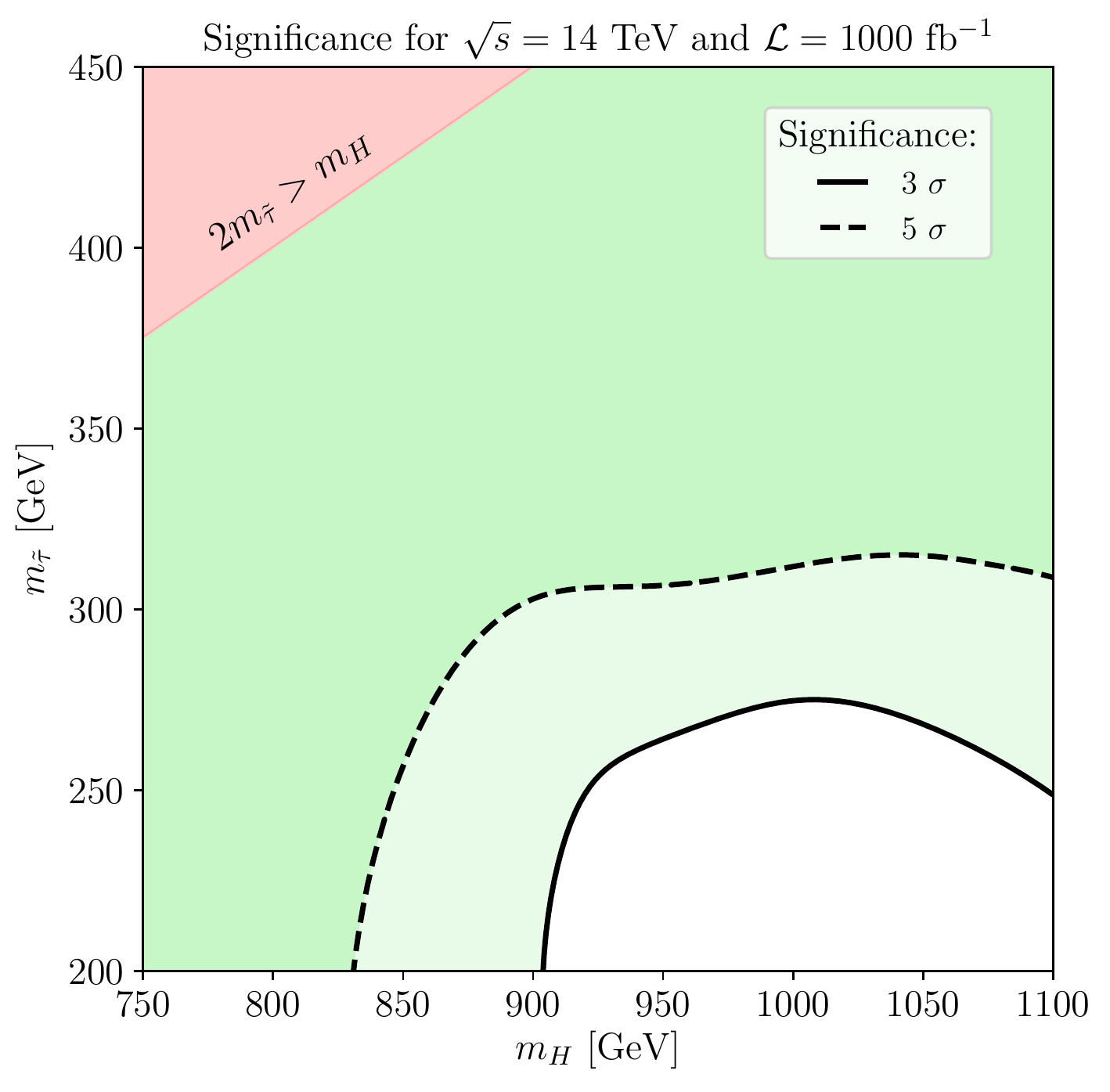}
		\end{tabular}
		\caption{Signal significance in the [$m_H$, $m_{\tilde \tau_1}$] plane, within Scenario I, for a center-of-mass energy of $\sqrt{s} =$ 14 TeV and total integrated luminosities of 100 fb$^{-1}$ (left panel) and 1000 fb$^{-1}$ (right panel). The dark gray area above the dashed line is the discovery level region ($\geq$ 5 standard deviations), while the light gray area above the solid line is the evidence level region ($\geq$ 3 standard deviations). Finally, the white area below the solid line corresponds to signal significances smaller than 3$\sigma$ and the red area is kinematically forbidden.}
		\label{fig:mstaumH-dis}
	\end{center}
\end{figure}

In Fig.~\ref{fig:mstaumH-dis} we present the same contour plot in the [$m_H$, $m_{\stau_{1}}$] plane as in Fig.~\ref{fig:ataumH-exc} but for the discovery prospects of the signal, for a total integrated luminosity of $\mathcal{L}=100$ fb$^{-1}$ (left panel ) and 1000 fb$^{-1}$ (right panel). We depict the evidence level (3$\sigma$) as a light green area limited by a solid black line whereas the discovery level (5$\sigma$) is shown as a darker green area limited by a dashed black line. From the left panel we see that for $\mathcal{L}=100$ fb$^{-1}$ the search strategy is sensitive to $m_H <$ 850 GeV regardless the value of the stau mass (within the considered range). For $m_H >$ 850 GeV the sensitivity is lost for stau masses below 300 GeV due to the same two reasons discussed before in the case of the exclusion plot: on the one hand, the signal cross section decreases considerably for large values of $m_H$, and on the other one, small stau masses produce a final state with less energetic tau leptons and lower $\Etmiss$, which in turn reduces the discrimination power of crucial kinematic variables as $m_{T2}$ or $m_{T}$. This high $m_H$ range can still be probed if larger values of the stau mass are considered.  In particular, the discovery level is reached for $m_{\stau_{1}} >$ 350-370 GeV. For an integrated luminosity of 1000 fb$^{-1}$, our analysis cover most of the considered area in the [$m_H$, $m_{\stau_{1}}$] plane. However, its sensitivity is not enough to reach the region with $m_H >$ 900 GeV and $m_{\stau_{1}} <$ 260 GeV. We see that this region of high $m_H$ and low $m_{\stau_{1}}$ is very challenging even within the context of the HL-LHC. 

\subsection{Scenario II $H/A\to \stau_{1,2}^{} \stau_{2,1}^*\to \tau^+\widetilde{\chi}_1^0\tau^-\widetilde{\chi}_1^0$}
\label{subsec:scenarioii}

The parameters involved in Scenario II are $m_H$, $m_A$, $\tan\beta$, $A_\tau$, $m_{\stau_{1}}$, and $m_{\stau_{2}}$. Thus, we have in this case an additional parameter arising from the stau sector, namely, $m_{\stau_{2}}$.  We select in this case 228 benchmark points. We explore the results obtained for each of them in terms of the parameters $A_\tau$, $\tan\beta$, and $m_A$ first and then in the stau sector, that we characterize by using the average of the two stau masses and their difference
\begin{eqnarray}
{\overline{m}_{\stau_{12}}}= \frac{m_{\stau_1} + m_{\stau_2}}{2},\quad \quad \Delta m = m_{\stau_2} - m_{\stau_1}.
\label{eq:stauvariables}
\end{eqnarray}
We choose $m_A$ instead of $m_H$ because it is a natural parameter in the MSSM and also one can obtain $m_H$ making use of $m_A$. Furthermore, as we move in the decoupling limit we find usually that $m_A$ $\sim$ $m_H$.

\begin{figure}[ht]
	\begin{center}
		\begin{tabular}{cc}
			\centering
			\hspace*{-3mm}
			\includegraphics[scale=0.40]{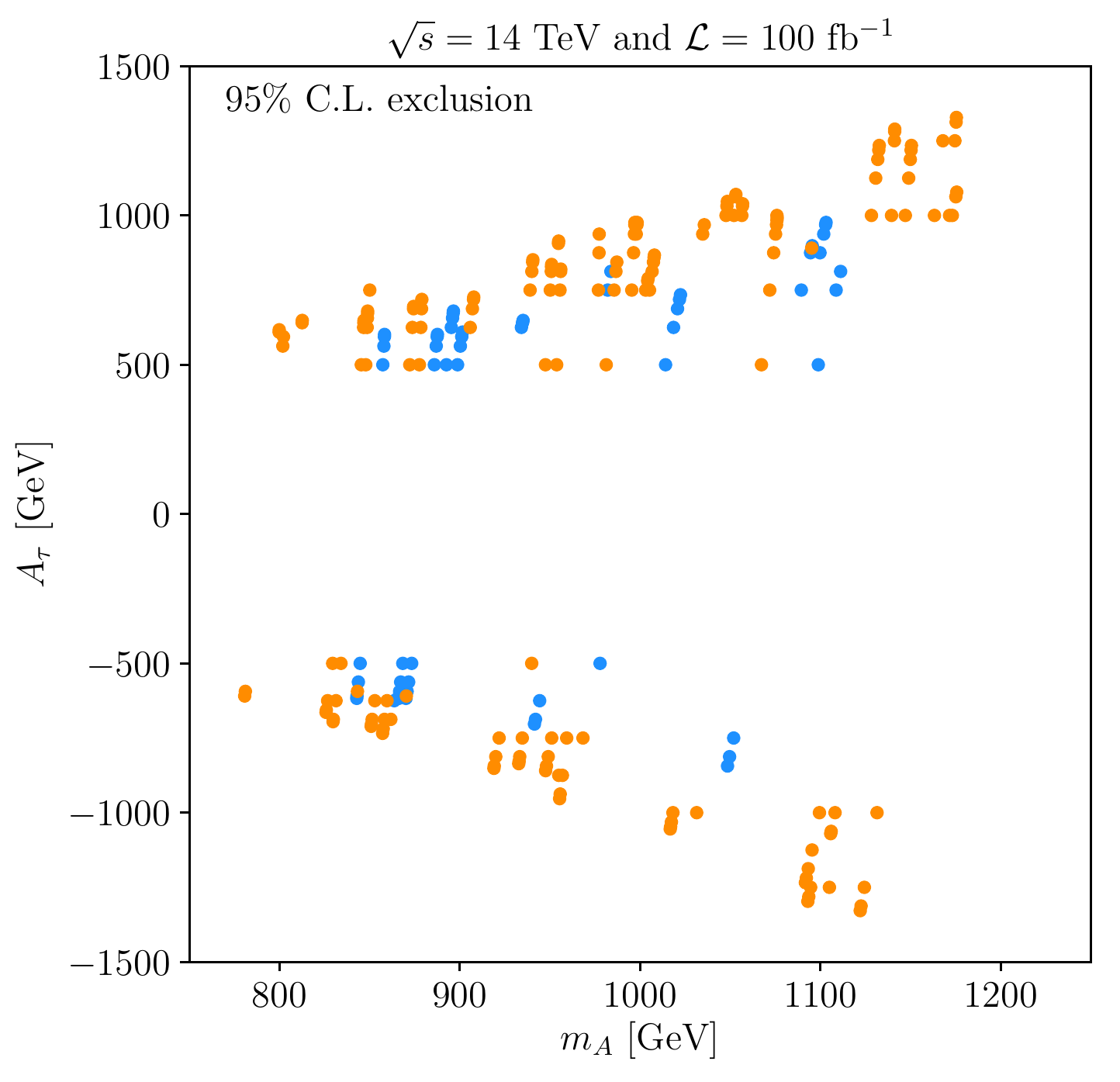} &
			\includegraphics[scale=0.40]{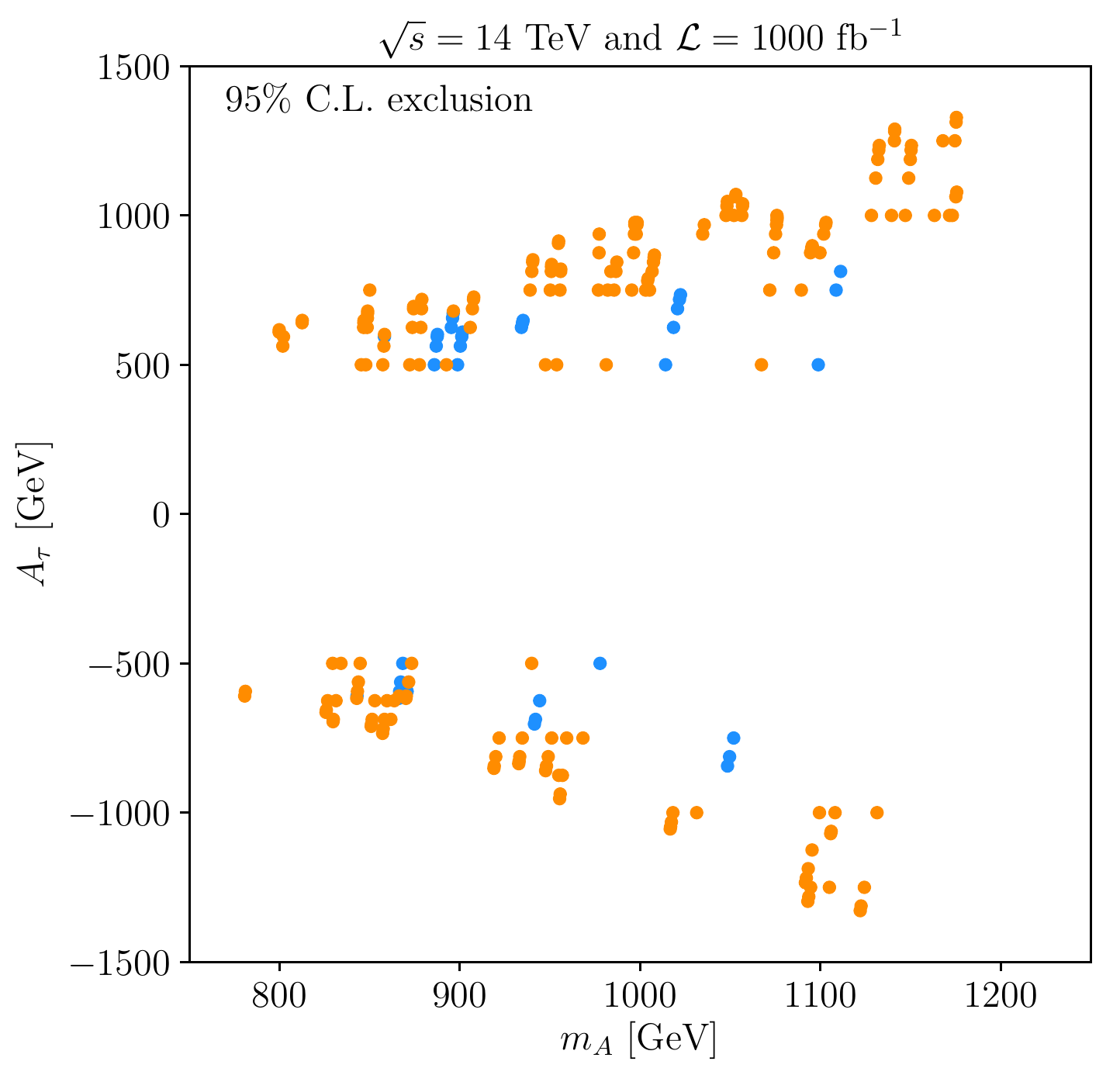} \\
			\includegraphics[scale=0.40]{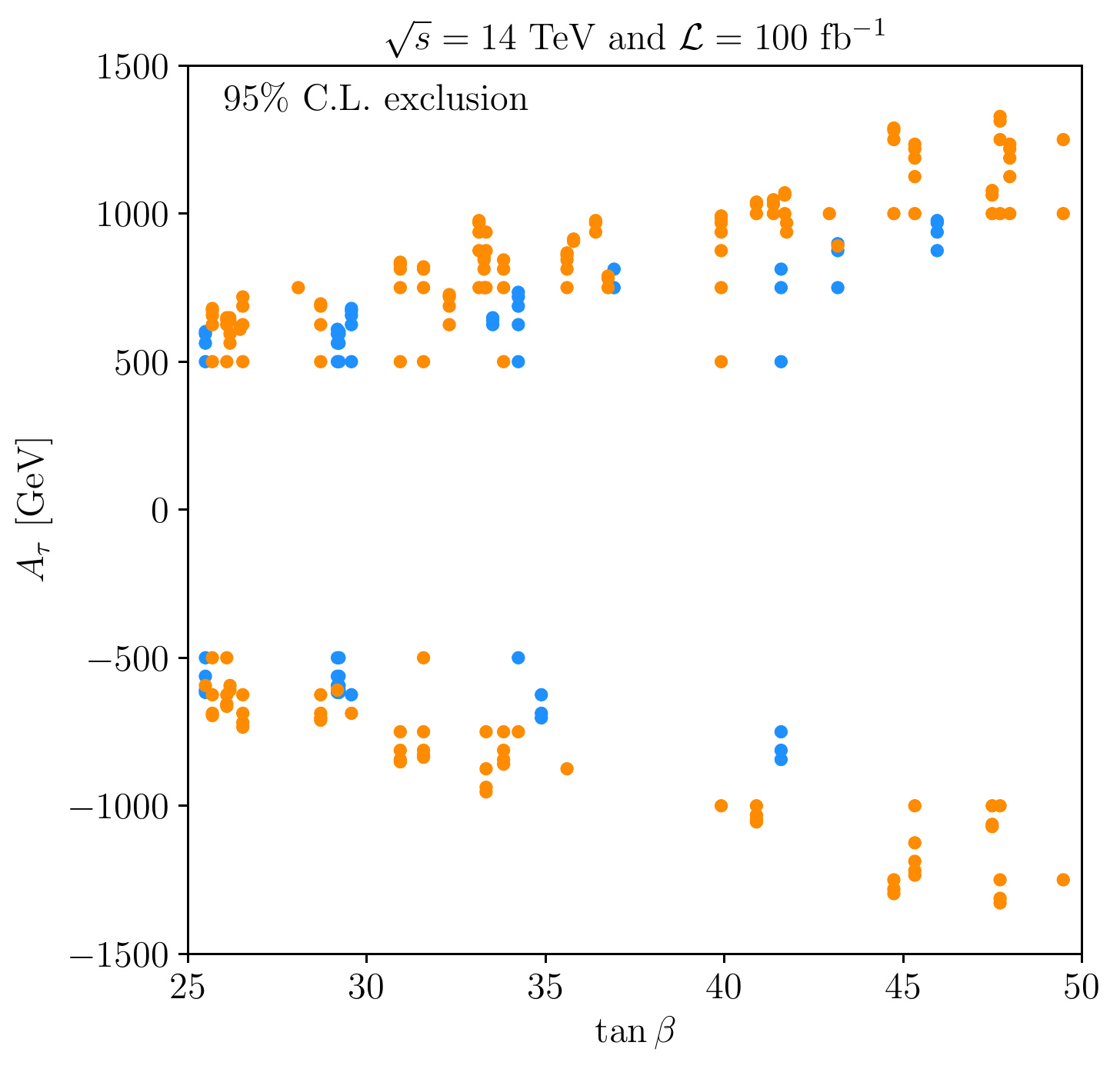} &
			\includegraphics[scale=0.40]{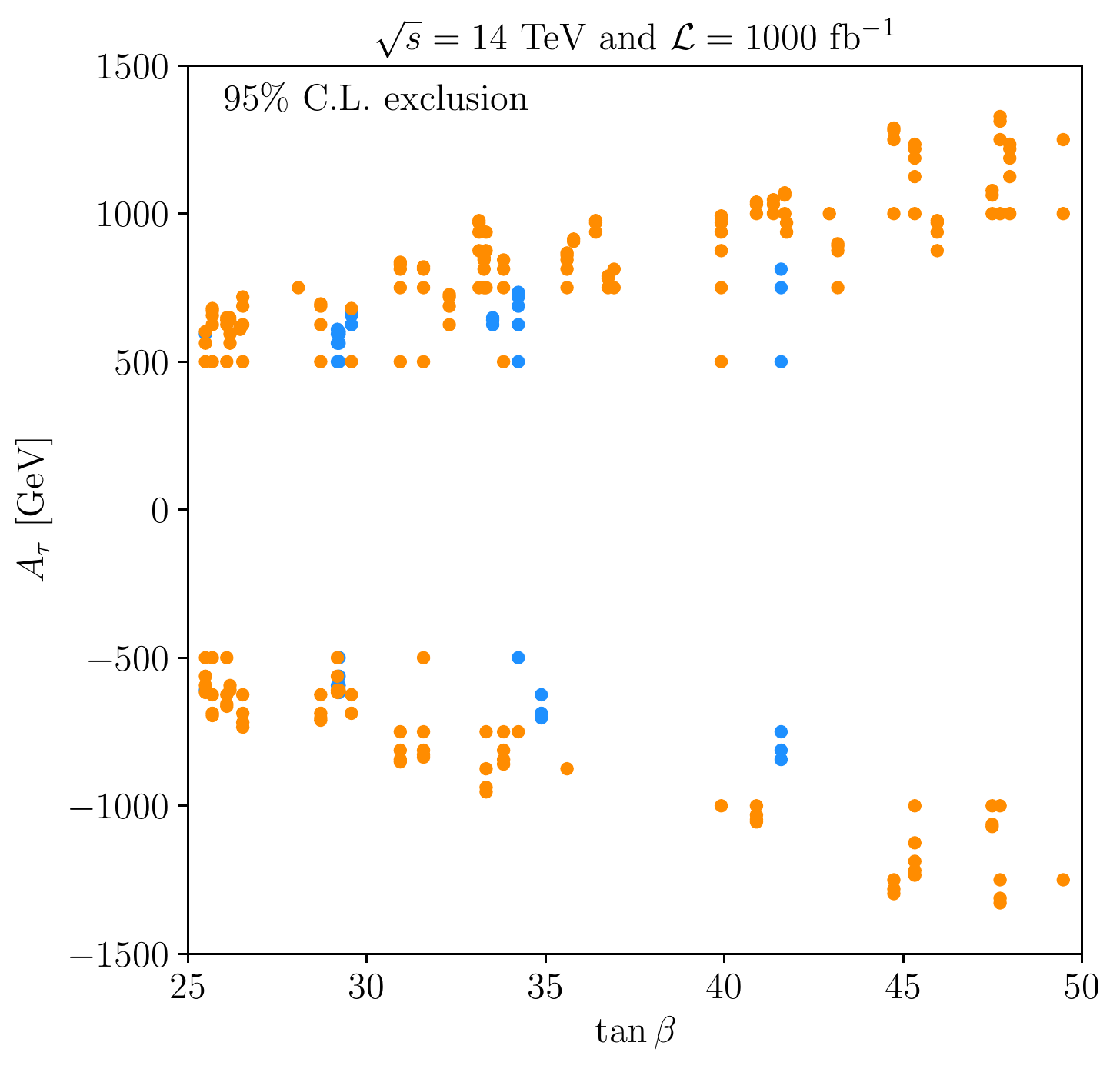}
		\end{tabular}
		\caption{Potential exclusion at 95$\%$ C.L. in the [$m_A$, $A_{\tau}$] plane (top plots) and [$\tan\beta$, $A_{\tau}$] plane (bottom plots), within Scenario II, for a center-of-mass energy of $\sqrt{s} =$ 14 TeV and total integrated luminosities of 100 fb$^{-1}$ (left panels) and 1000 fb$^{-1}$ (right panels). Benchmarks excluded by the analysis are shown in orange, while the allowed ones are shown in blue.}
		\label{fig:ataumAtanb-exc-12}
	\end{center}
\end{figure}

In Fig.~\ref{fig:ataumAtanb-exc-12} we show the results of applying the 95\% C.L. exclusion condition of Eq.~(\ref{eq:exclusion}) to the Scenario-II benchmarks in the [$m_A$, $A_{\tau}$] (top) and [$\tan\beta$, $A_{\tau}$] (bottom) planes for total integrated luminosities of 100 fb$^{-1}$ (left) and 1000 fb$^{-1}$ (right). From the plots on the top panel we see that the exclusion power of the search strategy extends to masses up to 1200 GeV. Again, for a given mass, the sensitivity increases for higher values of $|A_\tau|$. For $\mathcal{L}=100$ fb$^{-1}$ all the points with $m_A\lesssim 840$ GeV are excluded even for the lowest values of $A_{\tau}$ considered here ($|A_\tau|$ $\sim$ 500 GeV). For $\mathcal{L}=1000$ fb$^{-1}$ this conclusion is valid for masses below 860 GeV. Above these masses the sensitivity depends on the specific values of $A_\tau$ and $\tan\beta$. Regarding this last parameter, we see from the bottom panels that all the points with $\tan\beta \geq$ 47 ($\tan\beta \geq$ 42) are excluded for $\mathcal{L}=100$ fb$^{-1}$ (1000 fb$^{-1}$). The main reason for this behavior is the fact that larger values of $\tan\beta$ enhance the $b$-quark annihilation production cross section and the coupling with the staus at the same time, so that the search strategy is quite efficient even for benchmark points with large $m_A$ and $m_H$ and relatively low values of $A_\tau$ ($|A_\tau|$ $\sim$ 500 GeV).

\begin{figure}[ht]
	\begin{center}
		\begin{tabular}{cc}
			\centering
			\hspace*{-3mm}
			\includegraphics[scale=0.40]{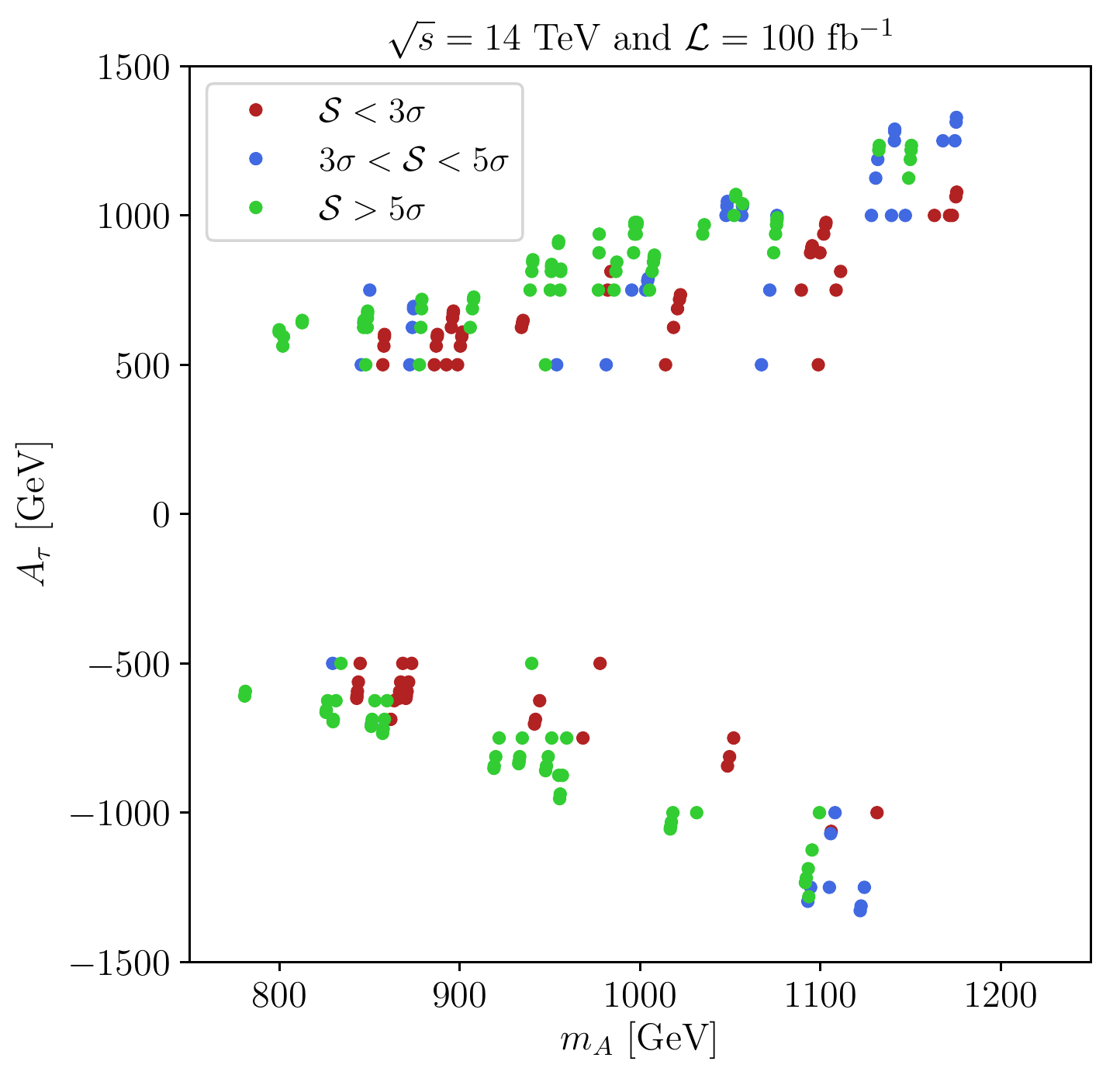} &
			\includegraphics[scale=0.40]{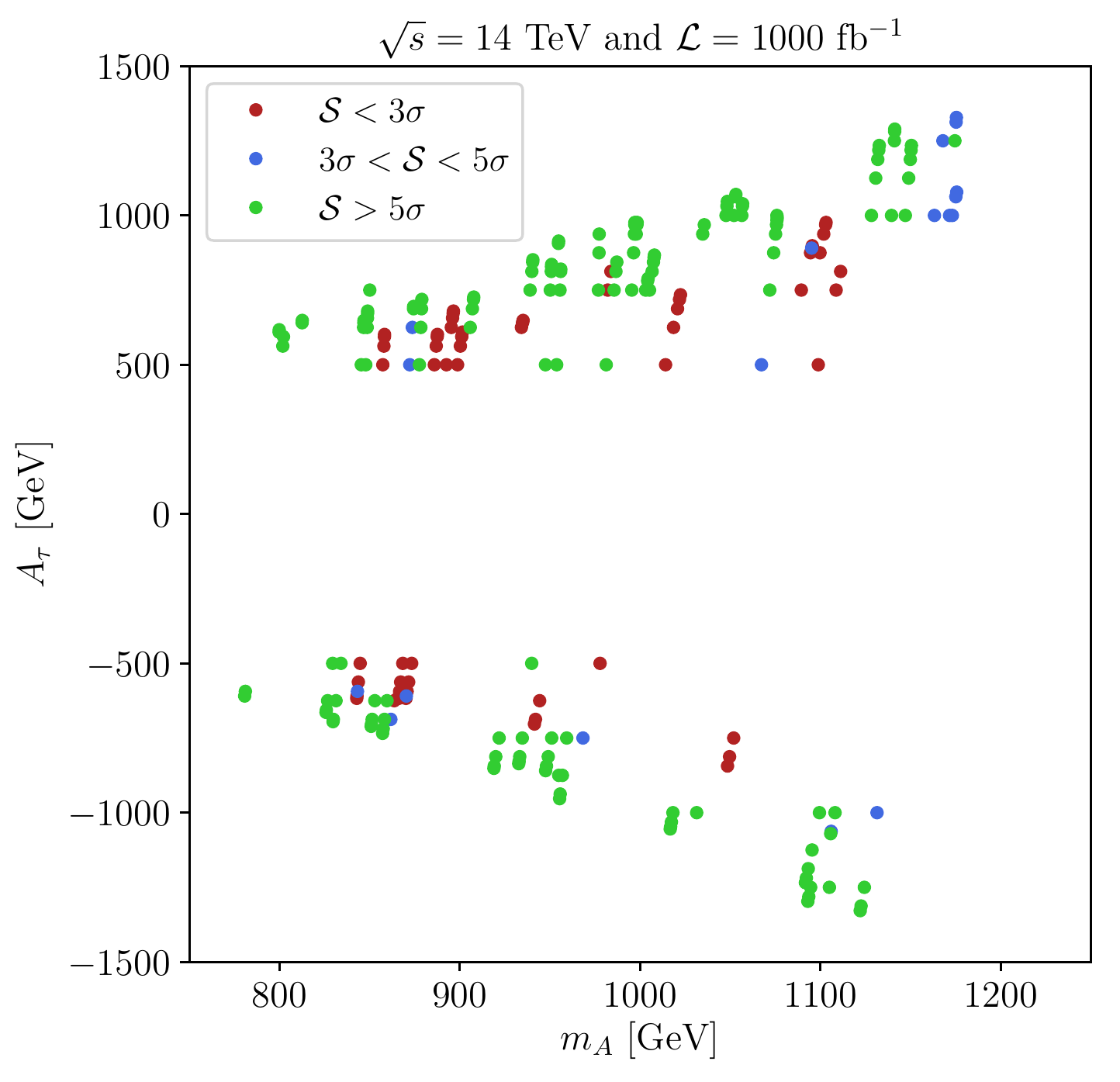}\\
			\includegraphics[scale=0.40]{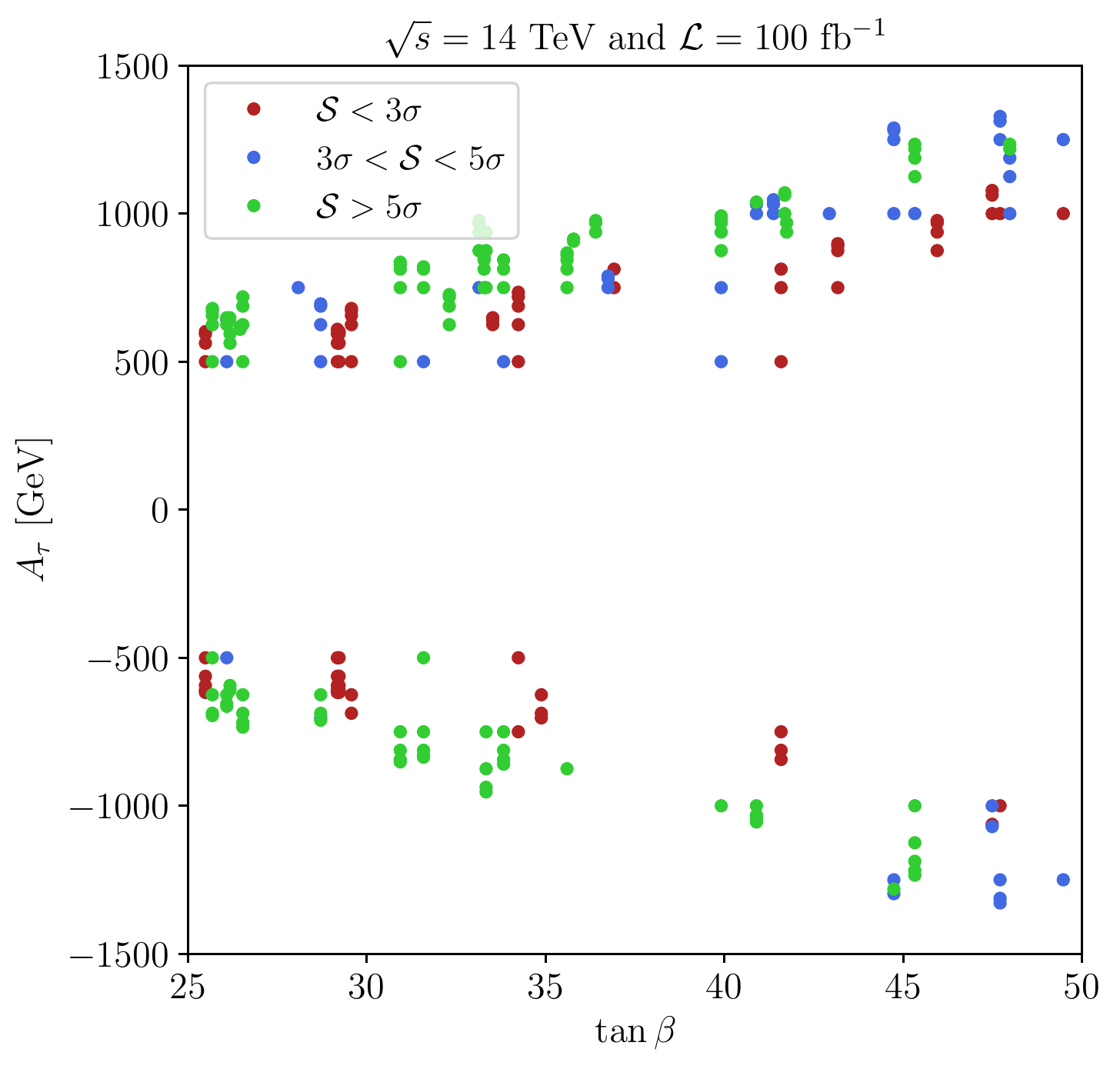} &
			\includegraphics[scale=0.40]{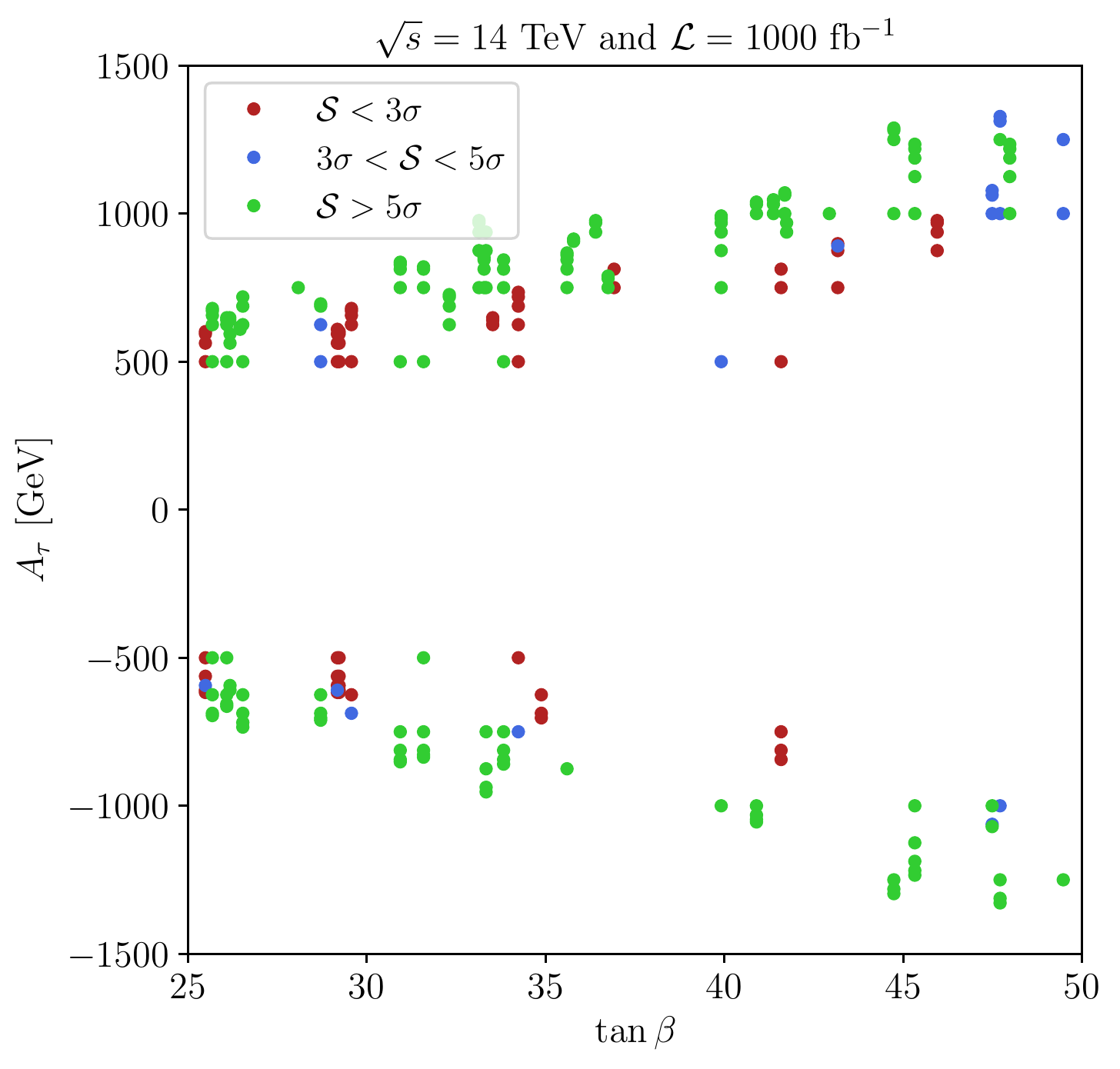}
		\end{tabular}
		\caption{Signal significance in the [$m_A$, $A_{\tau}$] plane (top plots) and [$\tan\beta$, $A_{\tau}$] plane (bottom plots), within Scenario II for a center-of-mass energy of $\sqrt{s} =$ 14 TeV and total integrated luminosities of 100 fb$^{-1}$ (left panels) and 1000 fb$^{-1}$ (right panels). Benchmarks with significances below the evidence level ($\mathcal{S}<3\sigma$), between the evidence level and the discovery level ($3\sigma <\mathcal{S}<5\sigma$) and above the discovery level ($\mathcal{S}>5\sigma$) are shown in red, blue and green, respectively.}
		\label{fig:ataumAtanb-dis-12}
	\end{center}
\end{figure}

In Fig.~\ref{fig:ataumAtanb-dis-12} we depict, in the same parameters planes as in Fig.~\ref{fig:ataumAtanb-exc-12}, the results corresponding to the signal significance prospects for $\mathcal{L}=100$ fb$^{-1}$ (left) and 1000 fb$^{-1}$ (right). Similarly to the results discussed above for the exclusion limits, benchmarks with higher values of $|A_\tau|$ are more likely to be detected by the search strategy. Specifically, all the benchmarks with $|A_\tau|$ $\geq$ 1125 GeV have signal significances above 3$\sigma$, with most of them reaching the discovery level. We note that for $\mathcal{L}=1000$ fb$^{-1}$ some of these benchmarks lie in the high mass region with $m_A$ $\geq$ 1110 GeV. On the other hand, for both luminosities the points with masses below 840 GeV exhibit significances at the discovery level in spite of the relatively small value of $A_\tau$ ($\sim$ 500 GeV). The same conclusions about the impact of the value of $A_\tau$ in the significance can be read off on the lower panels. In addition, we also see that for the case of $\mathcal{L}=1000$ fb$^{-1}$ all the points with $\tan\beta$ $\geq$ 46 reach significances above 3$\sigma$. In fact, more than half of the benchmarks lying in that region correspond to significances at the discovery level.

\begin{figure}[ht]
	\begin{center}
		\begin{tabular}{cc}
			\centering
			\hspace*{-3mm}
			\includegraphics[scale=0.40]{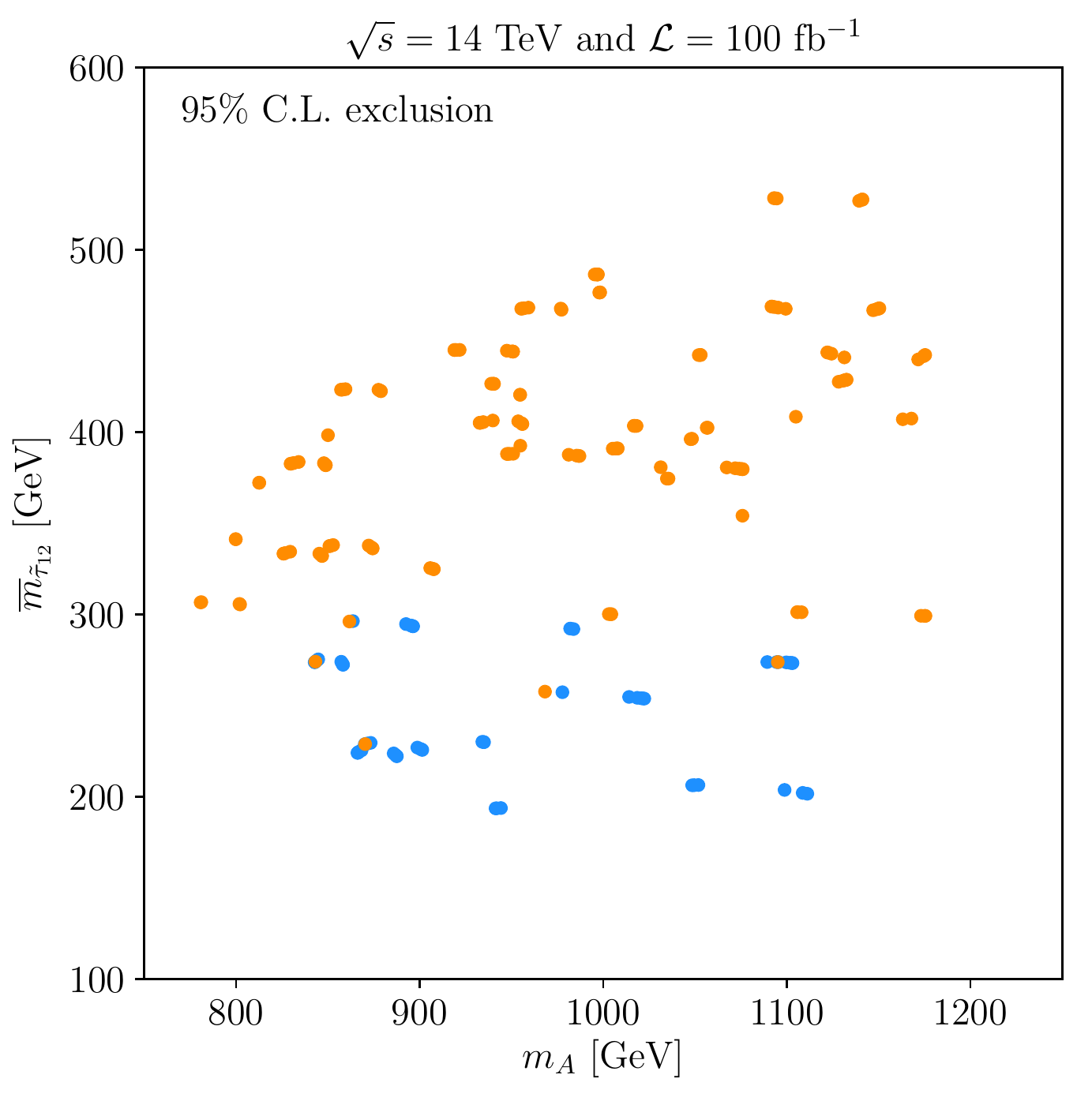} &
			\includegraphics[scale=0.40]{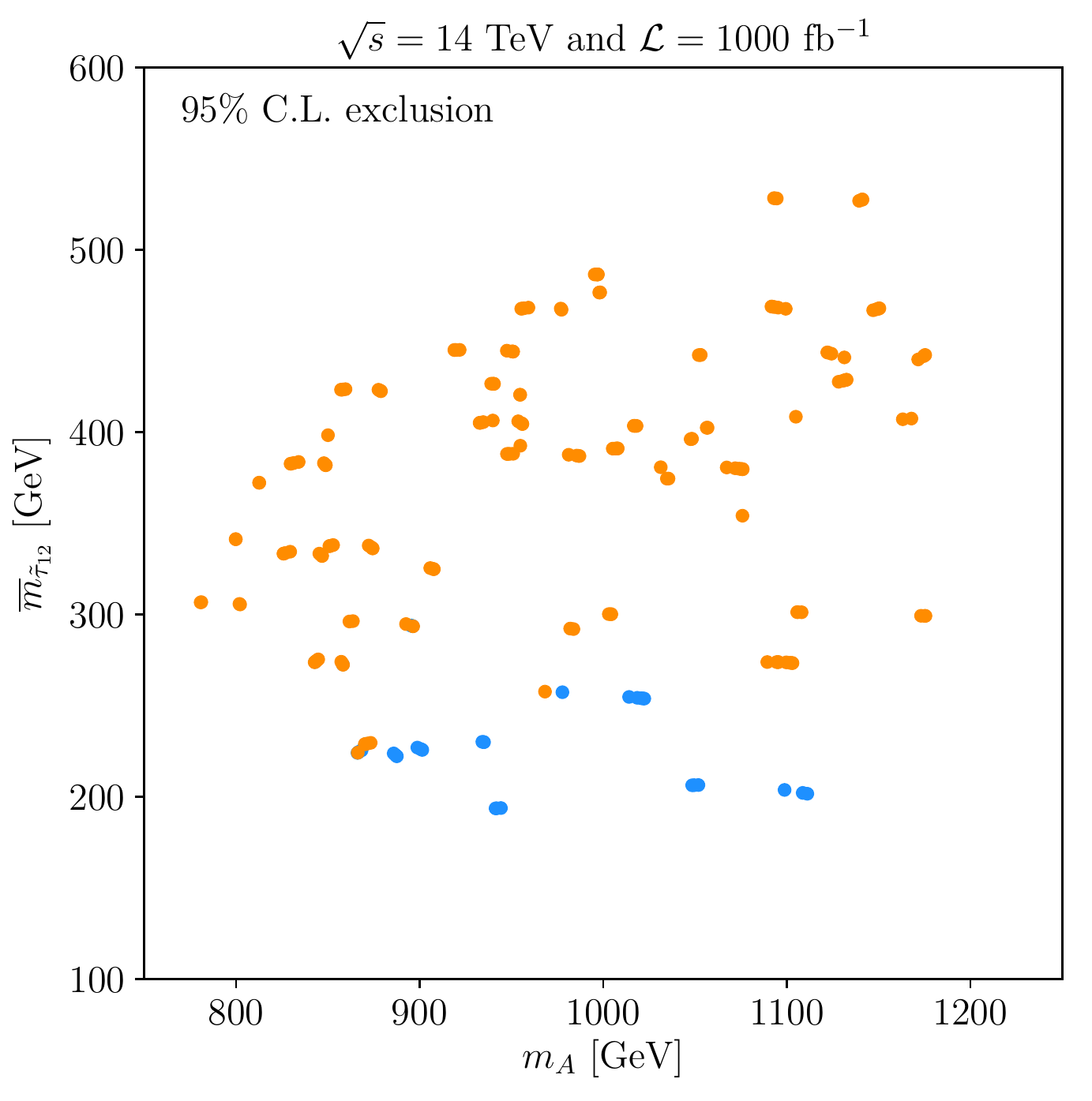}\\
			\includegraphics[scale=0.40]{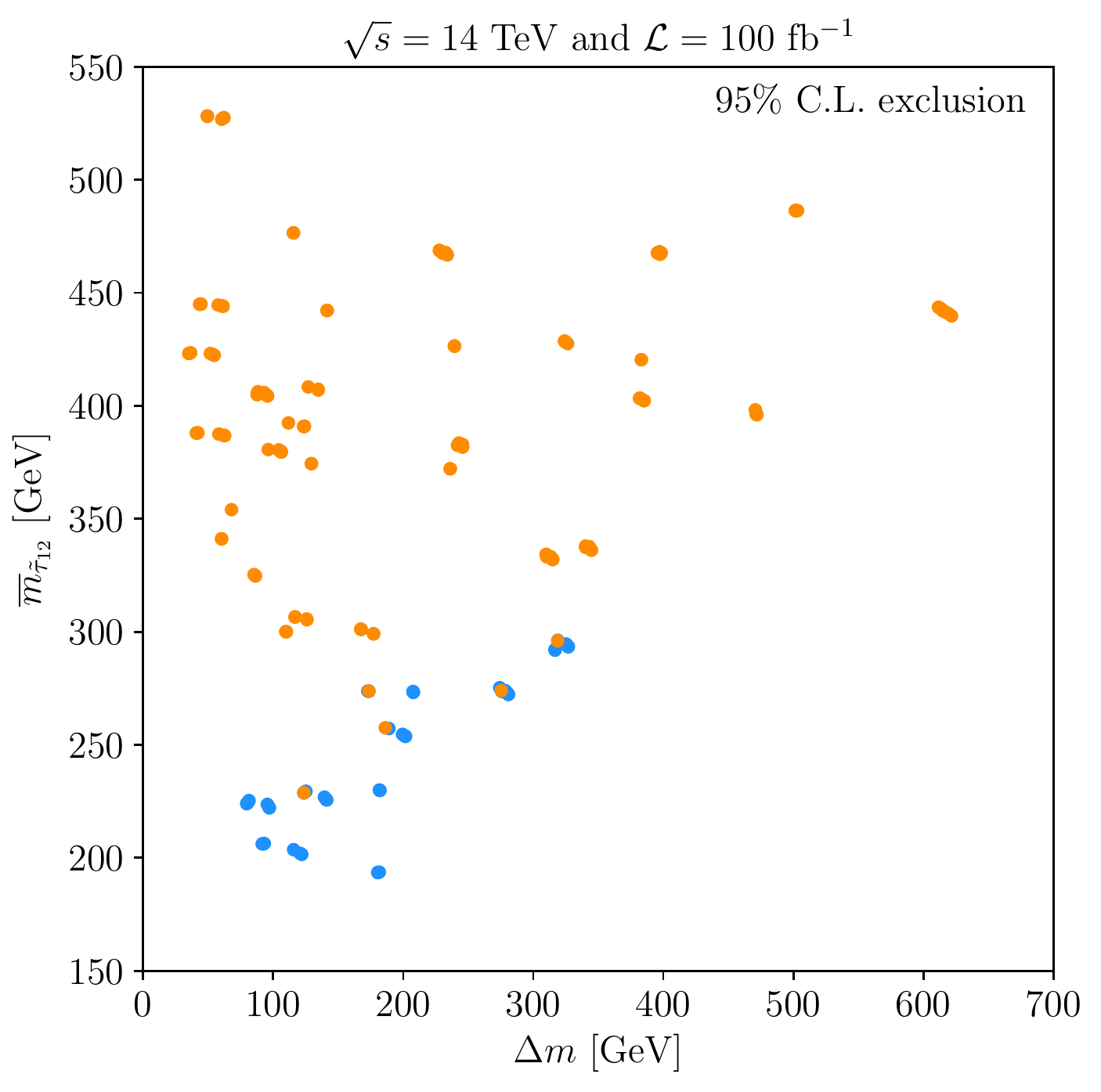} &
			\includegraphics[scale=0.40]{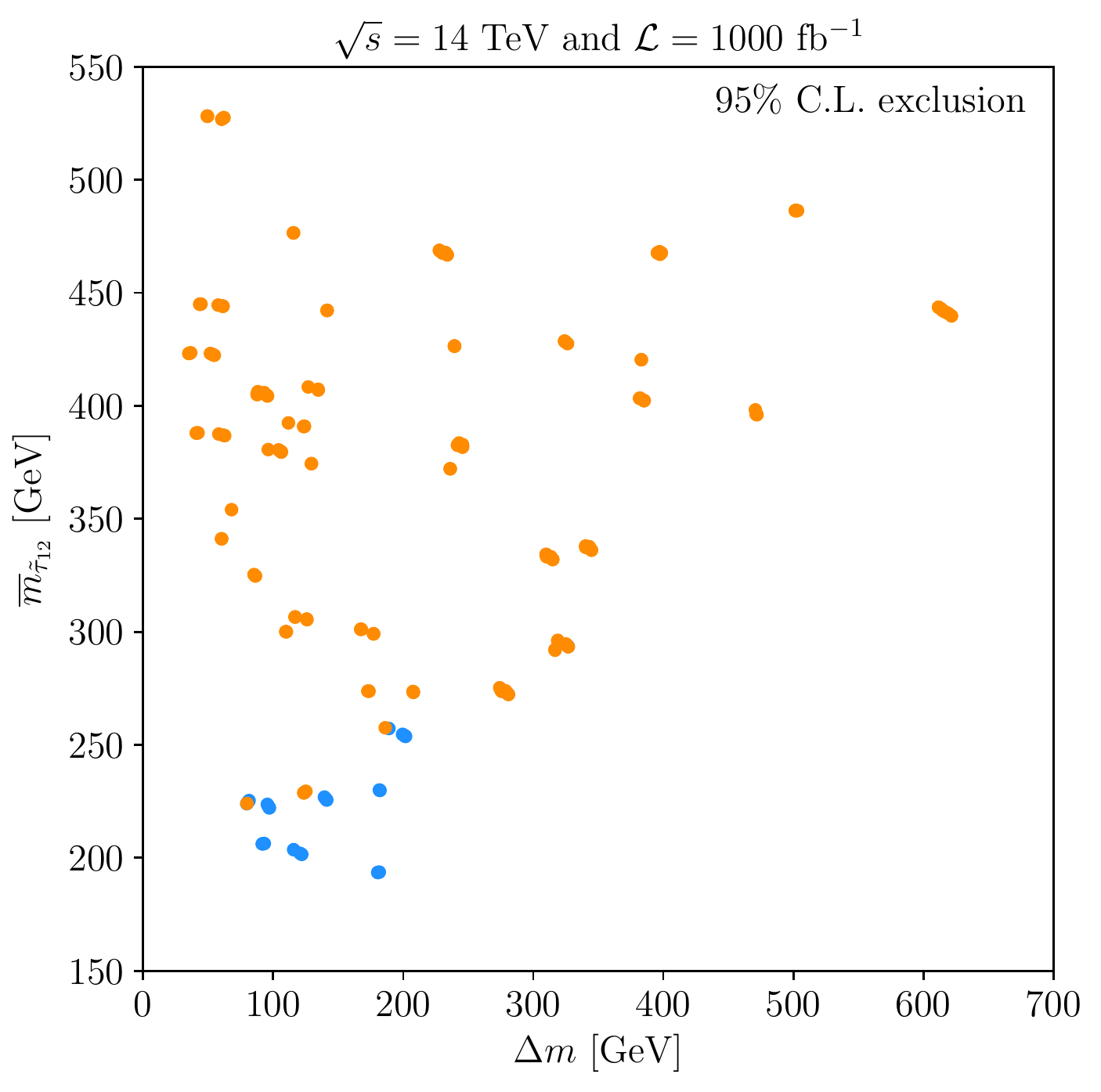}
		\end{tabular}
		\caption{Potential exclusion at 95$\%$ C.L. in the [$m_A$,  $\overline{m}_{\tilde \tau_{12}}$] plane (top) and [$\Delta m$,  $\overline{m}_{\tilde \tau_{12}}$] (bottom), within Scenario II, for a center-of-mass energy of $\sqrt{s} =$ 14 TeV and total integrated luminosities of 100 fb$^{-1}$ (left panels) and 1000 fb$^{-1}$ (right panels). Benchmarks excluded by the analysis are shown in orange, while the allowed ones are shown in blue.}
		\label{fig:m12mA-exc-12}
	\end{center}
\end{figure}

Let us turn now to the results in terms of the stau variables defined in Eq.~\eqref{eq:stauvariables}. These are shown in Fig.~\ref{fig:m12mA-exc-12} in the planes [$m_A$, $\overline{m}_{\stau_{12}}$] (top panels) and [$\Delta m$, $\overline{m}_{\stau_{12}}$] (bottom panels) for $\mathcal{L}=100$ fb$^{-1}$ (left) and $\mathcal{L}=1000$ fb$^{-1}$ (right). We see that all the points with $\avstau \geq$ 300 GeV are excluded in the case of $\mathcal{L}=100$ fb$^{-1}$, while this value decreases to $\avstau \geq$ 270 GeV for $\mathcal{L}=1000$ fb$^{-1}$. This conclusion is also visible in the [$\Delta m$, $\overline{m}_{\stau_{12}}$] plane, from which we also note that in the case of $\mathcal{L}=100$ fb$^{-1}$ the points with $\avstau \geq$ 300 GeV appear to be easily tested when the value of $\Delta m$ is smaller. The same conclusion can be drawn from the plots corresponding to $\mathcal{L}=1000$ fb$^{-1}$ for points with average stau masses below 270 GeV. This behaviour is due to the fact that the $m_{T2}$ cut is more efficient for smaller values of $\Delta m$ since this kinematic variable was originally designed to tag a pair of decaying particles with equal mass.

\begin{figure}[ht]
	\begin{center}
		\begin{tabular}{cc}
			\centering
			\hspace*{-3mm}
			\includegraphics[scale=0.40]{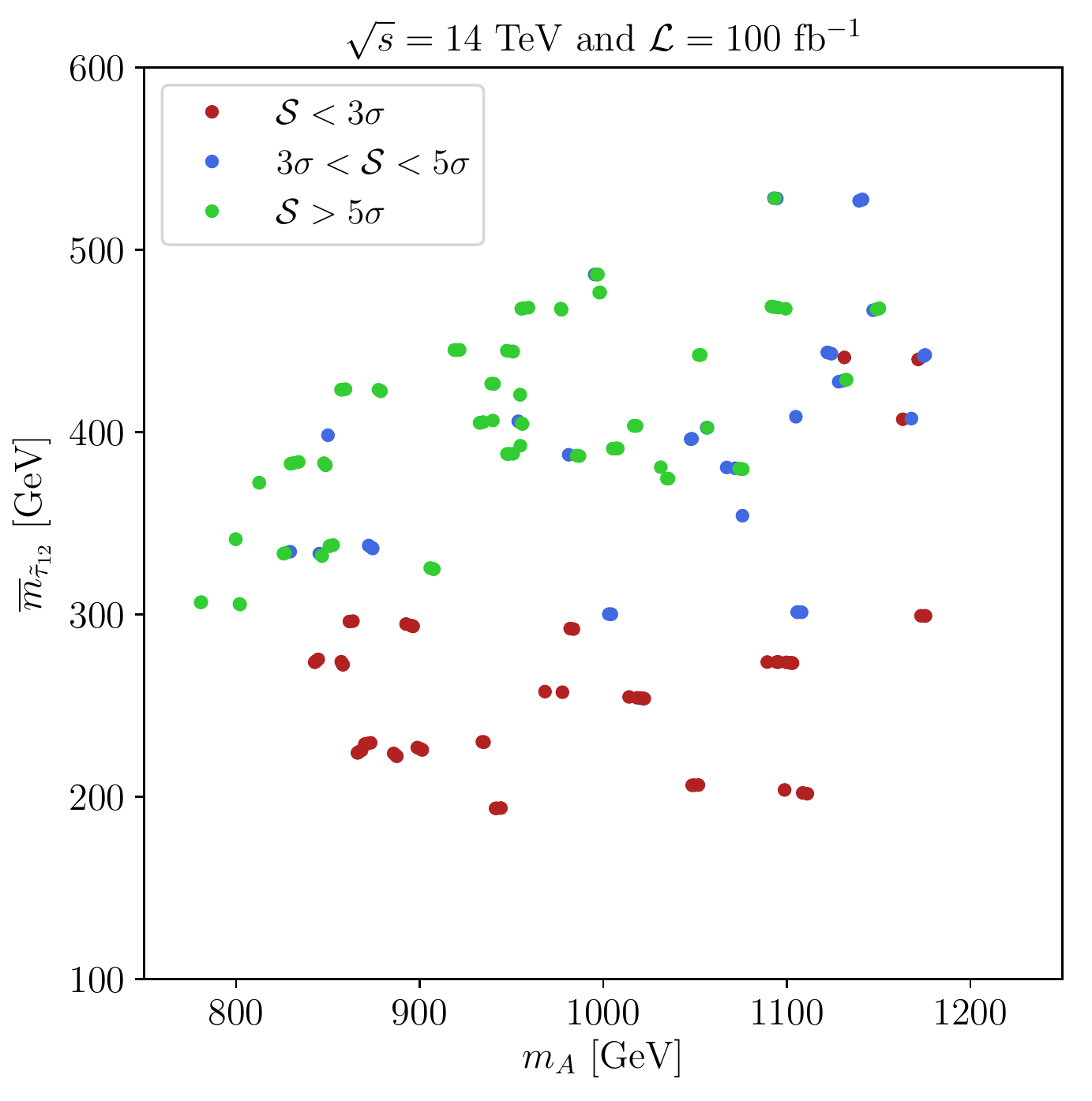} &
			\includegraphics[scale=0.40]{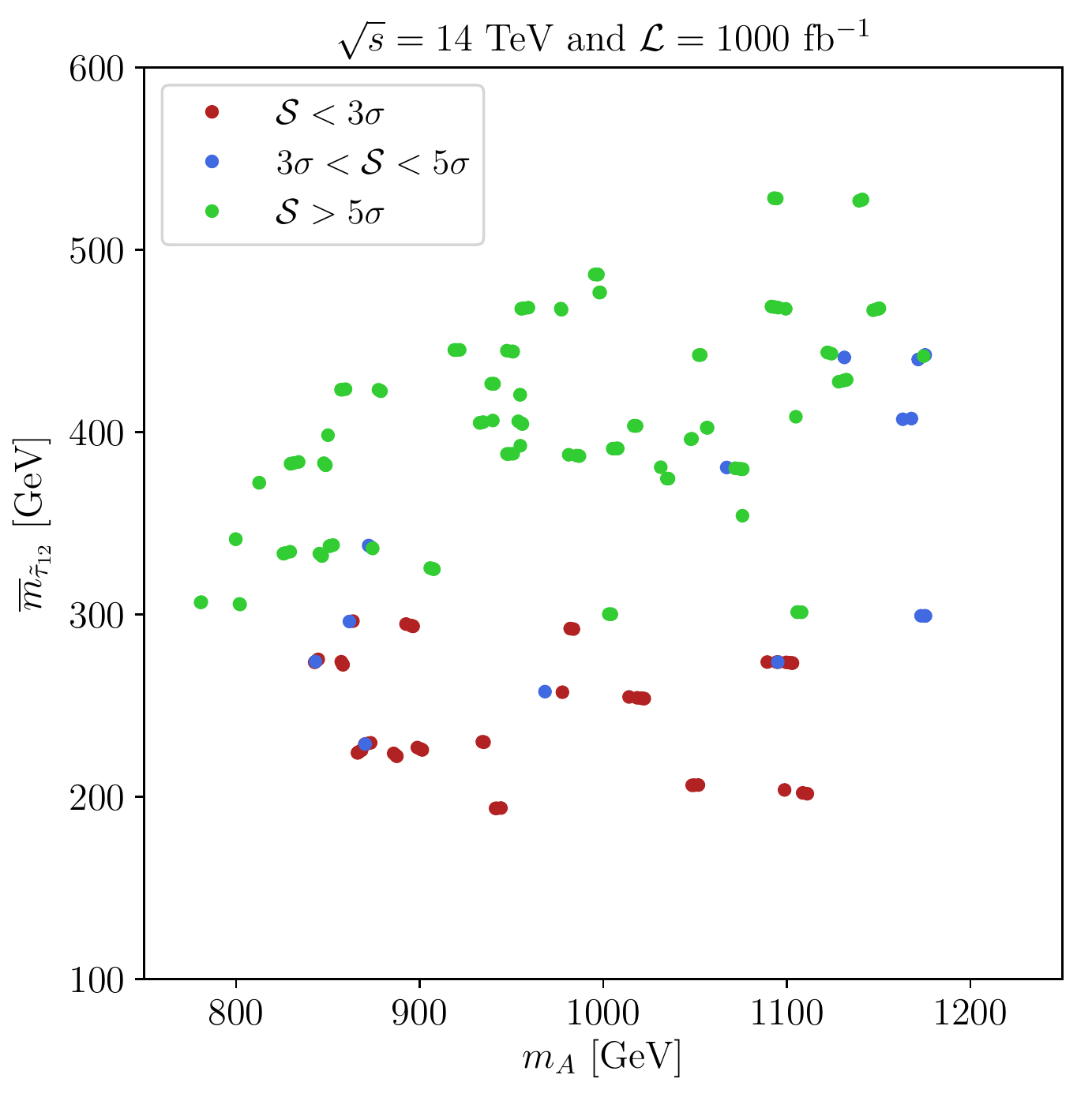}\\
			\includegraphics[scale=0.40]{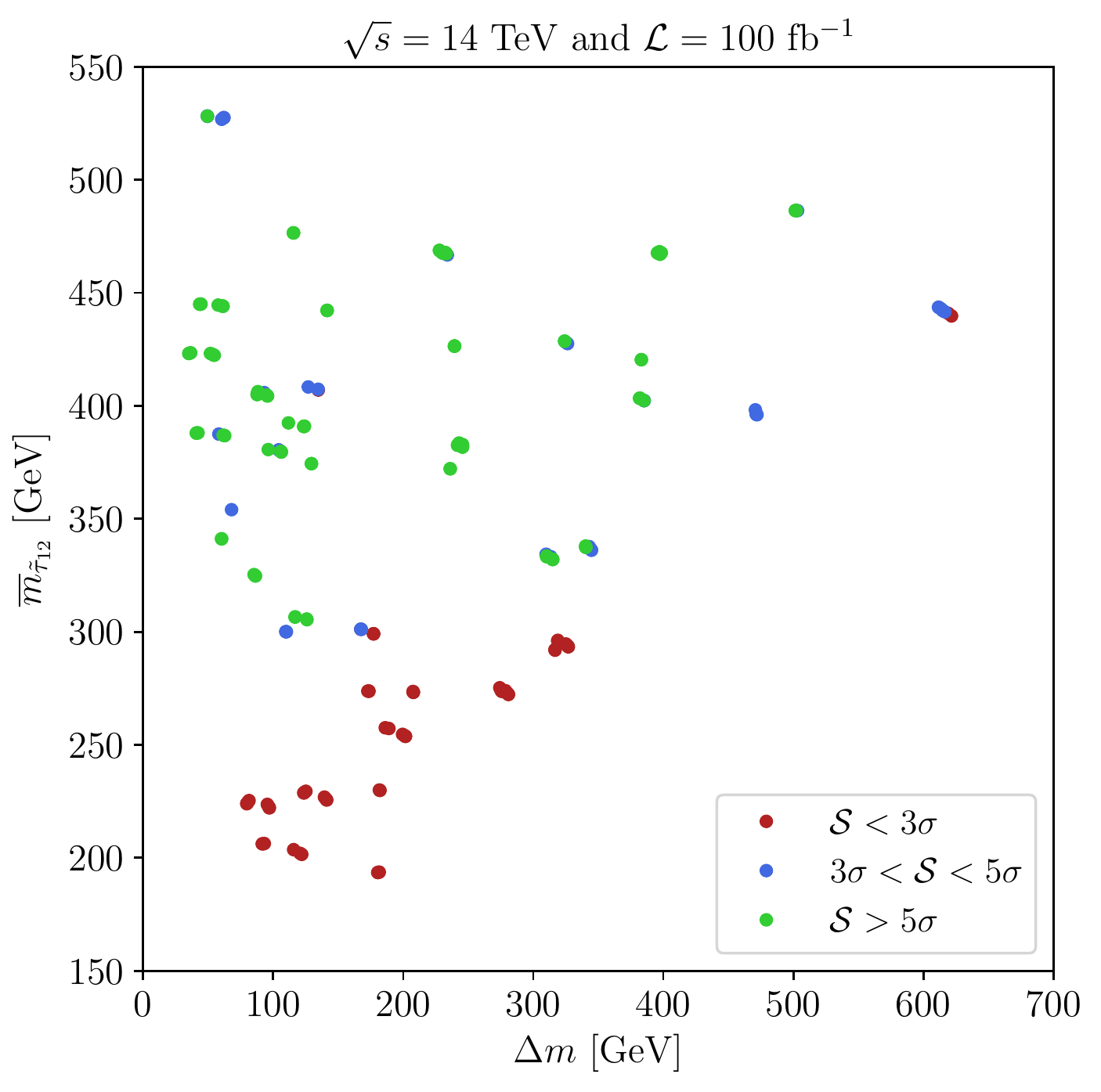} &
			\includegraphics[scale=0.40]{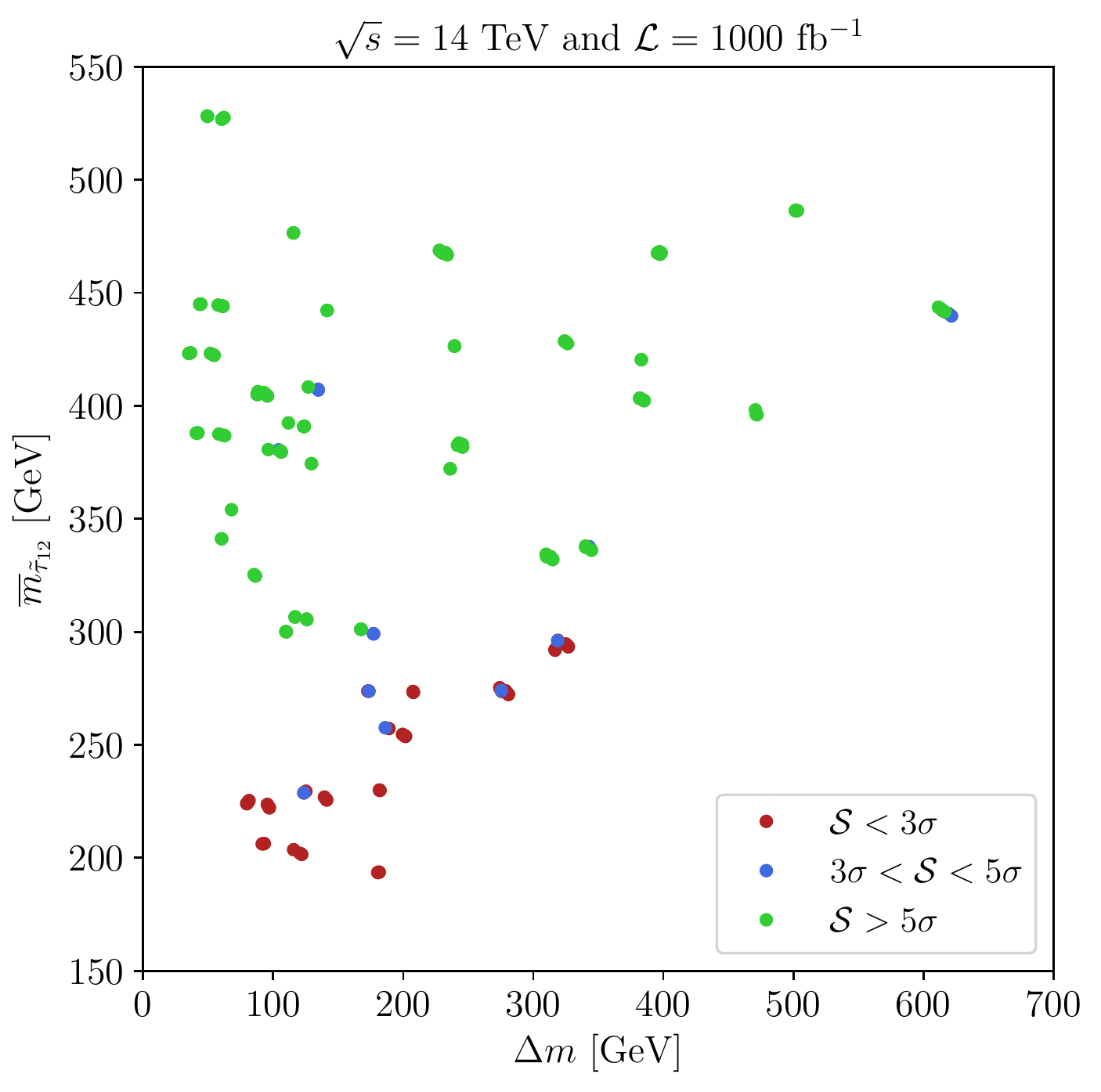}			
		\end{tabular}
		\caption{Signal significance in the [$m_A$,  $\overline{m}_{\tilde \tau_{12}}$] plane (top) and [$\Delta m$,  $\overline{m}_{\tilde \tau_{12}}$] (bottom), within Scenario II for a center-of-mass energy of $\sqrt{s} =$ 14 TeV and total integrated luminosities of 100 fb$^{-1}$ (left panels) and 1000 fb$^{-1}$ (right panels). Benchmarks with significances below the evidence level ($\mathcal{S}<3\sigma$), between the evidence level and the discovery level ($3\sigma <\mathcal{S}<5\sigma$) and above the discovery level ($\mathcal{S}>5\sigma$) are shown in red, blue and green, respectively.}
		\label{fig:m12mA-dis-12}
	\end{center}
\end{figure}

The results of the signal significance in terms of stau variables are shown in Fig.~\ref{fig:m12mA-dis-12}. In this case most of the points with $\avstau \geq$ 300 GeV reach the discovery level. In contrast, all the points with stau masses below this value cannot be probed with the analysis at $\mathcal{L}=100$ fb$^{-1}$. This situation improves only a bit at $\mathcal{L}=1000$ fb$^{-1}$, since in this case some points with $\avstau \leq$ 300 GeV show evidence level. However, there are no points reaching the discovery level in this region. The increase in luminosity from 100 fb$^{-1}$ to 1000 fb$^{-1}$ also makes that a considerable number of points in the mass region above 300 GeV with large values of $m_A$ or $\Delta m$ become accessible. In the case of the parameter $\Delta m$, the behaviour is the same as in the exclusion plots. For a given $\avstau$ value, the efficiency of the search strategy increases for smaller $\Delta m$ values.

\begin{figure}[ht]
	\begin{center}
		\includegraphics[scale=0.5]{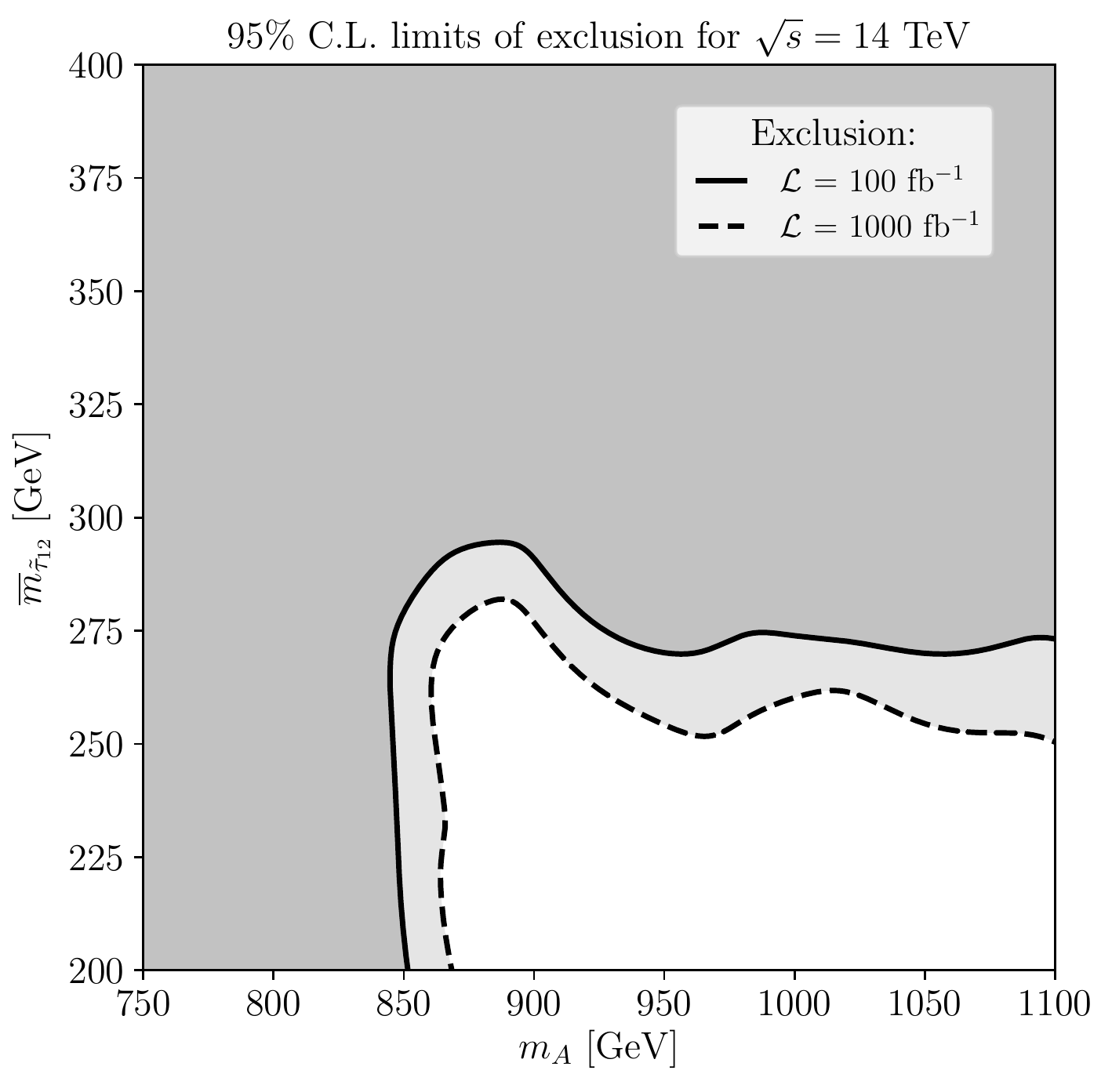} 
		\caption{Exclusion limits at 95\% C.L. in the [$m_A$, $\overline{m}_{\tilde \tau_{12}}$] plane, within Scenario II, for a center-of-mass energy of $\sqrt{s} =$ 14 TeV and total integrated luminosities of 100 fb$^{-1}$ (dark gray) and 1000 fb$^{-1}$ (light gray).}
		\label{fig:mstau12mA-exc}
	\end{center}
\end{figure}

As we did in Section~\ref{subsec:scenarioi}, we can interpolate the obtained results and show them in contour-line plots. In Fig.~\ref{fig:mstau12mA-exc} the contour-line plot of the exclusion potential at 95\% C.L. in the [$m_A$, $\overline{m}_{\tilde \tau_{12}}$] plane is depicted. In this figure the dark gray area represents the exclusion region for $\mathcal{L}=100$ fb$^{-1}$, while the light gray area corresponds to $\mathcal{L}=1000$ fb$^{-1}$. We can observe that the search strategy is able to exclude the region with $m_A\leq 850$ GeV. This region is slightly increased to $m_A\leq 870$ GeV for $\mathcal{L}=1000$ fb$^{-1}$. Above these masses, the search strategy excludes in general average stau masses that are greater than 275-290 GeV and 250-275 GeV for $\mathcal{L}=100$ fb$^{-1}$ and $\mathcal{L}=1000$ fb$^{-1}$, respectively. As in the case of Scenario I, the proposed search strategy is not sensitive to the region of low values of stau masses due to the specific kinematic variables that drive its discrimination power.

\begin{figure}[ht]
	\begin{center}
		\begin{tabular}{cc}
			\centering
			\hspace*{-3mm}
			\includegraphics[scale=0.5]{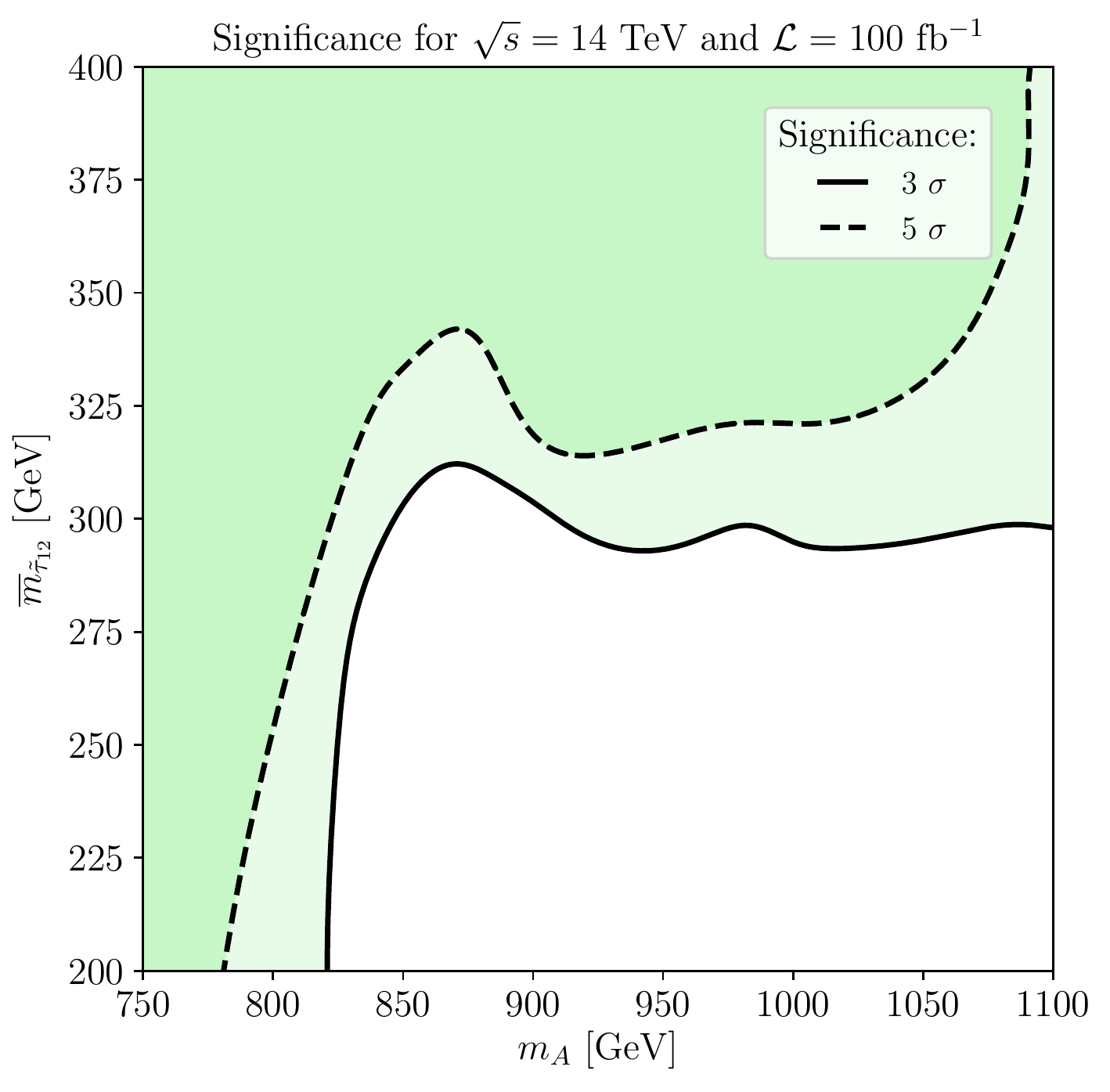} &
			\includegraphics[scale=0.5]{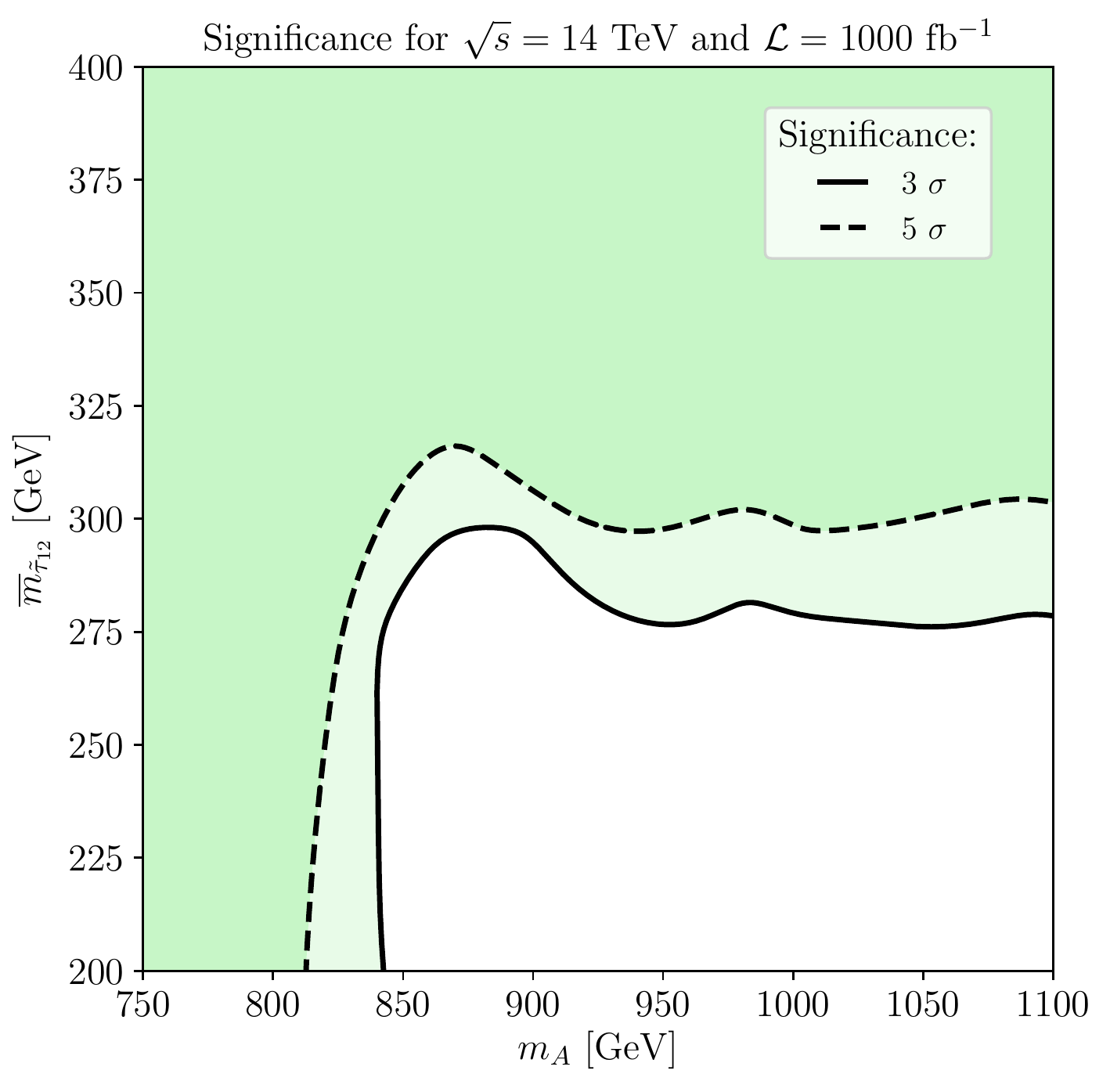}
		\end{tabular}
		\caption{Signal significance in the [$m_A$, $\overline{m}_{\tilde \tau_{12}}$] plane, within Scenario II, for a center-of-mass energy of $\sqrt{s} =$ 14 TeV and total integrated luminosities of 100 fb$^{-1}$ (left panel) and 1000 fb$^{-1}$ (right panel). The dark green and light green areas correspond to significances at the evidence level (3$\sigma$) and at the discovery level (5$\sigma$).}
		\label{fig:mstau12mA-dis}
	\end{center}
\end{figure}

In Fig.~\ref{fig:mstau12mA-dis} we show a similar contour-line plot as in Fig.~\ref{fig:mstau12mA-exc} but for the signal significances at $\mathcal{L}=100$ fb$^{-1}$ (left) and $\mathcal{L}=1000$ fb$^{-1}$ (right). The dark (light) gray area corresponds to significances at the discovery (evidence) level. For $\mathcal{L}=100$ fb$^{-1}$ the evidence level is reached for masses $m_A\leq$ 825 GeV regardless the value of $\avstau$, while for masses above 825 GeV the average stau mass needs to be larger than 300 GeV. For values of $m_A$ below $\sim 780$ GeV significances at the discovery level are obtained within all the considered $\avstau$ range. It is interesting to note that the discovery contour line drastically grows towards large values of $\avstau$ for $m_A$ $>$ 1090 GeV. This is because for such high values of $m_A$, stau masses above 450 GeV are required in order to reach 5$\sigma$ significances (see the upper left panel of Fig.~\ref{fig:m12mA-dis-12}). For $\mathcal{L}=1000$ fb$^{-1}$ the 3$\sigma$ region extends to  $m_A$ $\sim$ 850 GeV regardless the value of $\avstau$, and for $m_A$ $\geq$ 850 GeV significances at the evidence level are obtained for $\avstau$ above 275-290 GeV. By looking at the 5$\sigma$ contour line, we conclude that significances at the discovery level can be obtained for $m_A$ $<$ 815 GeV for any $\avstau$ within the range under study. Moreover, larger masses can still reach significances at the discovery level if $\avstau$ is approximately above 300-320 GeV. This is in contrast to the case of $\mathcal{L}=100$ fb$^{-1}$, where the region in which $m_A$ $>$ 1090 GeV is particularly challenging and requires quite larger values of $\avstau$ in order to reach the discovery level.

\subsection{Potential discrimination between stau mixing scenarios}
\label{sec:discrimination}

We explore now the possibility to distinguish the two mixing scenarios once they reach the discovery level with our search strategy and in the specific case in which they exhibit similar relevant mass spectra ($m_{H},m_{\widetilde{\tau}_1}$, and $m_{\widetilde{\tau}_2}$). As stated above, in Scenario I the tau leptons in the final state arise from the decay of a pair of $\widetilde{\tau}_1$, while in Scenario II they originate from the decay of the pair $\widetilde{\tau}_1\widetilde{\tau}^*_2$ or its conjugate ($\widetilde{\tau}_1^{*}\widetilde{\tau}_2$). Thus, the main difference between the signals associated to these scenarios relies on the difference between the stau masses. In this sense, one may expect that kinematic variables such as  $m^{\tau_1}_{T}$, $m^{\tau_2}_{T}$, and $m_{T2}$ will be sensitive to the mass splitting and therefore be well suited to discriminate between scenarios. 

\renewcommand{\arraystretch}{1.5}
\begin{table}[ht!]
	\begin{center}
	\begin{tabular}{cccc}
	\hline 
	Benchmark	&  $m_{H}$  &  $m_{\widetilde{\tau}_1}$ & $m_{\widetilde{\tau}_2}$\\ \hline
	SI-7	&  951 GeV & 367 GeV & 409 GeV  \\
	SI-20	&  1075 GeV & 320 GeV & 388 GeV    \\
	SII-47	&  951 GeV & 367 GeV & 409 GeV    \\
	SII-82	& 1099 GeV & 352 GeV & 583 GeV   \\ \hline
	\end{tabular}
	\end{center}
    \caption{List of the relevant parameters of the four benchmarks used for the comparison between Scenarios I and II. }
    \label{tab:BMs}
\end{table}

In order to establish the extent of the above statement, we will compare the two stau mixing scenarios by considering two benchmarks belonging to Scenario I and two ones corresponding to Scenario II. The relevant parameters of these four benchmarks are listed in Table~\ref{tab:BMs}. Note that the benchmarks SI-7 and SII-47 have the same relevant mass spectrum whereas this is not the case for SI-20 and SII-82. However, we can justify the use of this pair for the sake of comparison as follows. First of all, for Scenario I, the considerable difference in the value of $m_{\widetilde{\tau}_2}$ is not relevant since only the light stau contributes to the process, and then a benchmark belonging to it with exactly the same value of $m_{\widetilde{\tau}_2}$ than the benchmark SII-82 would have exactly the same distributions as the SI-20. Second, the differences in $m_H$ (24 GeV) and $m_{\widetilde{\tau}_1}$ (32 GeV) are significantly smaller than the mass splitting present in SII-82 (231 GeV) and then will not affect the main conclusions arising from the comparison between the distributions.

\begin{figure}[H]
	\begin{center}
		\begin{tabular}{cc}
			\centering
			\hspace*{-3mm}
			\includegraphics[scale=0.5]{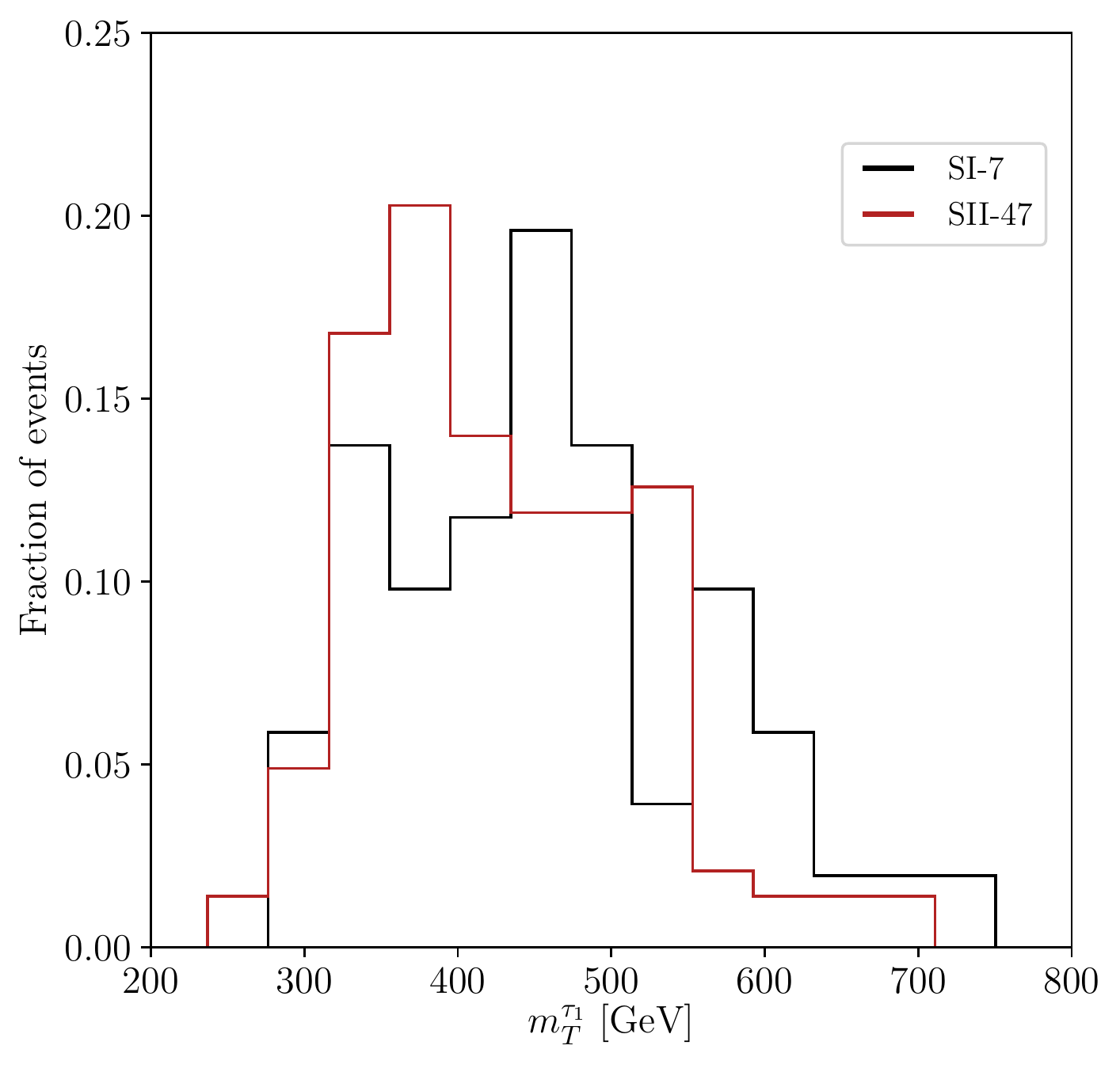} &
			\includegraphics[scale=0.5]{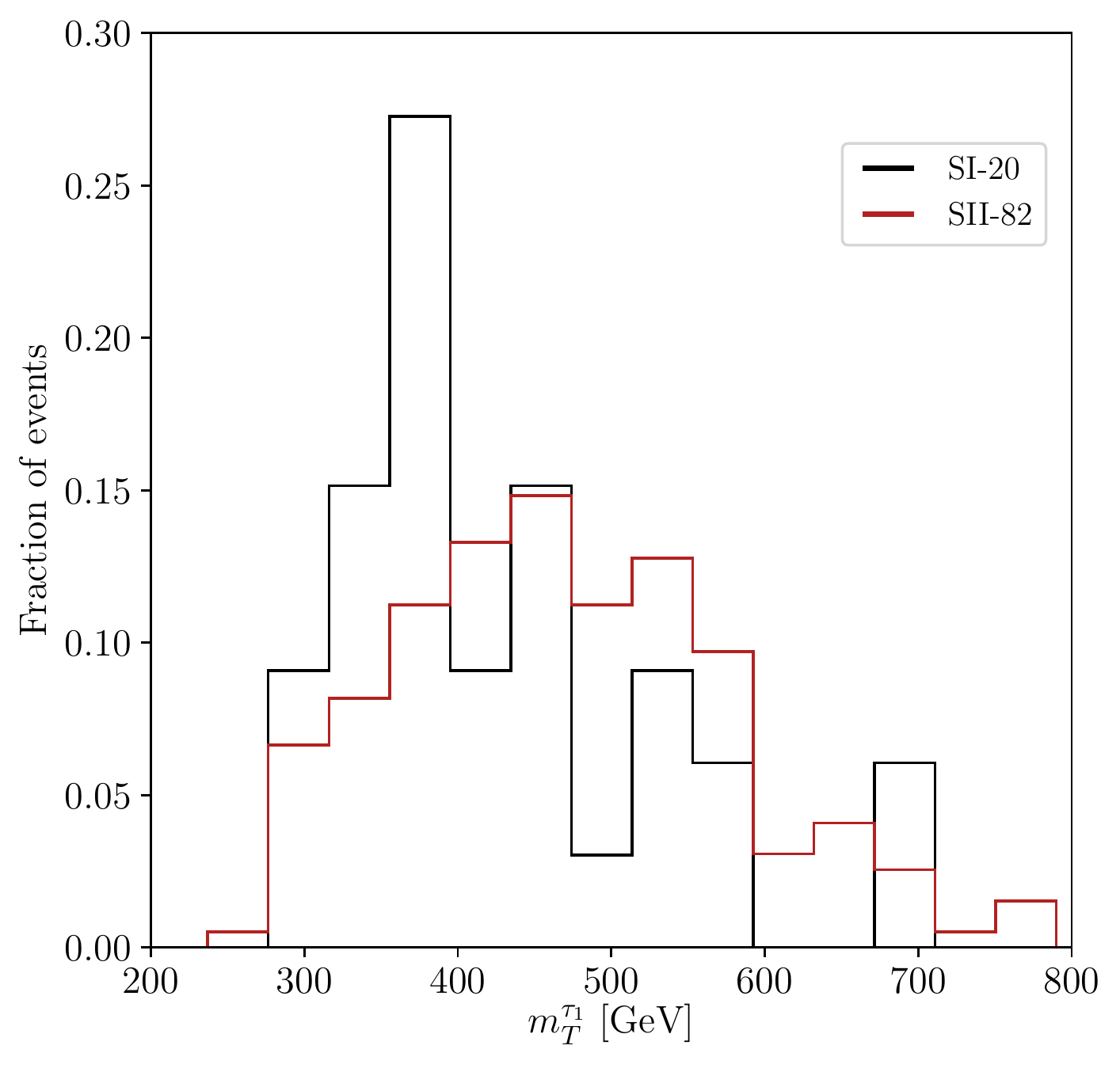} \\
			\includegraphics[scale=0.5]{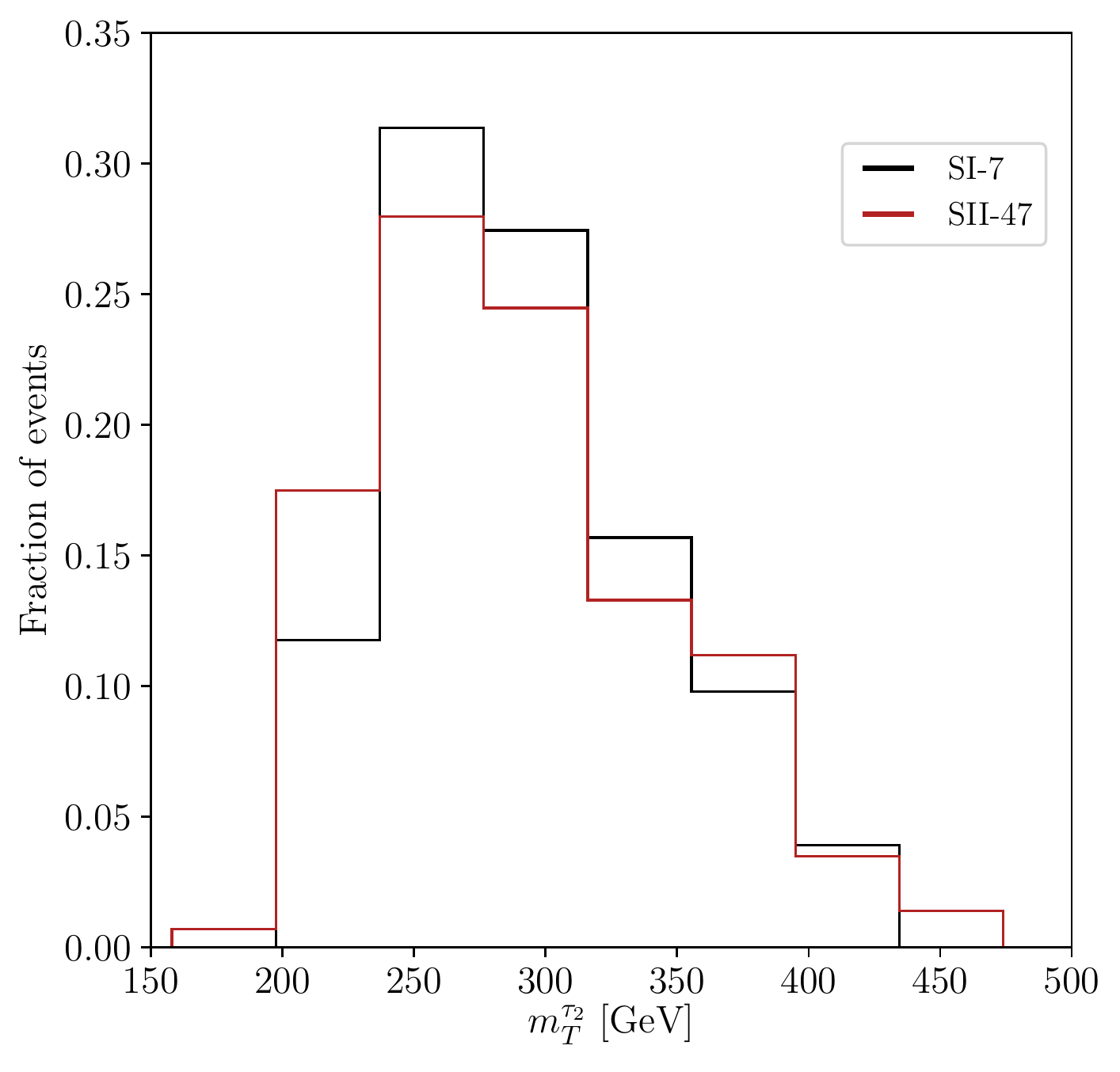} &
			\includegraphics[scale=0.5]{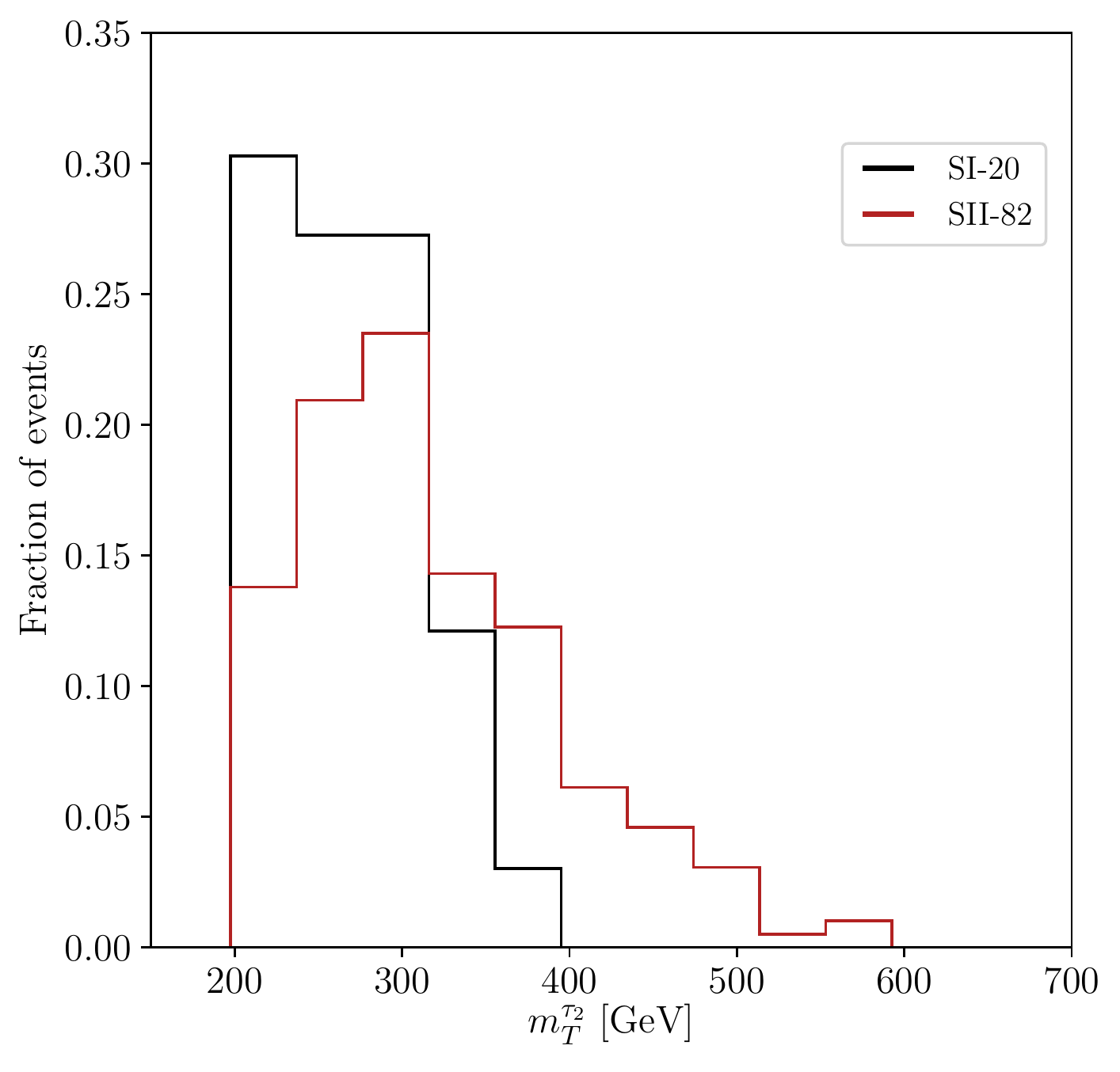} \\
			\includegraphics[scale=0.5]{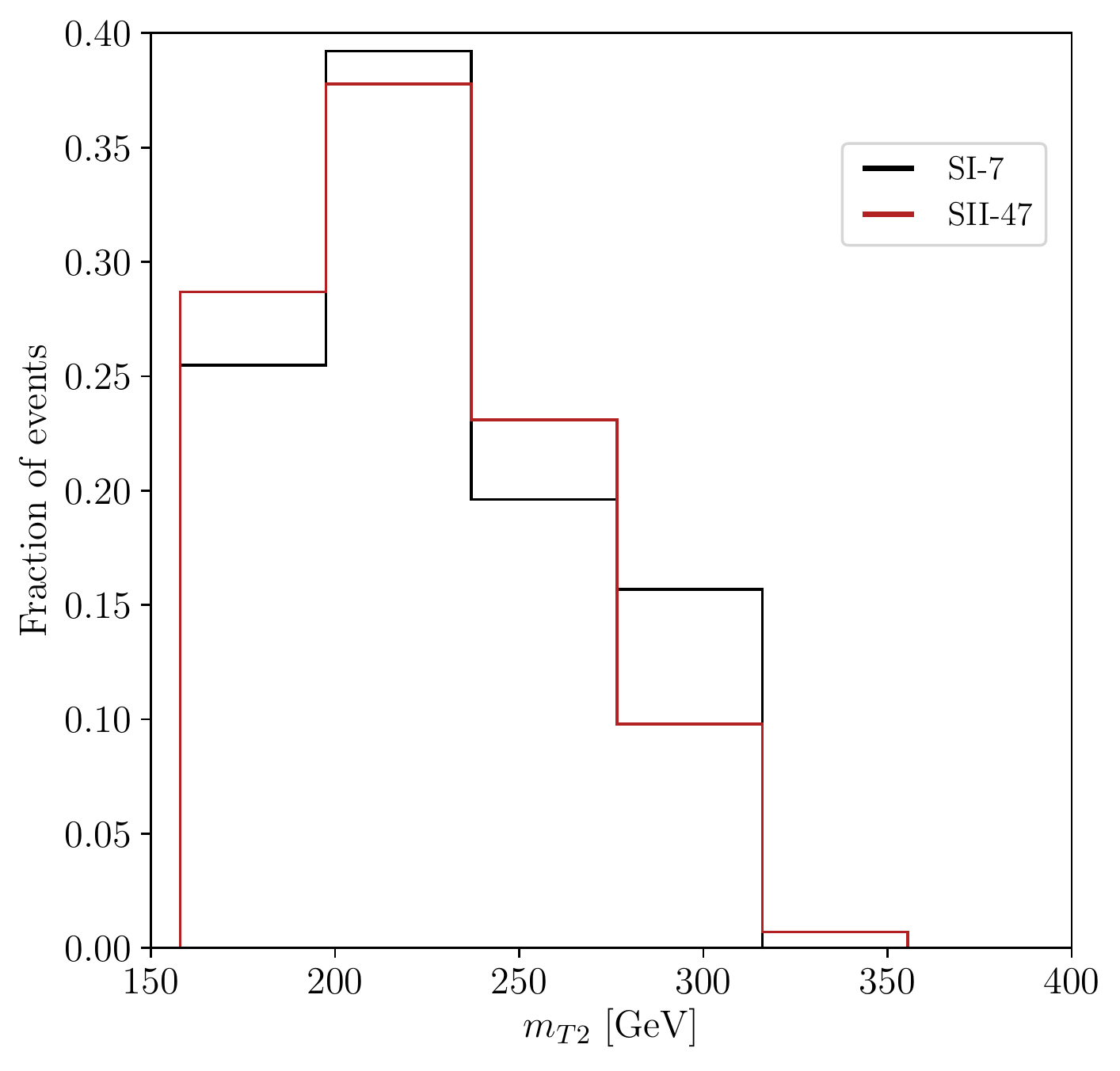} &
			\includegraphics[scale=0.5]{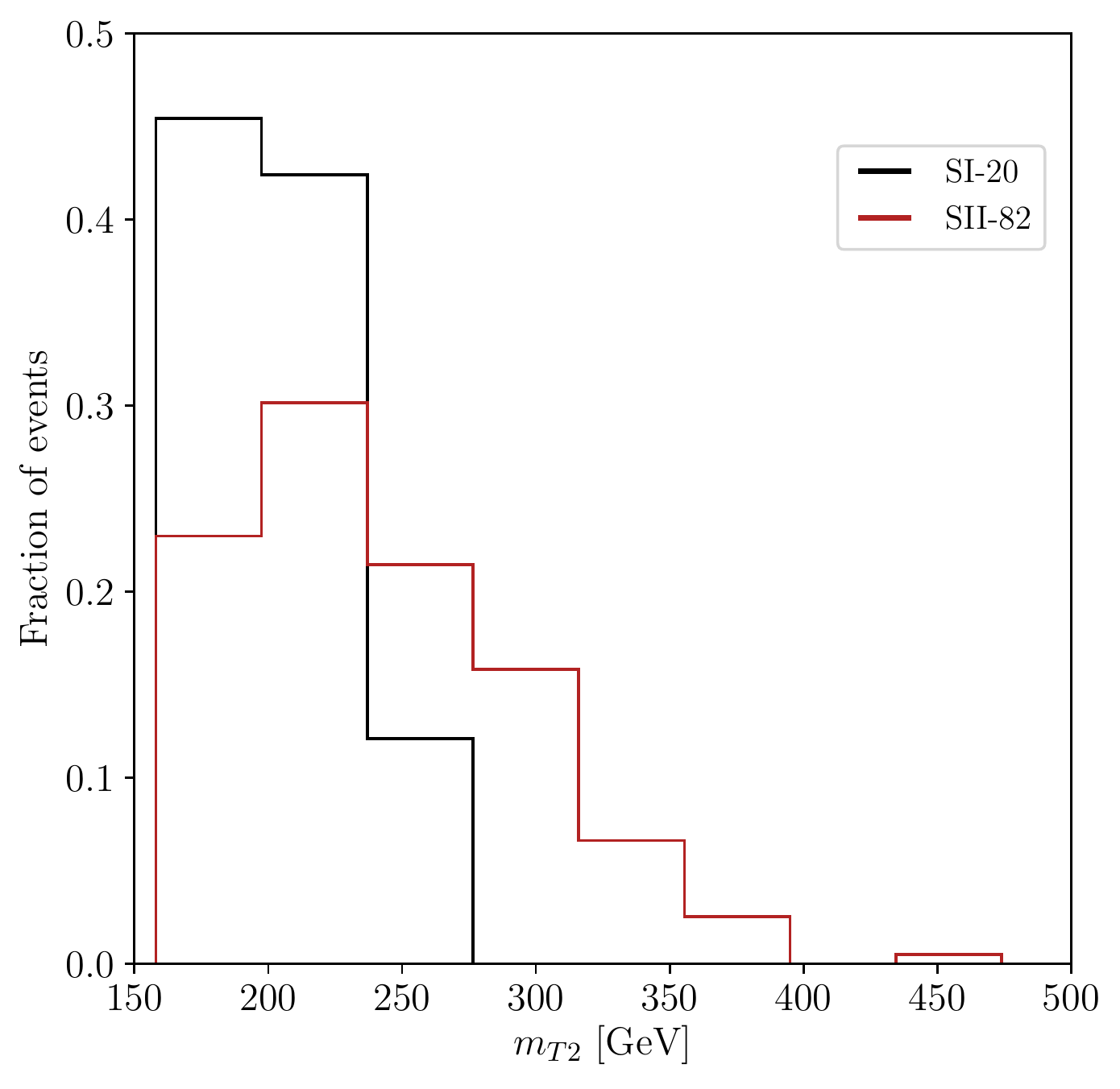} 
		\end{tabular}
		\caption{Distributions of the kinematic variables $m^{\tau_1}_{T},m^{\tau_2}_{T}$ and $m_{T2}$ (from top to bottom) after applying the cuts listed in Table~\ref{tab:cuts} for benchmarks SI-7 and SII-47 (left panels) and SI-20 and SII-82 (right panels).}
		\label{fig:comparison}
	\end{center}
\end{figure}

In Fig.~\ref{fig:comparison} we show the distributions corresponding to $m^{\tau_1}_{T}$, $m^{\tau_2}_{T}$, and $m_{T2}$ after applying the cuts of our search strategy (see Table ~\ref{tab:cuts}). On the left panels (right panels) we compare the distributions of the benchmarks SI-7 and SII-47 (SI-20 and SII-82). In the case of benchmarks SI-7 and SII-47, we see that out of the three considered variables\footnote{For the two comparisons between scenarios presented in this section we have also explored many other distributions of variables such as $E_T^\text{miss},|p_T^{\tau_2}/p_T^{\tau_1}|,m_{\tau\tau}$ or $\Delta R(\tau_1,\tau_2)$. Since none of these distributions has proven to be useful to discriminate between the stau mixing scenarios we do not include any results in this regard.} only the $m^{\tau_1}_{T}$ exhibits some sensitivity to the mixing pattern, with the peaks of the distributions of SI-7 and SII-47 shifted by approximately 80 GeV. The difficulty to distinguish these two benchmarks comes from the fact that the splitting between the stau masses in SII-47 ($\Delta m = 42$ GeV) is too small to produce traceable changes in distributions based on the tau leptons in the final state. The case of the benchmarks SI-20 and SII-82 is more promising since now the mass splitting is significantly higher ($\Delta m = 231$ GeV). In fact, as we can see from the right panels of Fig.~\ref{fig:comparison}, not only the $m^{\tau_1}_T$ distributions are shifted but also both the $m^{\tau_2}_T$ and $m_{T2}$ distributions present different endpoints according to the benchmark. The $m^{\tau_2}_T$ distribution for SI-20 has an endpoint in $\sim 400$ GeV, while for SII-82 the distribution extends until $\sim 600$ GeV. Thus, a cut such as $m^{\tau_2}_T>350$ GeV rejects the majority of SI-20 events while retaining a significant number of SII-82 events. The same behaviour occurs in the $m_{T2}$ distributions, with the endpoint being $\sim 275$ GeV for SI-20 and around $475$ GeV for SII-82. Again, we see that by means of requiring $m^{\tau_2}_T$ to be above 275 GeV, we are able to get rid off all the SI-20 events while still keeping a substantial amount of SII-82 events. As expected, we see that the higher the stau mass splitting the better the chance of distinguishing between mixing patterns through the inspection of kinematic distributions in the proposed signal region.

\section{Conclusions}
\label{sec:conclusions}

In this paper we have proven that our stau pair search strategy, developed in~\cite{Arganda:2018hdn} and applied to a type of MSSM scenario in the large-$\tan\beta$ regime with large stau mixing, dominated by decays of the heavy CP-even Higgs $H$ to a pair of lightest staus, $\widetilde{\tau}_1\widetilde{\tau}_1^{*}$ (Scenario I), is also very efficient in the complementary scenario in which decays of both the CP-even and CP-odd heavy Higgs contribute mainly to the production of $\widetilde{\tau}_1\widetilde{\tau}_2^{*} \;+\; c.c$ pairs (Scenario II), and focusing also on the stau decays that drive to final states  made up of a $\tau$-lepton pair and a large amount of missing transverse energy.

This search strategy, with a luminosity of $\mathcal{L}=100$ fb$^{-1}$, allows us to set exclusion limits at the 95\% C.L. for most of the [$m_H$, $m_{\stau_{1}}$] parameter space of Scenario I, if $m_H <$ 930 GeV and $m_{\stau_{1}} >$ 260 GeV. With a HL-LHC luminosity of 1000 fb$^{-1}$ the search strategy is able to exclude the whole Scenario-I area comprised by heavy-Higgs masses between 750 GeV and 1100 GeV and stau masses between 200 GeV and 450 GeV. On the other hand, for $\mathcal{L}=100$ fb$^{-1}$, we can reach signal significances at the evidence level if $m_H <$ 850 GeV regardless the value of the stau mass (within the considered range). If one requires discovery level significances, stau masses above 350 GeV are needed. For $\mathcal{L}=1000$ fb$^{-1}$, our analysis is sensitive to most of the considered area in the [$m_H$, $m_{\stau_{1}}$] plane, although not enough to reach the region with $m_H >$ 900 GeV and $m_{\stau_{1}} <$ 260 GeV.

With regard to Scenario II, the search strategy excludes the region with $m_A\leq 850$ GeV at the 95\% C.L with $\mathcal{L}=100$ fb$^{-1}$. This region extends slightly to $m_A\leq 870$ GeV for $\mathcal{L}=1000$ fb$^{-1}$. Above these masses, the search strategy sets 95\% C.L exclusion limits for average stau masses that are greater than 275-290 GeV (250-275 GeV) for $\mathcal{L}=100$ fb$^{-1}$ ($\mathcal{L}=1000$ fb$^{-1}$). Considering a luminosity of 100 fb$^{-1}$, significances at the discovery level are obtained for masses $m_A\leq$ 780 GeV regardless the value of $\avstau$, while for masses above 780 GeV the average stau mass needs to be larger than 300-320 GeV or even higher ($\sim 450$ GeV) when $m_A$ is above 1090 GeV. In the case of the HL-LHC with $\mathcal{L}=1000$ fb$^{-1}$, 5$\sigma$ significances can be reached for $m_A$ $<$ 815 GeV for any $\avstau$ within the range under study. In addition, larger masses can still give rise to discovery-level significances if $\avstau$ is approximately above 300-320 GeV.

Finally, under the assumption that the LHC will be able to discover staus by means of our search strategy, we have outlined the potential for discriminating between the two possible scenarios of stau mixing within the large-$\tan\beta$ regime, when they share the same relevant mass spectrum and both reach 5$\sigma$ significances with our search strategy. We have shown that kinematic cuts in variables sensitive to the stau mass splitting, such as  $m^{\tau_1}_{T}$, $m^{\tau_2}_{T}$, and $m_{T2}$, may be useful to discern which of these two types of MSSM scenario is realized in nature. By comparing two pairs of benchmarks, one with $\Delta m=42$ GeV and the other with $\Delta m=231$ GeV, we have illustrated the fact that the discrimination power of these variables depends essentially on how large the mass splitting is.

As a main conclusion, we can say that our search strategy is really efficient for the discovery or for the exclusion of heavy (scalar or pseudoscalar) Higgs bosons decaying into a stau pair. From a more general point of view, our collider analysis could be applied to any process at the LHC with the resonant production of a pair of charged scalars which decay into a tau lepton and a dark-matter candidate, resulting in final states with a $\tau$-lepton pair plus a large amount of $E_T^\text{miss}$.

\section*{Acknowledgments}
The work of EA is supported by the ``Atracci\'on de Talento'' program (Modalidad 1) of the Comunidad de Madrid (Spain) under the grant number 2019-T1/TIC-14019. This work has been also partially supported by CONICET and ANPCyT under projects PICT 2016-0164 (EA, AM, NM), PICT 2017-2751 (EA, AM, NM), PICT 2017-2765 (EA), and PICT 2017-0802 (AM). VML acknowledges support by the Deutsche Forschungsgemeinschaft (DFG, German Research Foundation) under Germany‘s Excellence Strategy – EXC 2121“Quantum Universe” – 390833306. VML thanks warmly IFLP in La Plata for kind hospitality hosting him during the completion of this work. 

%%%%%%%
%%%%%%%
%%%%%%%

\bibliographystyle{JHEP}
\bibliography{lit}
\end{document}